\documentclass[12pt,preprint]{./aastex}
\usepackage{epsfig}
\shorttitle{H$\alpha$ imaging of A2151}
\shortauthors{Cedr\'es et al.}
\begin{document}
\title{Star forming galaxies in the Hercules cluster: H$\alpha$ imaging of A2151}
\author{Bernab\'e Cedr\'es\altaffilmark{1}, Jorge Iglesias-P\'aramo\altaffilmark{1}, Jos\'e Manuel V\'{\i}lchez\altaffilmark{1}, Daniel Reverte\altaffilmark{2,1}, Vasiliki Petropoulou\altaffilmark{1} and Jonathan Hern\'{a}ndez-Fern\'{a}ndez\altaffilmark{1}} 
\email{bce@iaa.es, jiglesia@iaa.es, jvm@iaa.es, daniel.reverte@gtc.iac.es, vasiliki@iaa.es, jonatan@iaa.es}

\altaffiltext{1}{Instituto de Astrof\'{\i}sica de Andaluc\'{\i}a -- C.S.I.C., E-18008 Granada, Spain}
\altaffiltext{2}{GRANTECAN S.A., Centro de Astrof\'{\i}sica de La Palma, La Palma, Spain}

\begin{abstract}

This paper presents the first results of an H$\alpha$ imaging survey of galaxies in the central regions of the A2151 cluster. A total of 50 sources were detected in H$\alpha$, from which 41 were classified as secure members of the cluster and 2 as likely members based on spectroscopic and photometric redshift considerations. The remaining 7 galaxies were classified as background contaminants and thus excluded from our study on the H$\alpha$ properties of the cluster.
The morphologies of the 43 H$\alpha$ selected galaxies range from grand design spirals and interacting galaxies to blue compacts and tidal dwarfs or isolated extragalactic H{\sc ii} regions, spanning a range of magnitudes of $-21 \leq M_{B} \leq -12.5$~mag.
From these 43 galaxies, 7 have been classified as AGN candidates. 
These AGN candidates follow the $L$(H$\alpha$) vs. $M_{B}$ relationship of the normal galaxies implying that the emission associated with the nuclear engine has a rather secondary impact on the total H$\alpha$ emission of these galaxies.
 
A comparison with the clusters Coma and A1367 and a sample of field galaxies has shown the presence of cluster galaxies with $L$(H$\alpha$) lower than expected for their $M_{B}$, a consecuence of the cluster environment. This fact results in differences in the
 $L$(H$\alpha$) vs. $EW$(H$\alpha$) and $L$(H$\alpha$) distributions of the clusters with respect to the field, and in cluster to cluster variations of these quantities, which we propose are driven by a global cluster property as the total mass.
In addition, the cluster H$\alpha$ emitting galaxies tend to avoid the central regions of the clusters, again with different intensity depending on the cluster total mass.
For the particular case of A2151, we find that most H$\alpha$ emitting galaxies are located close to the regions with the higher galaxy density, offset from the main X-ray peak.
Overall, we conclude that both, the global cluster environment as well as the cluster merging history play a non negligible role in the integral star formation properties of clusters of galaxies.

\end{abstract}
\keywords{Galaxies: clusters}

\section{Introduction}

The Hercules supercluster is probably among the most massive structures in the local Universe (Chincarini et al. 1981; Barmby \& Huchra 1998). 
This supercluster is composed of three smaller clusters: A2151, A2147 and A2152, that are probably bound (Barmby \& Huchra 1998). 
Optical and X-ray studies suggest that A2151 could still be in the process of collapsing, given the lack of HI deficiency in the spiral population (Giovanelli \& Haynes 1985; Dickey 1997) and the irregular distribution of the hot ICM and its low X-ray flux (Magri et al. 1988; Huang \& Sarazin 1996).
On top of that A2151 has at least three distinct subclusters with varying dwarf to giant galaxy ratios (Bird et al. 1995; S\'{a}nchez-Janssen et al. 2005). 

The main aim of this paper is to study the population of galaxies with recent star formation in A2151, as traced by their H$\alpha$ luminosities. 
It has been largely discussed in the literature the effect of the environment on the H$\alpha$ emission of galaxies, the commonly accepted conclusion being that the total H$\alpha$ luminosity of galaxies is lowered in high density environments (e.g. Gavazzi et al. 2002; Dom\'{\i}nguez et al. 2002; Lewis et al. 2002; G\'{o}mez et al. 2003; Balogh et al. 2004; Tanaka et al. 2004; Rines et al. 2005). 
However, most of these studies suffer from two main drawbacks: (a) some of them use information on the H$\alpha$ emission of galaxies from fiber spectroscopy, as it is the case of those works based on SDSS and 2dF data. It is known that there is a bias on the total H$\alpha$ flux measured in fiber surveys for low redshift galaxies, imposed by the size of the fiber on the sky.
(b) in most cases the samples of galaxies were selected on the basis of their optical fluxes and miss star forming dwarf galaxies. 
The inclusion of these dwarf galaxies in the large surveys devoted to the study of the star formation in cluster galaxies imposes new challenges to the models of evolution of galaxies in dense environments (Reverte et al. 2007).

For these reasons, we stress the relevance of deep H$\alpha$ imaging surveys that allow to capture the totality of the H$\alpha$ emission of galaxies and to include all the H$\alpha$ emitting sources down to a given H$\alpha$ flux, disregarding their continuum emission. 
Our previous work in this direction (Iglesias-P\'{a}ramo et al. 2002) highlights the necessity of expanding the available databases of H$\alpha$ selected samples of clusters of galaxies in order to allow a proper comparison with field samples and to improve our understanding of the role played by the environment on the activity of star formation of galaxies.

In this paper we present the first results of an H$\alpha$ survey of the A2151 cluster and a comparison of the H$\alpha$ properties of similar galaxy samples in other clusters and in the local field. 
A total of 50 galaxies with H$\alpha$ emission were detected.
In section 2 we describe the observations. 
In section 3 we explain the data reduction procedure. Section 4 describes the selection of the sample. 
Section 5 shows the main results derived from the observations related to the H$\alpha$ properties of the sample galaxies and compared with other samples from the literature.
Finally, in section 6 we discuss the implications of these results and present the main conclusions.
Throughout the paper we assume a cosmology with $H_{0} = 73$~km~s$^{-1}$~Mpc$^{-1}$, $\Omega_{M} = 0.27$ and $\Omega_{\Lambda} = 0.73$.

\section{Observations}

The data were obtained during two observational runs. The first one was from 6 to 7 of May 2003 at the 256~cm Nordic Optical Telescope (NOT) located at the Roque de los Muchachos Observatory. 
The instrumentation used was the ALFOSC spectrograph in direct imaging mode with a redshifted H$\alpha$ filter centered at $\lambda_c=6803 {\rm \AA}$ and with FWHM of $\Delta\lambda=50 {\rm \AA}$ for the emission line (ON-band frames), and a Gunn-$r$ centered at $\lambda_c=6800 {\rm \AA}$ and with FWHM of $\Delta\lambda=1020 {\rm \AA}$ broadband filter for the continuum (OFF-band frames). 
The field of view was a circle of approximately 6~arcmin in diameter with a plate scale of 0.188~arcsec~pix$^{-1}$. In this run, five pointings were obtained, covering an area of about 141~arcmin$^2$. 

The second run was during 4 and 5 of June 2003 at the 420~cm William Herschel Telescope (WHT), located also at the Roque de los Muchachos Observatory. 
The instrumentation used was the Prime Focus (PF) Camera with two redshifted H$\alpha$ filters centered at $\lambda_c=6785$ and $6834 {\rm \AA}$, and with FWHM of $\Delta\lambda=51$ and $58 {\rm \AA}$ respectively for the emission line (ON-band frames), and a Harris-$R$ centered at $\lambda_c=6408 {\rm \AA}$ with a FWHM of $\Delta\lambda=1562 {\rm \AA}$ broadband filter for the continuum (OFF-band frames). 
The nominal field of view was 16.2$\times$16.2~arcmin$^{2}$ with a plate scale of 0.24 arcsec~pix$^{-1}$. Unfortunately, for the narrowband filters, the convergence of the PF beam reduced the useful field of view to approximately 33~arcmin$^2$. During this run 11 pointings were obtained, covering a total area of about 363~arcmin$^2$. 
In Table~\ref{fil} we present a log of the pointings, the filters employed at each telescope and the corresponding exposure times of the frames. 
The average seeing measured for all the frames varied between 1''--1.3''.

The total coverage of our sample is 0.15~deg$^{2}$ (equivalent to about 7\% of the total Abell area assuming an Abell radius for A2151 of 0.82 deg from Solanes, 2001) including the areas of higher density of galaxies.

\section{Data reduction \label{data_red}}

The data were reduced employing the IRAF package. All images were bias subtracted and flat-field corrected employing twilight sky flats. For the majority of the galaxies detected, three or more exposures were obtained. Then, these images were combined employing the IRAF task IMCOMBINE with the median operation. This was done in order to increase the signal to noise ratio and to wipe out the cosmic rays.
For few galaxies with less than three exposures, the cosmic rays were removed employing the IRAF task CREDIT.

Net H$\alpha$ frames were produced by subtracting the OFF-band from the ON-band frames. For this, all the OFF-band frames were scaled to the ON-band frames under the assumption that the resulting average net flux of field stars is null.
Given that some stellar spectra may show Balmer absorption lines or other features that could affect the scaling procedure, a minimum of ten field stars per frame were used.

ON-band exposures of several spectrophotometric standard stars from Oke (1990) were obtained each night in both observational runs in order to flux calibrate our data. The flux calibration was carried out following Barth et al. (1994). 
The fluxes were corrected for Galactic extinction using the reddening maps of Schlegel et al. (1998) and the extinction law of Cardelli et al. (1989).

\begin{figure}
\includegraphics[width=16cm]{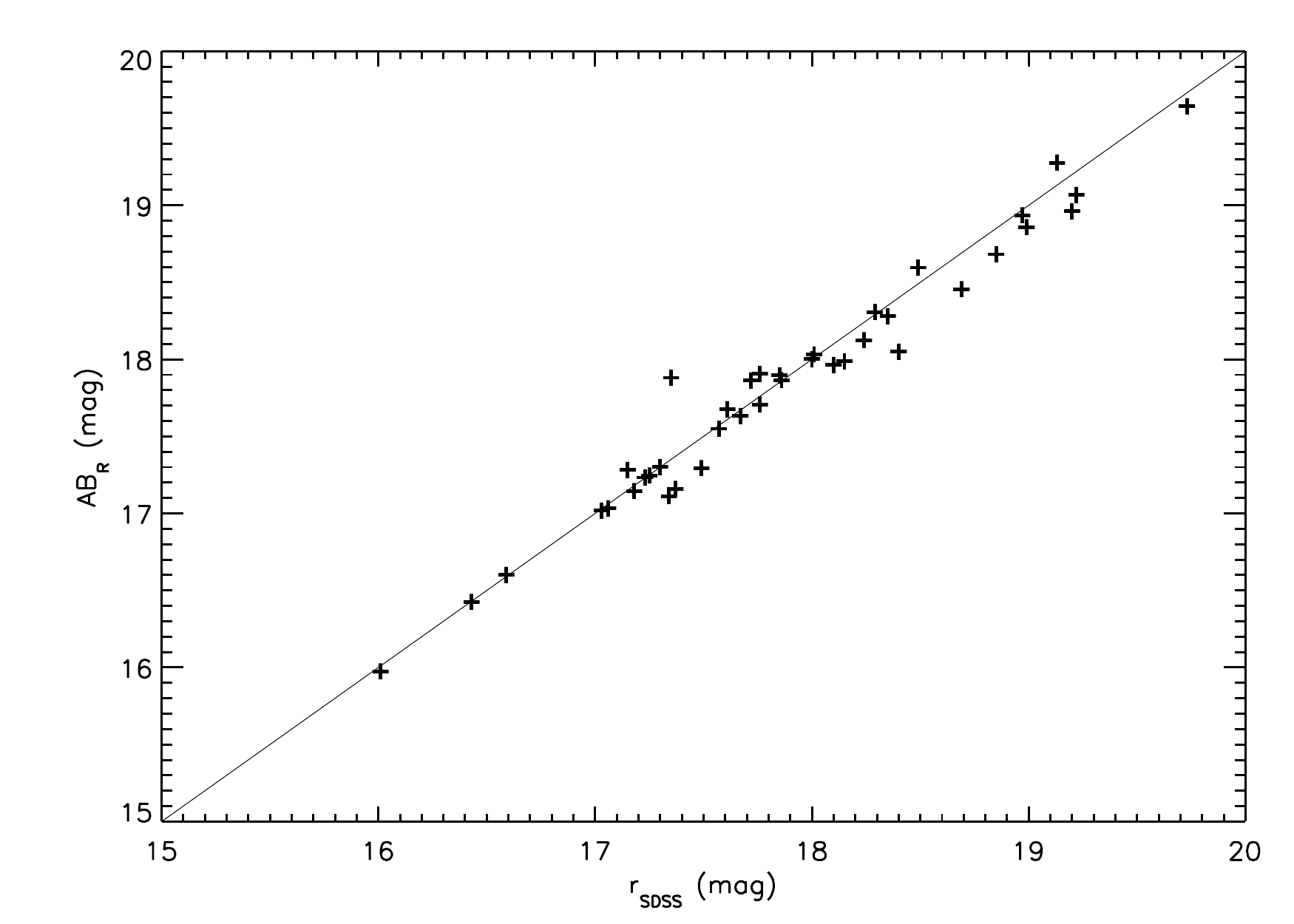}
\caption{SDSS-$r'$ versus AB$_{R}$ magnitudes for a list of field stars in our frames. The solid line shows a 1:1 correspondence}
\label{cal}
\end{figure}

In order to verify the quality of our photometric calibration, we make use of the SDSS database and of the fact that the effective wavelengths of the broad band filters used in this work are close enough to each other and to the one corresponding to the SDSS-$r'$ filter\footnote{The effective wavelengths for our broad band filters are 6800\AA\ for the Gunn-$r$, 6390\AA\ for the Harris-$R$ and 6250\AA\ for the SDSS-$r'$ filters respectively.}. Thus, under the assumption of a nearly flat shape of the red continuum of the galaxies the magnitudes obtained with the three filters in a given photometric system are directly comparable. 
As a first steep we derive AB magnitudes\footnote{AB$_{\nu} = -2.5 \log f_{\nu} - 48.6$, where $\nu$ is the effective frequency of each of the broad band filters and $f_{\nu}$ is the flux density at this frequency in erg~s$^{-1}$~cm$^{-2}$~Hz$^{-1}$.} for a sample of non-saturated stars in our OFF-band frames and compare them to their SDSS-$r'$ magnitudes. 
We show in Figure~\ref{cal} the comparison between our AB magnitudes and the SDSS-$r'$ ones for a sample of stars present in our frames. 
The median of the difference between the two sets of data is about 0.03~mag, with a standard deviation of 0.1~mag. 
This result illustrates that the zero point of our calibration fairly agrees with that of SDSS.

The fluxes for each emitting galaxy were extracted employing the FOCAS package, that runs under IRAF. 
We applied this code to the NET H$\alpha$ frames and detected objects with flux larger than 3$\sigma$ over the sky background and within an area larger than 30~pix (approximately equivalent to the seeing disk). Given the typical values of the $\sigma$ of the sky background in our frames, the limiting surface brightness for detection in this work is approximately $4\times10^{-19}$~erg~cm$^{-2}$~s$^{-1}$~arcsec$^{-2}$. 

\section{Selection of the sample}

\begin{figure}
\includegraphics[width=16cm]{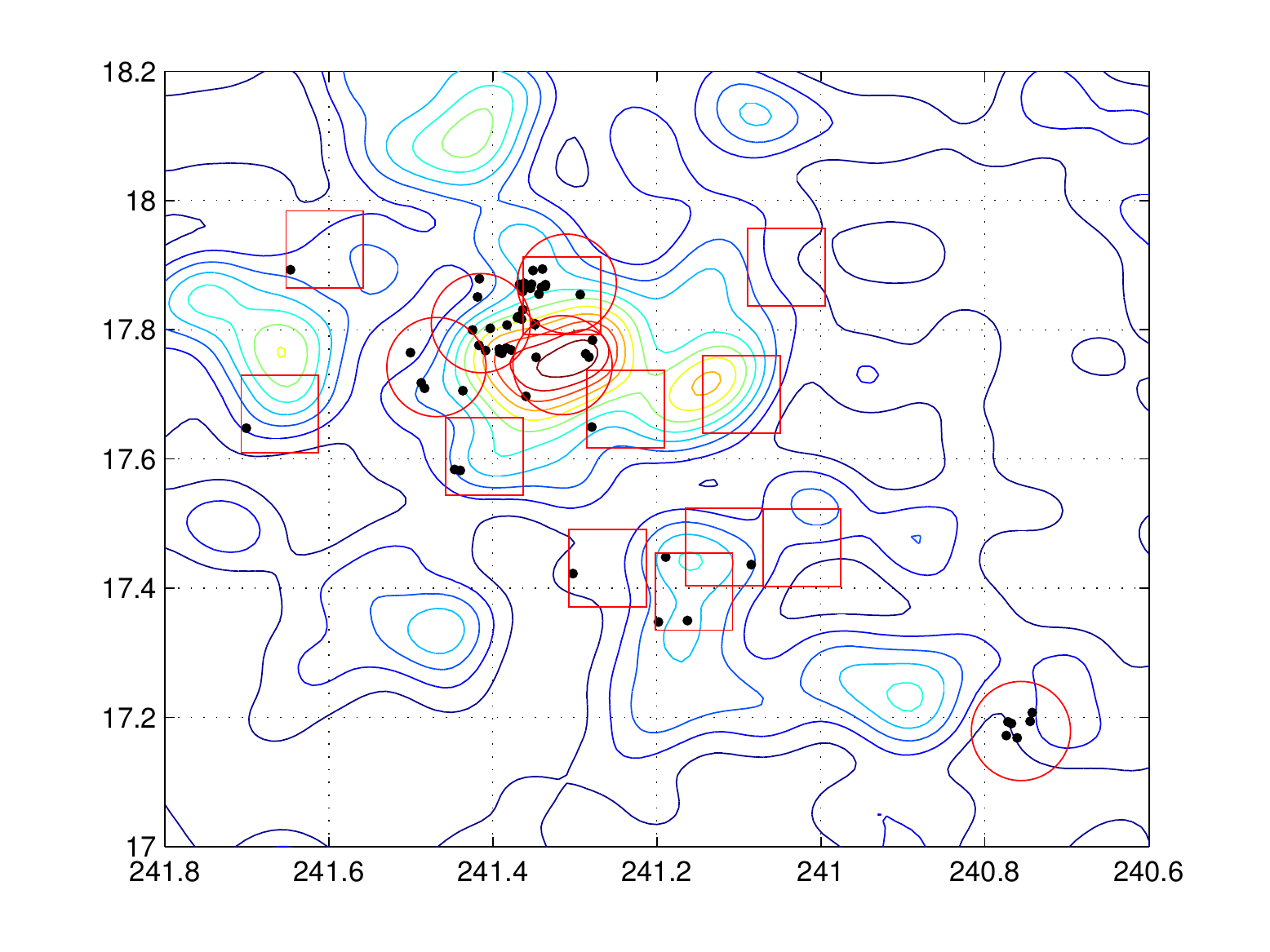}
\caption{Emission line galaxies detected in A2151 (black dots). The approximate position and size of the field of view of the instruments employed are the large circles (NOT) and the large squares (WHT). The contours indicate the density of galaxies and are extracted from 2MASS database.}
\label{espacial}
\end{figure}

This work is part of a more ambitious project devoted to the complete coverage in H$\alpha$ of the central parts of a sample of clusters representative of the local Universe. With this aim, we begun an H$\alpha$ imaging follow-up of galaxies in A2151 for which the radial velocities, extracted from NED, were consistent with the known velocity distribution of this cluster\footnote{We assume that a galaxy belongs to A2151 if its radial velocity ranges between $8900 \leq v_{r} \leq 13000$~km~s$^{-1}$, which corresponds to 3$\sigma$ around the systemic velocity of the cluster.}. 
During the realization of this work the database of radial velocities of galaxies in this cluster was enlarged, mainly due to the contribution of the SDSS. 
Thus, the H$\alpha$ flux could be measured for a larger number of galaxies present in our frames. 
This fact turned our pointed observations into an H$\alpha$ volume limited survey. 
The resulting total area covered by our observations corresponds to approximately 0.2 square degrees and is mostly concentrated in an area within the inner 15~arcmin (in radius) of the cluster, as it is shown in Figure~\ref{espacial}.

\begin{figure}
\includegraphics[width=16cm]{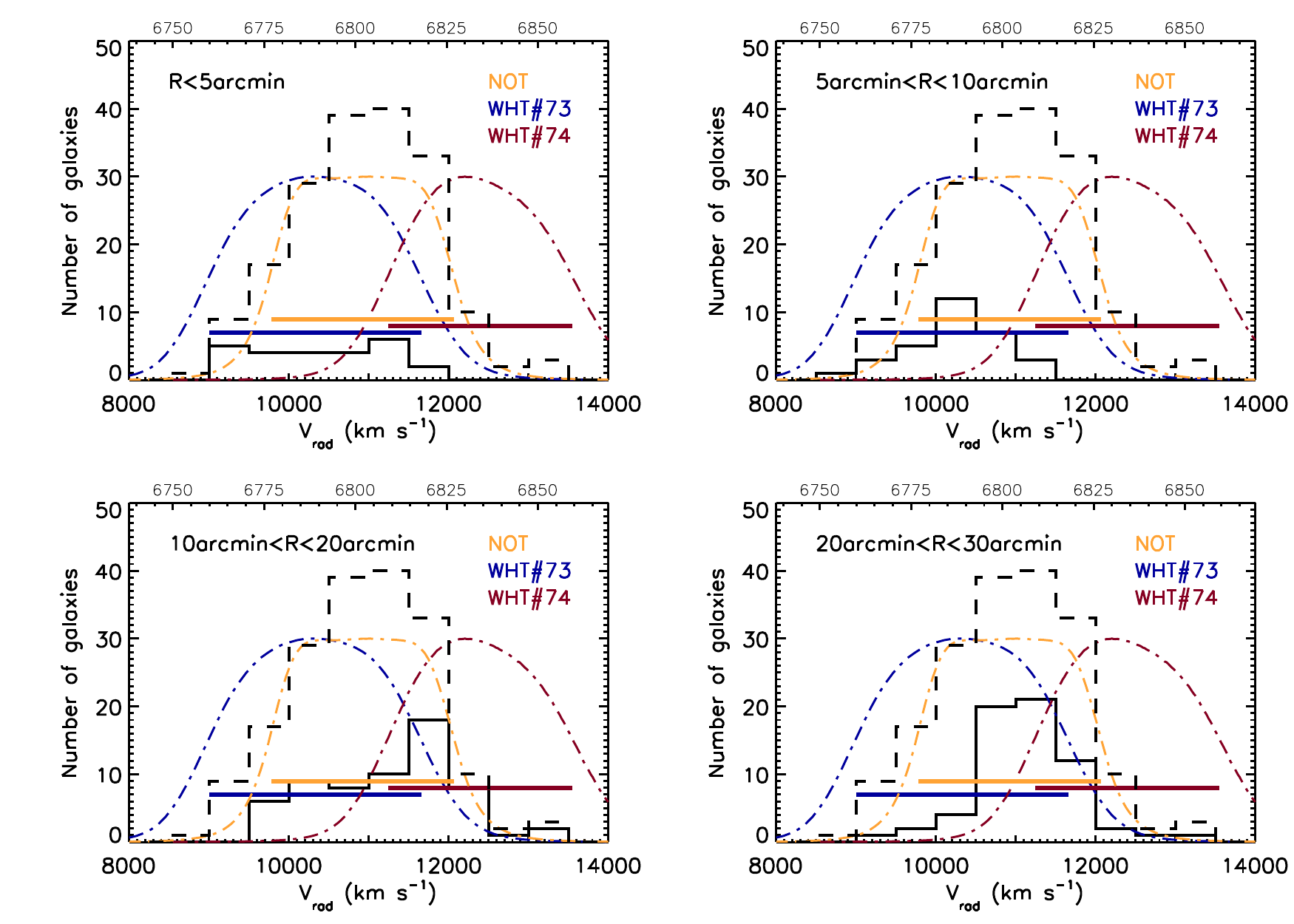}
\caption{Velocity distribution of the SDSS galaxies in the central regions of A2151. The solid histograms represent the SDSS galaxies. The dashed histogram corresponds to the whole ($R < 30~arcmin$) SDSS sample. The dot-dashed lines correspond to the transmittance profiles of the narrow band filters used in this work. 
The horizontal lines represent the interval where the transmittance is larger than 50\% of the peak transmitance for each filter. 
The upper X-axis indicates the wavelength (in \AA) of the redshifted H$\alpha$ line corresponding to the radial velocity indicated in the lower X-axis.}
\label{histo_velo}
\end{figure}

The H$\alpha$ flux can only be measured here for galaxies whose H$\alpha$ line is redshifted within the transmittance window of the narrow band filter. 
For this reason in each frame we initially restrict our sample to galaxies with available spectroscopic redshift and for which the H$\alpha$ line is shifted into the wavelength range where the transmittance of the filter is higher than 50\% of the peak transmitance, thus those galaxies detected in H$\alpha$ as described in the previous section are included in our sample. 
This restriction results in a degree of incompleteness of the sample of galaxies for which an H$\alpha$ flux was measured. 
In order to estimate this incompleteness effect for the central regions of A2151, we present in Figure~\ref{histo_velo} the velocity distribution for all the galaxies with measured radial velocity from spectroscopic redshift in SDSS within different regions around the center of A2151. 
The sample is split into four subsamples according to the distance to the center of the cluster. 
As it can be seen in Figure~\ref{histo_velo}, the distribution of galaxies in A2151 is broad and peaked around 10000~km~s$^{-1}$ for galaxies within the inner 5~arcmin of the center. In the annulus between 5 and 10~arcmin, the distribution gets narrower and still centered at 10000~km~s$^{-1}$. When we move to galaxies in the annulus between 10 and 20~arcmin, the velocity distribution becomes again broader and peaked at 12000~km~s$^{-1}$. Finally, in the annulus between 20 and 30~arcmin from the center of the cluster the velocity distribution of galaxies gets again narrower and centered at 11000~km~s$^{-1}$. 
This implies that for the fields observed at the NOT the H$\alpha$ flux of some cluster galaxies is lost because of our transmittance filter restriction. These galaxies (approximately 22\% of the total SDSS sample) are the ones populating the low velocity tail of the distribution corresponding to the inner 10~arcmin (in radius).
On the other hand, this problem is not present for the fields observed with the WHT because most of the cluster galaxies fulfill the transmittance filter criterion imposed above. 
Since most of the outer fields were observed with the WHT (see Figure~\ref{espacial}), 
we conservatively estimate the incompleteness of our H$\alpha$ sample to be less than 20\%.

In addition to these galaxies for which the radial velocity is known from spectroscopic observations, our A2151 sample also includes those sources detected in emission as explained in Section~\ref{data_red} but whose radial velocity is not known. 
The fact that these objects show up in our NET H$\alpha$ frames as positive detections suggest that they are very likely galaxies belonging to A2151. 
However, we cannot exclude the possibility that they are background objects for which we are detecting redshifted [O{\sc iii}], H$\beta$, [O{\sc ii}] or Ly$\alpha$ emission. 
One way to disentangle whether these are cluster or background (contaminant) objects is using the SDSS photometric redshift (photo-z), which is available for all of them. At the mean redshift of A2151 ($z \approx 0.037$), the emission line contaminants must be at $z \gtrsim 0.36$ (at this value of $z$, the [O{\sc iii}]5007\AA\ line shifts to 6800\AA, which corresponds to the wavelength of H$\alpha$ at the redshift of A2151). 
It is well known that the accuracy of the photo-z is much lower than that of the spec-z. In particular, the typical $\sigma$ declared for the SDSS photo-z is of the order of 0.05\footnote{See http://www.sdss.org/dr6/}, which is larger than the FWHM of our narrow band filters. However, given that the expected contamination comes from galaxies at redshifts $z \gtrsim 0.36$, we can use the SDSS photo-z information to identify the contaminants included in our sample.
For this, we make use of the three SDSS photo-z estimations available for most galaxies of our sample\footnote{From the SDSS tables {\em PhotoZ} and {\em PhotoZ2}. The reader can take a look for details on these photo-z estimations at http://www.sdss.org/dr6/}.
Given the already mentioned accuracy of the SDSS photo-z and the expected redshift of the contaminants, we assume that all galaxies of our sample with photo-z$\geq 0.1$ must be background contaminants. 
As for all except two of our galaxies (one with and the other without available spec-z) we have three different photo-z estimations, we will split our sample into four categories: 
(a) those galaxies with spectroscopic $z$ within 3$\sigma$ of the systemic velocity of the cluster, are flagged ``0'' and, by default, considered secure members of the cluster;
(b) those with no spectroscopic $z$ and at least two photo-z estimations consistent with photo-z $\leq 0.1$, are flagged ``1'' and considered secure members of the cluster; 
(c) those with no spectroscopic $z$ and just one photo-z estimation consistent with photo-z $\leq 0.1$, are flagged ``2'' and considered likely members of the cluster; and 
(d) those with no spectroscopic $z$ and three photo-z estimations consistent with photo-z $\geq 0.1$ are flagged ``3'' and considered contaminants. 
The only galaxy with no spectroscopic $z$ and with just one available photo-z estimation, was classified as background contaminant. 
In order to assess the robustness of our photo-z based criterion to identify the contaminants we tested it with the subsample of galaxies with available spec-z. The result was that all these galaxies were classified as secure members of the cluster (as it was found from their spectroscopic radial velocities).

\begin{figure}
\includegraphics[width=16cm]{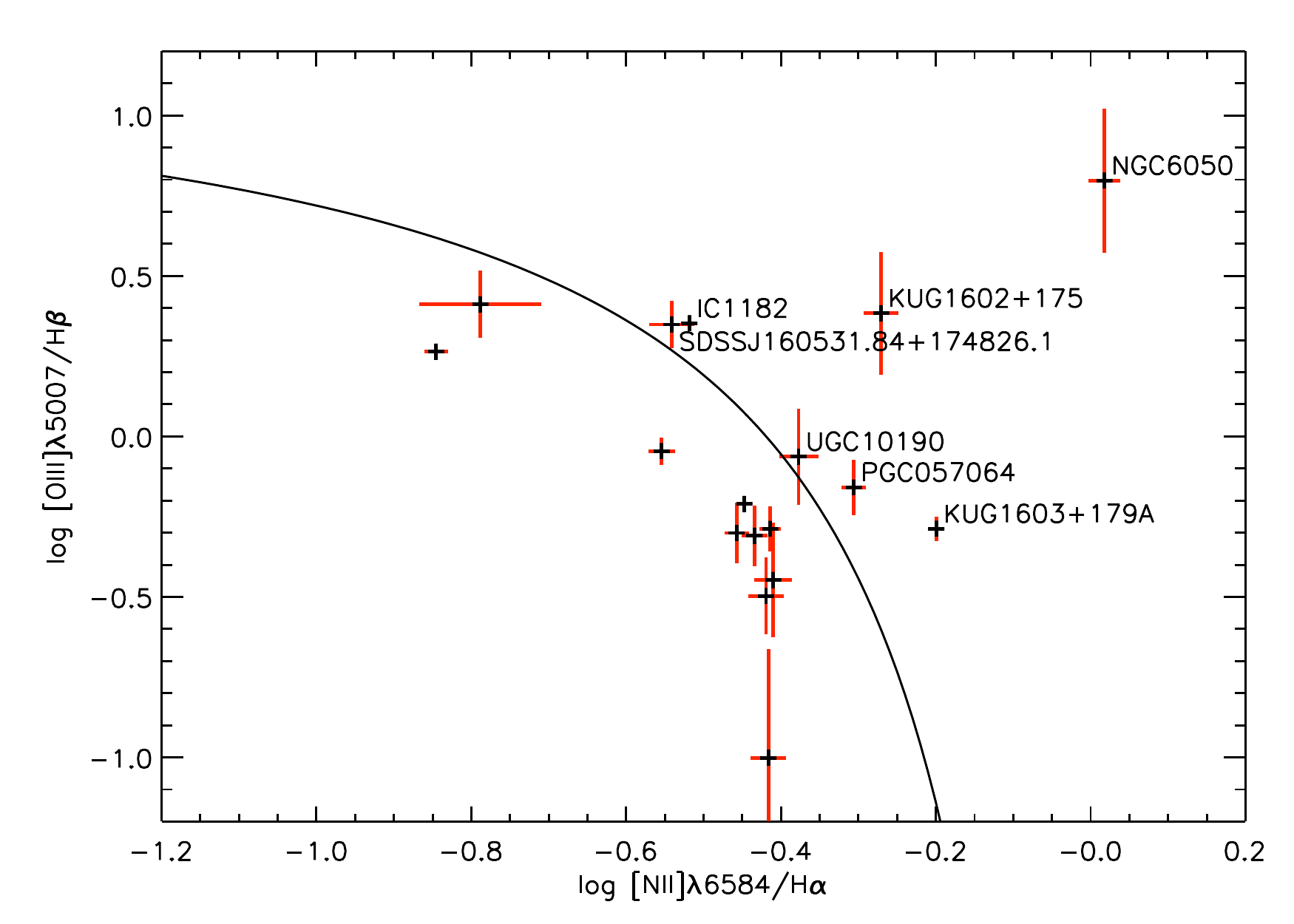}
\caption{Diagnostic diagram [N{\sc ii}]6584/H$\alpha$ vs. [O{\sc iii}]5007/H$\beta$ for the H$\alpha$ emitting galaxies with SDSS spectra. Only galaxies for which the four lines are detected in emission are plotted. The solid line corresponds to the separation between starbursts and AGNs of Kauffmann et al. (2003). The names of the seven galaxies located to the right of the line are indicated.}
\label{agn}
\end{figure}

The SDSS spectra allow us to look for the presence of AGNs in our sample. We show in Figure~\ref{agn} the classical diagnostic diagram [N{\sc ii}]6584/H$\alpha$ vs. [O{\sc iii}]5007/H$\beta$ (initially proposed by Baldwin, Phillips \& Terlevich 1981) for our galaxies with the line separating starbursts and AGNs proposed by Kauffmann et al. (2003). 
Only those galaxies of the sample for which these four emission lines were detected in SDSS are shown in the plot. 
A total of five galaxies lie clearly above the sequence of Kauffmann et al. -- KUG1602+175 (\#9), NGC6050 (\#21), IC1182 (\#39), PGC057064 (\#29) and KUG1603+179A (\#33) -- and two of them only marginally -- UGC10190 (\#27) and SDSSJ160531.84+174826.1 (\#34) --. 
Also, broad Balmer emission lines are apparent in IC1182 (\#39) and PGC057064 (\#29).
In addition IC1182 (\#39) and SDSSJ160531.84+174826.1 (\#34) are reported to be X-ray sources in NED, which could be indicative of the presence of an AGN. 
However, as shown by Rosa-Gonz\'{a}lez et al. (2007), the evidence of X-ray emission does not imply unequivocally the presence of an AGN since it may come from a recent stellar population. 
For this reason, we prefer not to include the detection in X-ray as one of our criteria to flag the AGN candidates in our sample.
IC1182 (\#39) was studied in detail by Moles et al. (2004) and they concluded that although the line ratio diagnostics are not conclusive about the nature of IC1182 (\#39), they discard the active nature of the nuclear emission of this galaxy based on other criteria.
SDSSJ160531.84+174826.1 (\#34) has been reported to host an intermediate-mass black hole, besides being a dwarf disk galaxy (Dong et al. 2007).
After these considerations, 
galaxies in our sample fulfilling any of the two criteria mentioned above are flagged as AGN candidates in Table~\ref{lista}.

Finally, our sample contains a total of 50 galaxies detected in H$\alpha$ whose main properties are shown in Table~\ref{lista}:
(1) Id. number; 
(2) Galaxy name from NED database or from SDSS; 
(3) Right Ascension (J2000.0); 
(4) Declination (J2000.0); 
(5) H$\alpha$~$+$~[N{\sc ii}] flux in units of 10$^{-15}$~erg~cm$^{-2}$~s$^{-1}$ uncorrected for Galactic extinction; 
(6) H$\alpha$~$+$~[N{\sc ii}] equivalent width in \AA; 
(7) Radial (spectroscopic) velocity in km~s$^{-1}$ from NED or SDSS, when available; 
(8) $M_{B}$ in mag, from the SDSS $g'$ magnitudes, assuming a distance to A2151 of 158.3~Mpc and an average correction of $g' - B \approx -0.3$ (Fukugita et al. 1995);
(9) Flag accounting for the cluster membership: ``0'' for galaxies with spectroscopic radial velocity, secure members of the cluster; ``1'' for secure members of the cluster from the photo-z criterion; ``2'' for likely members of the cluster from the photo-z criterion; ``3'' for contaminants from the photo-z criterion;
(10) Flag for AGN candidates;
(11) Hubble type from LEDA (those galaxies with no available Hubble type in LEDA have been classified as ``Comp''\footnote{In fact, these galaxies appear really compact in the optical frames. Their average effective radius, $r_{50}$ from the SDSS database, is almost 0.5 times compared to that of the total sample.} because of their compact appearance in the optical frames).

Figure~\ref{mosaic} shows the optical frames of the 50 H$\alpha$ selected galaxies in the same order as they appear listed in Table~\ref{lista}. We describe in some lines the most remarkable features regarding the H$\alpha$ emission properties of the sample: 
\begin{itemize}
\item The reported velocities for the galaxies NGC6050 (\#21) and UGC10190 (\#27) are near the detection limits for the FWHM of our filters, so it is possible that the asymmetric distribution of the H$\alpha$ emission of these galaxies could be an effect of their rotation velocity fields, that make the H$\alpha$ line lay outside the transmission curves of the filters.
\item PGC057077 (\#38) is classified in LEDA as E. However, its SDSS spectrum is typical of a star forming galaxy, with strong emission line features, the H$\alpha$ emission extending well over the whole galaxy.
\item Two of the galaxies are associated to tidal features -- [D97]ce-060 (\#43) and IC1182[S72]d (\#42) -- and because of their low luminosities they can be considered tidal dwarf candidates. 
\item Some galaxies can be associated to galaxy interactions because of their disturbed morphologies and their close projected distance to other cluster members: NGC6045 (\#13), SDSSJ160508.81+174545.3 (\#14), NGC6050 (\#21), SDSSJ160305.24+171136.1 (\#4), SDSSJ160304.20+171126.7 (\#5), PGC057064 (\#29), LEDA084724 (\#44),\\ SDSSJ160547.17+173501.6 (\#45).
\item Two objects could be isolated H{\sc ii} regions associated to brighter galaxies:\\ SDSSJ160528.84+174906.2B (\#31), SDSSJ160523.67+174828.8 (\#22).
\item Finally, there is a candidate of star formation induced by the ram pressure exerted by the IGM: LEDA 1543586 (\#15). The H$\alpha$ emission of this galaxy is clearly assymetric (see Figure~5) with most knots in the side facing the cluster center and almost no emission in the opposite side. This pattern of H$\alpha$ emission has been already observed in cluster galaxies (Gavazzi et al. 2001) and is associated with the ram pressure exerted by the intergalactic medium on the interstellar medium of disk galaxies.
\end{itemize}

A detailed investigation of the structural properties of the A2151 emission line galaxies will be presented in a forthcoming paper.

\section{Results}

\subsection{H$\alpha$ related properties of the A2151 galaxy sample}

In this section we discuss on the main statistical properties of the H$\alpha$ emission for our A2151 H$\alpha$ emitting galaxies. 
As it has been shown in the previous section, some of the selected emission line galaxies are very likely background contaminants. 
Given that we are interested in the properties of the cluster members, those galaxies considered contaminants (i.e. flagged ``3'' in Table~\ref{lista}) will be discarded in the subsequent analysis. 
Thus, hereafter we will refer to the subsample of secure (flags ``0'' and ``1'') and likely (flag ``2'') members of the cluster as the cluster galaxies.
For clarity, hereafter we will refer to H$\alpha +$[N{\sc ii}] as H$\alpha$, although we warn that no correction for [N{\sc ii}] contamination was performed for our sample galaxies (neither for the other samples used throughout this work).
However, if required for other purposes an efficient correction for the [N{\sc ii}] contamination can be succesfully performed (e.g. following Reverte (2008)).

Figure~\ref{lha_ewha} shows the H$\alpha$ luminosities vs. H$\alpha$ equivalent widths for our sample galaxies. 
The points appear spread in the plot occupying the locus limited by $38.5 \leq \log L($H$\alpha) \leq 42$~erg~s$^{-1}$ and $30 \leq EW($H$\alpha) \leq 200$~\AA. 
The minimum value of $EW$(H$\alpha$) measured for our sample is 2\AA\ (see also Table~2). We stress the importance of this point since this value has been used in the literature as the limit between active and passive star forming galaxies (Rines et al. 2005), thus meaning that our sample contains all the active star forming galaxies in the surveyed regions of A2151.
The candidate AGN galaxies are on average more H$\alpha$ luminous than the total sample, reflecting the low fraction of low luminosity AGNs with respect to high luminosity ones.

Figure~\ref{mb_lha} shows a linear relation between $M_{B}$ and $L$(H$\alpha$), similar to what has been found for Virgo cluster (see Figure~8 of Iglesias-P\'{a}ramo et al. 2002, with data taken from GOLDMine, Gavazzi et al. 2003). Again the candidate AGN galaxies follow the same trend as the rest.
The similar behavior of the candidate cluster AGNs and the rest of the galaxies suggests that their nuclear H$\alpha$ emission, resulting from the non thermal UV radiation, would be small compared to the emission of the gas ionized by the stellar UV radiation. 
This plot also gives an idea about the sensitivity of our survey. As it can be seen, we detect galaxies down to a value of $\log L$(H$\alpha$)/(erg~s$^{-1}$) $\approx$ 39 and $M_{B} \approx -14$~mag, implying that our H$\alpha$ selected sample contains not only the bright spirals of A2151 but also a non negligible fraction of star forming dwarf galaxies.

\subsection{Comparison with galaxies in different environments}

For the subsequent discussion we make use of H$\alpha$ fluxes and equivalent widths for other volume limited samples of galaxies, namely the samples of Coma and A1367\footnote{The list of H$\alpha$ emitters of A1367 in Iglesias-P\'{a}ramo et al. (2002) contains several galaxies for which no information of the redshift was available by that time. 
Given that photo-z information is currently available from the SDSS database, we have applied to those galaxies the same criterion of membership to the cluster as applied to our A2151 galaxies in this paper. 
For this reason, the sample of H$\alpha$ emitters of A1367 used in this paper is smaller than the one in Iglesias-P\'{a}ramo et al. 
Further details on this point will be given in a forthcoming paper.} (from Iglesias-P\'{a}ramo et al. 2002) and the H$\alpha$ survey of the local 11~Mpc volume (from Kennicutt et al. 2008, hereafter K08). 
The cluster samples will serve as a representation of other high density environments in the nearby Universe, and the K08 sample will be taken as a control field sample and it does not contain massive clusters since the two most massive structures of the very local Universe (i.e. the Virgo and Ursa Major clusters) have been excluded. 

Figure~\ref{lha_mb_all} shows the $L$(H$\alpha$) vs. $M_{B}$ for the four galaxy samples under consideration. 
The K08 galaxies, delineate an (almost linear) relatively tight relation with the scatter increasing at low luminosities. The cluster samples follow the same relation although showing a larger scatter.
This scatter is mainly due to the presence of cluster galaxies showing,  for a given $M_{B}$,  a lower $L$(H$\alpha$) than the one expected from the K08 sample. 
The fraction of such galaxies appears maximum for Coma and minimum for A2151. 
They are mostly bright galaxies, although at least one of them shows magnitude as faint as $M_{B} \simeq -17$~mag.
The K08 sample reaches luminosity values as low as $\log L($H$\alpha)$/(erg~s$^{-1}$)$= 36.5$, as shown in Figure~\ref{lha_mb_all}. The three cluster samples reach luminosities as low as  $\log L($H$\alpha)$/(erg~s$^{-1}$)$= 39$,
though few cluster dwarfs have been meassured with $L$(H$\alpha$) lower than this value. Therefore in order to make a robust comparison with K08 we limit our analysis to  $\log L($H$\alpha)$/(erg~s$^{-1}$)$\geq$ 39.
 
Figure~\ref{contour} shows the contours of the $L$(H$\alpha$) vs. $EW$(H$\alpha$) plot for the three clusters, overimposed to those of the K08 sample. 
As it can be seen, the K08 galaxies occupy a large region in the plot with a maximum between $39 \leq \log L($H$\alpha) \leq 40$~erg~s$^{-1}$ and $\log EW($H$\alpha) \approx 1.5$~\AA. 
In contrast, the cluster samples show different topologies in this figure. The distribution of Coma presents a well defined maximum, peaking at $EW$(H$\alpha$)$\approx$ 1.3\AA \ and $\log L$(H$\alpha$)/(erg~s$^{-1}$) $\approx$ 40.6.
This peak -- slightly shifted to higher $L$(H$\alpha$) -- can be also identified in the distributions of A1367 and A2151 as a relative maximum. In addition, these two clusters show central maxima near that of the K08 sample, a feature that is missing for Coma.
Besides the fact that K08 sample extends towards lower values of $L$(H$\alpha$), we remark the different topology of the distributions presented by the clusters with respect to the field sample within the range of $L$(H$\alpha$) and $EW$(H$\alpha$) common to the four samples. In the discussion, we argue that these different topologies in fact can reflect a direct imprint of the environment.

Figure~\ref{histo_lha} shows the histogram of the $L$(H$\alpha$) distributions of the four samples. 
The total histogram of the three clusters, made out of a combination of the three original histograms weighted each one by the inverse of the 
surveyed volume\footnote{For each cluster the surveyed volume was approximated by the area subtended in the sky by the survey times the Abell radius.}, 
is shown in the bottom panel (black dashed line) together with the histogram of the K08 sample (continuous line).
The histogram of Coma shows a broad maximum between $40 \leq \log L($H$\alpha)/$erg~s$^{-1} \leq 41$ whereas those of A1367 and A2151 show a plateau in this range and a peak at $39.5 \leq \log L($H$\alpha)/$erg~s$^{-1} \leq 40$. 
The histogram of the K08 sample shows a monotonic increase until it reaches a maximum at $\log L($H$\alpha)/$erg~s$^{-1} = 39.5$ and then it shows a decrease for lower luminosities. 
In order to compare the combined cluster distribution to that of the K08 sample, the cluster distribution has been normalized to the K08 in the range $41 \leq \log L($H$\alpha)/$erg~s$^{-1} \leq 42$. 
It can be seen that the cluster distribution shows a different shape from K08 in the range  $39 \leq \log L($H$\alpha)/$erg~s$^{-1} < 41$, reflecting a relative deficit of cluster star-forming galaxies with respect to the field sample. 
Taking a look to figure~\ref{lha_mb_all}, this range of $L$(H$\alpha$) is mainly populated by star-forming dwarfs and outliers from the K08 relation, these last being a likely analog of the spirals with weak H$\alpha$ emission detected in Virgo cluster by 
Koopmann \& Kenney (2004).

Figure~\ref{dist_cum} shows the cumulative distribution of the H$\alpha$ detections versus the distance to the center of the cluster for Coma, A1367 and A2151. 
The center of each cluster has been taken from NED and is coincident with the corresponding maximum in the galaxy distribution.
Galaxies are separated in three categories: (a) $EW$(H$\alpha$) $> 40$\AA; (b) $20$\AA\ $<$ $EW$(H$\alpha$) $< 40$\AA; and (c) $EW$(H$\alpha$) $<$ 20\AA ~(hereafter EW40, EW20 and EW0 galaxies respectively). 
We have restricted the samples to those galaxies inside the inner 1~Mpc of the center of each cluster since the observations cover reasonably well these regions for the three clusters.
As can be seen from the figure, the three clusters show very different behaviors:
For A2151, the three categories follow each other from 0.15~Mpc to 0.5~Mpc from the center. Beyond this distance there are no EW40 galaxies whereas EW20 and EW0 galaxies still can be found until distances of 1~Mpc.
For Coma, the inner 0.3~Mpc are almost completely devoid of star forming galaxies. Most of the EW40 and EW20 galaxies can be found at distances larger than 0.7~Mpc from the center. On the contrary, the EW0 galaxies are homogeneously distributed between 0.3~Mpc and 1~Mpc. 
For A1367 the distribution of galaxies is almost homogeneous from 0.15~Mpc to 1~Mpc for the three categories and it shows an intermediate behavior between Coma and A2151. 
Summarizing, our results show that the EW40 galaxies (those that most actively form stars) are centrally concentrated in A2151 whereas avoid the inner regions in Coma and A1367. 
Galaxies with intermediate star formation activity (the EW20 ones) tend to follow the EW40 ones in the three clusters, at least for radii shorter than $\lesssim 0.5$~Mpc. 
The EW0 galaxies are the most homogeneously distributed among the three categories for the three clusters.

\section{Discussion and conclusions}

In the previous section we have seen how the environment affects the H$\alpha$ emission of cluster galaxies with respect to those in less dense environments. The existence of cluster galaxies with low H$\alpha$ emission has already been reported in the literature 
(e.g. Kennicutt 1983; Koopmann \& Kenney 2004), and has been atrributed to mechanisms like galaxy interactions or ram pressure stripping among others (Boselli \& Gavazzi 2006, and references therein). In fact, it has been found a dependence of star formation rate on local density that holds even at distances will outside the virialized region of the cluster (e.g. Lewis et al. 2002; G\'{o}mez et al. 2003; Balogh et al. 2004; Rines et al. 2005) in the sense that the higher the local density of galaxies the lower fraction of star forming galaxies. This effect, known as the star formation - density relation seems to be independent of the global environment considered (cluster, loose group, compact group, etc). 
In addition, Poggianti et al. (2006) have found a relation between the fraction of emission line galaxies and the cluster mass within the virialized region. 

From a general point of view, it is not unexpected the existence of an imprint of the global cluster environment on properties related to the H$\alpha$ luminosity of cluster galaxies: in the previous section a relative deficit of cluster star-forming galaxies with 
respect to the field sample has been found. In addition, some cluster to cluster differences have been reported: we have shown in Figure~\ref{contour} that the well defined topology of K08 distribution, showing a well defined single maximum, is not reproduced by 
the cluster samples which show different topologies.
These results can be interpreted assuming a sequence parameterized by the total mass of the clusters: as the mass of the cluster increases and the intergalactic medium 
becomes denser and hotter, the ram pressure stripping  becomes more efficient at removing the gas component from galaxies. Since dwarf galaxies present smaller escape velocities, the ram pressure stripping is expected to be
more efficient for them than for giant galaxies (e.g. Mori \& Burkert 2000). 
The measured X-ray luminosities are 40.5$\times$10$^{44}$/4.5$\times$10$^{44}$/0.9$\times$10$^{44}$~erg~s$^{-1}$ for Coma/A1367/A2151 (Magri et al. 1988)\footnote{A value of H$_{0}$ = 50~km~s$^{-1}$~Mpc$^{-1}$ was assumed by these authors to estimate the 
X-ray luminosities.}, and the velocity dispersions are 1008/879/691~km~s$^{-1}$ for Coma/A1367/A2151 (Struble \& Rood 1999). According to this sequence, as cluster mass increases, an increasing fraction of low luminosity star-forming galaxies should be devoided 
of gas (Mori \& Burkert 2000; Boselli et al. 2008) and the ram pressure stripping can be observed in more luminous galaxies. Thus, for very massive clusters few (if any) low luminosity star-forming galaxies would be observed in H$\alpha$, and 
the H$\alpha$ emission of high luminosity galaxies would suffer from severe gas stripping. This scenario is qualitatively illustrated by Figure~\ref{lha_mb_all}, where in the case of Coma the outliers from the K08 relation are mainly bright (non dwarf) galaxies. 
For A1367 and A2151 these outliers shift towards lower luminosity. This rearrangement of the  H$\alpha$ emission in cluster galaxies is reflected also in the change of topolgy and distribution shown in Figures~\ref{contour} and \ref{histo_lha}, respectively.

The spatial distribution of star forming galaxies in the clusters shows a clear effect of the cluster environment on their recent star formation activity (Figure~\ref{dist_cum}): for Coma and A1367 the EW40 and EW20 galaxies avoid the central regions of the 
clusters and most of them are located beyond 0.5~Mpc from the cluster center, illustrating the existence of mechanisms quenching the star formation.  
However, this does not hold for A2151: in this cluster all the star forming galaxies are distributed in a similar way with clustercentric distance and most of them can be located at distances lower than 0.5~Mpc from the cluster center. 

Interestingly, for A2151 we can find most of the strong H$\alpha$ emitting galaxies (EW40) at very short distances from the cluster center, contrary to what is seen in the other two clusters.
Taking a look to the X-ray map (Magri et al. 1988, their Plate~6) and to the 2MASS and SDSS galaxy distributions of this cluster, we can see that the maximum of the cluster galaxy distribution is not coincident with the primary X-ray maximum, but with a secondary one located $\approx 10-15$~arcmin to the East of the primary maximum. This implies that most A2151 galaxies are not embedded in a hot X-ray intergalactic medium. 
In the case of Coma and A1367, despite they also show some degree of X-ray structure, the maxima of their galaxy distributions are located well inside the X-ray emitting regions.
The optical and X-ray structure of A2151 suggests that an evolved structure, coincident with the primary X-ray maximum, is incorporating a less evolved structure, coincident with the maximum of the galaxy distribution and the secondary X-ray maximum, thus
giving rise to a cluster merger event. Most H$\alpha$ emitters in A2151 are located around the secondary X-ray maximum, in the regions of higher density of A2151 galaxies. This last point gives support to the idea that cluster mergers could trigger episodes of star formation (Owen et al. 1999; Poggianti et al. 2004; Johnston-Hollitt et al. 2008).

Putting all these results together, we conclude that both, the global cluster environment as well as the cluster merging history play a non negligible role in the integral star formation properties of clusters of galaxies.
Further observations are required to confirm these points. 
In a near future a larger and homogeneous H$\alpha$ survey covering a larger and homogeneous sample of nearby clusters will help to shed light on these and other points regarding the recent star formation history of cluster galaxies.

\acknowledgements
We want to thank the anonymous referee for her/his interesting comments and suggestions that have contributed to the improvement of this manuscript. 
Thanks are given to P. Papaderos for his help and A. Boselli for many useful comments and suggestions in the initial stages of this work.
We acknowledge financial support by the Spanish Ministerio de Ciencia e Innovaci\'on grants AYA2007-67645-C03-02.
This research has made use of the NASA/IPAC Extragalactic Database (NED) which is operated by the Jet Propulsion Laboratory, California Institute of Technology, under contract with the National Aeronautics and Space Administration.
This publication makes use of data products from the Two Micron All Sky Survey, which is a joint project of the University of Massachusetts and the Infrared Processing and Analysis Center/California Institute of Technology, funded by the National Aeronautics and Space Administration and the National Science Foundation.
IRAF is distributed by the National Optical Astronomy Observatories, which are operated by the Association of Universities for Research in Astronomy, Inc., under cooperative agreement with the National Science Foundation.
The WHT is operated on the island of La Palma by the Isaac Newton Group in the Spanish Observatorio del Roque de los Muchachos of the Instituto de Astrof\'{\i}sica de Canarias. 
The data presented here have been taken using ALFOSC, which is owned by the Instituto de Astrof\'{\i}sica de Andaluc\'{\i}a (IAA) and operated at the Nordic Optical Telescope under agreement between IAA and the NBIfAFG of the Astronomical Observatory of Copenhagen.

\newpage

\begin{table}
\begin{tabular}{cccccc}
\hline
RA  & DEC & Telescope & H$\alpha$ filter & Cont. filter & Int. time \\
J2000 (hh mm ss) & J2000 (dd mm ss) & & $\lambda_c$ (\AA) & & for H$\alpha$ (s)\\
\hline
16 05 52.5 & 17 44 32.6 & NOT & 6803 & $r'$ (SDSS)  & 3$\times$1200 \\
16 05 13.6 & 17 45 55.1 & NOT & 6803 & $r$ (Gunn)   & 3$\times$1200 \\
16 05 14.5 & 17 52 14.9 & NOT & 6803 & $r$ (Gunn)   & 3$\times$1200 \\
16 05 39.6 & 17 47 15.7 & NOT & 6803 & $r$ (Gunn)   & 3$\times$1200 \\
16 03 01.5 & 17 10 44.4 & NOT & 6803 & $r$ (Gunn)   & 3$\times$1200 \\
16 05 15.9 & 17 51 11.9 & WHT & 6785 & $r'$ (SDSS)  & 3$\times$600 \\
16 04 57.1 & 17 40 39.0 & WHT & 6785 & $R$ (Harris) & 2$\times$600 \\
16 04 23.2 & 17 42 01.4 & WHT & 6785 & $R$ (Harris) & 3$\times$600 \\
16 04 10.2 & 17 53 50.6 & WHT & 6785 & $R$ (Harris) & 3$\times$600 \\
16 05 04.1 & 17 27 35.6 & WHT & 6785 & $R$ (Harris) & 3$\times$600 \\
16 04 56.3 & 17 25 52.7 & WHT & 6785 & $R$ (Harris) & 3$\times$600 \\
16 04 32.7 & 17 23 17.5 & WHT & 6785 & $R$ (Harris) & 3$\times$600 \\
16 05 33.9 & 17 36 14.8 & WHT & 6834 & $R$ (Harris) & 3$\times$600 \\
16 06 16.5 & 17 55 25.3 & WHT & 6785 & $r'$ (SDSS)  & 5$\times$600 \\
16 04 05.4 & 17 27 46.8 & WHT & 6785 & $r'$ (SDSS)  & 3$\times$600 \\
16 06 35.2 & 17 40 13.1 & WHT & 6785 & $r'$ (SDSS)  & 3$\times$600\\

\hline
\end{tabular}
\caption{Pointings of the A2151 survey:
(1) and (2) R.A. and Dec. of the approximate central position of the pointings;
(3) Telescope used to obtain the data: NOT (Nordic Optical Telescope), WHT (William Herschell Telescope);
(4) Central wavelength (\AA) of the filter used to measure the H$\alpha$ emission;
(5) Broadband filter employed to measure the continuum emission;
(7) Total exposure time of the H$\alpha$ frames.
}
\label{fil}
\end{table}

\newpage

\begin{deluxetable}{ccccccccccc}
\tabletypesize{\scriptsize}
\rotate
\tablewidth{210mm}
\tablecaption{ \scriptsize Detected galaxies with emission in our survey: 
(1) Identification number; 
(2) Galaxy name; 
(3) R.A.; 
(4) Dec.; 
(5) H$\alpha$+[N{\sc ii}] flux in units of $10^{-15}$~erg~cm$^{-2}$~s$^{-1}$ (uncorrected for Galactic extinction); 
(6) H$\alpha$+[N{\sc ii}] equivalent width in \AA; 
(7) Radial (spectroscopic) velocity in km~s$^{-1}$ from NED; 
(8) $M_{B}$ in mag, from the SDSS $g'$ magnitudes, assuming a distance to A2151 of 158.3~Mpc and an average correction of $g' - B \approx -0.3$ mag (Fukugita et al. 1995);
(9) Flag accounting for membership to the cluster: ``0'' and ``1'' for secure members, ``2'' for likely members, ``3'' for background galaxies; 
(10) Flag for AGN candidates; 
(11) Hubble type of the galaxies obtained from LEDA database. Galaxies with no available Hubble type are classified as ``Comp'' (compact).
\label{lista}}
\startdata
\hline
Id.\# & Galaxy & RA         & DEC        & $f$(H$\alpha$+[N{\sc ii}])         & EW$_{\alpha}$ & $V_{rad}$     & $M_{B}$ & \multicolumn{2}{c}{Flags} & Hubble \\
Number& Name   & J2000.0    & J2000.0    & ($10^{-15}$erg~cm$^{-2}$~s$^{-1}$) & (\AA)                 & (km~s$^{-1}$) & (mag) & Member & AGN & Type\\
      &        & (hh mm ss) & (dd mm ss) &                                    &                       &          &     &      &     &     \\
\hline
\#1 & SDSSJ160258.11+171227.8	& 16 02 58.1 & 17 12 27.9 & 0.42$\pm$0.03 	&   49$\pm$5	&  ---  & $-14.73$ & 3 & - & Comp\\
\#2 & SDSSJ160258.70+171138.1	& 16 02 58.7 & 17 11 38.2 & 0.20$\pm$0.02 	&   30$\pm$4	&  ---  & $-14.06$ & 2 & - & Comp\\
\#3 & SDSSJ160302.54+171007.0	& 16 03 02.5 & 17 10 07.1 & 0.18$\pm$0.07 	&   5$\pm$2	&  ---  & $-15.26$ & 3 & - & Comp\\
\#4 & SDSSJ160305.24+171136.1	& 16 03 05.2 & 17 11 36.2 & 1.3$\pm$0.5		&   4$\pm$1	& 9930  & $-18.22$ & 0 & - & Comp\\
\#5 & SDSSJ160304.20+171126.7	& 16 03 04.2 & 17 11 26.7 & 9.8$\pm$0.6		&   31$\pm$2	& 10830 & $-17.42$ & 0 & - & Comp\\
\#6 & LEDA084703		& 16 03 05.7 & 17 10 20.4 & 27$\pm$1		&   53$\pm$2	& 10093 & $-18.79$ & 0 & - & Sm\\
\#7 & LEDA084710        	& 16 04 20.4 & 17 26 11.2 & 37$\pm$2		&   23$\pm$6	& 10723 & $-19.23$ & 0 & - & Sb\\
\#8 & KUG1602+174A		& 16 04 39.0 & 17 20 59.9 & 18$\pm$3		&   11$\pm$2	& 10626 & $-19.24$ & 0 & - & Sb\\
\#9 & KUG1602+175		& 16 04 45.4 & 17 26 54.3 & 134$\pm$8		&   32$\pm$2	& 10638 & $-20.57$ & 0 & Y & Sc\\
\#10 & KUG1602+174B		& 16 04 47.6 & 17 20 52.0 & 56$\pm$5		&   22$\pm$2	& 10581 & $-19.70$ & 0 & - & Sb\\
\#11 & $[$D97$]$ce-200		& 16 05 06.8 & 17 47 02.0 & 4.6$\pm$0.3		&   29$\pm$2	& 9927  & $-17.74$ & 0 & - & S\\
\#12 & LEDA084719		& 16 05 07.1 & 17 38 57.8 & 15$\pm$3		&   8$\pm$1	& 10162 & $-19.20$ & 0 & - & Sbc\\
\#13 & NGC6045                 & 16 05 07.9 & 17 45 27.6 & 192$\pm$13	        &   22$\pm$2	& 9986  & $-21.04$ & 0 & - & SBc\\
\#14 & SDSSJ160508.81+174545.3	& 16 05 08.8 & 17 45 45.4 & 1.7$\pm$0.1		&   36$\pm$3	&  ---  & $-16.56$ & 1 & - & Comp\\
\#15 & LEDA1543586		& 16 05 10.5 & 17 51 16.1 & 8.7$\pm$0.6		&   30$\pm$2	& 9905  & $-18.46$ & 0 & - & Sc\\
\#16 & IC1173			& 16 05 12.6 & 17 25 22.3 & 98$\pm$10		&   14$\pm$2	& 10431 & $-20.79$ & 0 & - & SABc\\
\#17 & SDSSJ160520.64+175201.5	& 16 05 20.6 & 17 52 01.5 & 2.3$\pm$0.2		&   32$\pm$2	&  ---  & $-15.02$ & 2 & - & Comp\\
\#18 & SDSSJ160520.58+175210.6	& 16 05 20.6 & 17 52 10.7 & 2.4$\pm$0.2		&   20$\pm$2	& 10350 & $-18.51$ & 0 & - & Comp\\
\#19 & SDSSJ160521.50+175337.8	& 16 05 21.5 & 17 53 37.8 & 0.36$\pm$0.06 	&   12$\pm$2	&  ---  & $-16.03$ & 1 & - & Comp\\
\#20 & $[$D97$]$ce-143		& 16 05 21.7 & 17 51 56.3 & 1.4$\pm$0.2		&   11$\pm$2	& 11587 & $-17.56$ & 0 & - & Comp\\
\#21 & NGC6050			& 16 05 23.4 & 17 45 25.8 & 38.$\pm$7		&   6$\pm$1	& 9572  & $-20.87$ & 0 & Y & SABc\\
\#22 & SDSSJ160523.67+174828.8	& 16 05 23.7 & 17 48 28.9 & 0.6$\pm$0.2		&   29$\pm$9	&  ---  & $-14.58$ & 1 & - & Comp\\
\#23 & SDSSJ160523.66+174832.3	& 16 05 23.7 & 17 48 32.3 & 2.8$\pm$0.6		&   27$\pm$6	&  ---  & $-16.76$ & 1 & - & Comp\\
\#24 & SDSSJ160524.27+175329.3	& 16 05 24.3 & 17 53 29.4 & 1.19$\pm$0.08 	&   46$\pm$3	&  ---  & $-16.16$ & 1 & - & Comp\\
\#25 & SDSSJ160524.65+175213.0	& 16 05 24.7 & 17 52 13.0 & 0.42$\pm$0.07 	&   81$\pm$16	&  ---  & $-12.47$ & 3 & - & Comp\\
\#26 & $[$DKP87$]$160310.21+175956.7 & 16 05 25.0 & 17 51 50.6 & 3$\pm$1	&   22$\pm$6	&  ---  & $-17.41$ & 1 & - & Comp\\
\#27 & UGC10190  		& 16 05 26.3 & 17 41 48.6 & 14$\pm$3		&   8$\pm$1	& 11077 & $-19.21$ & 0 & Y & Sc\\
\#28 & SDSSJ160526.83+175218.6	& 16 05 26.8 & 17 52 18.7 & 1.5$\pm$0.3		&   36$\pm$8	&  ---  & $-14.95$ & 3 & - & Comp\\
\#29 & PGC057064         	& 16 05 27.2 & 17 49 51.6 & 50$\pm$12		&   21$\pm$7	& 10254 & $-18.65$ & 0 & Y & Sab\\
\#30 & SDSSJ160528.22+175212.3	& 16 05 28.2 & 17 52 12.3 & 0.7$\pm$0.3		&   15$\pm$7	&  ---  & $-15.08$ & 1 & - & Comp\\
\#31 & SDSSJ160528.84+174906.2B  & 16 05 28.6 & 17 49 13.7 & 0.51$\pm$0.06 	&   190$\pm$30  &  ---  & $-12.68$ & 1 & - & Comp\\
\#32 & SDSSJ160528.84+174906.2	& 16 05 28.8 & 17 49 06.2 & 1.8$\pm$0.9		&   11$\pm$6	&  ---  & $-17.09$ & 1 & - & Comp\\
\#33 & KUG1603+179A		& 16 05 30.6 & 17 46 07.2 & 98$\pm$4		&   54$\pm$2	& 11189 & $-20.05$ & 0 & Y & Sb\\
\#34 & SDSSJ160531.84+174826.1	& 16 05 31.8 & 17 48 26.2 & 3.2$\pm$0.2		&   26$\pm$2	& 9402  & $-17.23$ & 0 & Y & Comp\\
\#35 & SDSSJ160532.06+174617.8	& 16 05 32.1 & 17 46 17.9 & 0.8$\pm$0.1		&   12$\pm$2	&  ---  & $-16.51$ & 3 & - & Comp\\
\#36 & SDSSJ160533.36+174548.3	& 16 05 33.4 & 17 45 48.3 & 2.4$\pm$0.1		&   41$\pm$3	&  ---  & $-17.00$ & 1 & - & Comp\\
\#37 & SDSSJ160533.98+174554.1	& 16 05 34.0 & 17 45 54.1 & 1.0$\pm$0.1		&   15$\pm$2	&  ---  & $-16.93$ & 1 & - & Comp\\
\#38 & PGC057077 		& 16 05 34.2 & 17 46 11.8 & 87$\pm$2		&   148$\pm$5	& 10233 & $-19.11$ & 0 & - & E\\
\#39 & IC1182			& 16 05 36.8 & 17 48 07.5 & 375$\pm$11		&   111$\pm$3	& 10240 & $-20.98$ & 0 & Y & S0/a \\
\#40 & SDSSJ160539.93+175243.9	& 16 05 39.9 & 17 52 43.9 & 0.6$\pm$0.3		&   16$\pm$7	&  ---  & $-14.95$ & 3 & - & Comp\\
\#41 & SDSSJ160540.47+175102.3	& 16 05 40.5 & 17 51 02.4 & 0.3$\pm$0.2		&   18$\pm$8	&  ---  & $-14.34$ & 3 & - & Comp\\
\#42 & IC1182:[S72]d		& 16 05 41.9 & 17 47 58.4 & 21.4$\pm$0.6  	&   195$\pm$7	& 9983  & $-17.73$ & 0 & - & Comp\\
\#43 & $[$D97$]$ce-060		& 16 05 44.8 & 17 42 20.2 & 0.08$\pm$0.06 	&   2$\pm$2	& 11100 & $-16.87$ & 0 & - & Comp\\
\#44 & LEDA084724		& 16 05 45.5 & 17 34 56.7 & 30$\pm$1		&   30$\pm$4	& 12206 & $-18.72$ & 0 & - & Sbc\\
\#45 & SDSSJ160547.17+173501.6	& 16 05 47.2 & 17 35 01.7 & 1.0$\pm$0.3		&   6$\pm$2	&  ---  & $-16.62$ & 1 & - & Comp\\
\#46 & LEDA3085054		& 16 05 56.0 & 17 42 33.9 & 2.4$\pm$0.1		&   69$\pm$4	& 11145 & $-16.43$ & 0 & - & S\\
\#47 & SDSSJ160556.98+174304.1	& 16 05 57.0 & 17 43 04.1 & 4.6$\pm$0.2		&   60$\pm$3	&  ---  & $-17.09$ & 1 & - & Comp\\
\#48 & LEDA140568		& 16 06 00.1 & 17 45 52.0 & 1.1$\pm$0.1		&   15$\pm$2	& 11959 & $-17.42$ & 0 & - & S\\
\#59 & CGCG108-149		& 16 06 35.3 & 17 53 33.2 & 35$\pm$3		&   10$\pm$4	& 11049 & $-20.14$ & 0 & - & SABa\\
\#50 & PGC057185 		& 16 06 48.2 & 17 38 51.6 & 11$\pm$2		&   6$\pm$2	& 11246 & $-19.51$ & 0 & - & Sbc\\
\enddata
\end{deluxetable}

\newpage
\clearpage

\newpage
\clearpage

\newpage
\clearpage

\newpage
\clearpage

\newpage
\clearpage

\begin{figure}
\includegraphics[width=6cm]{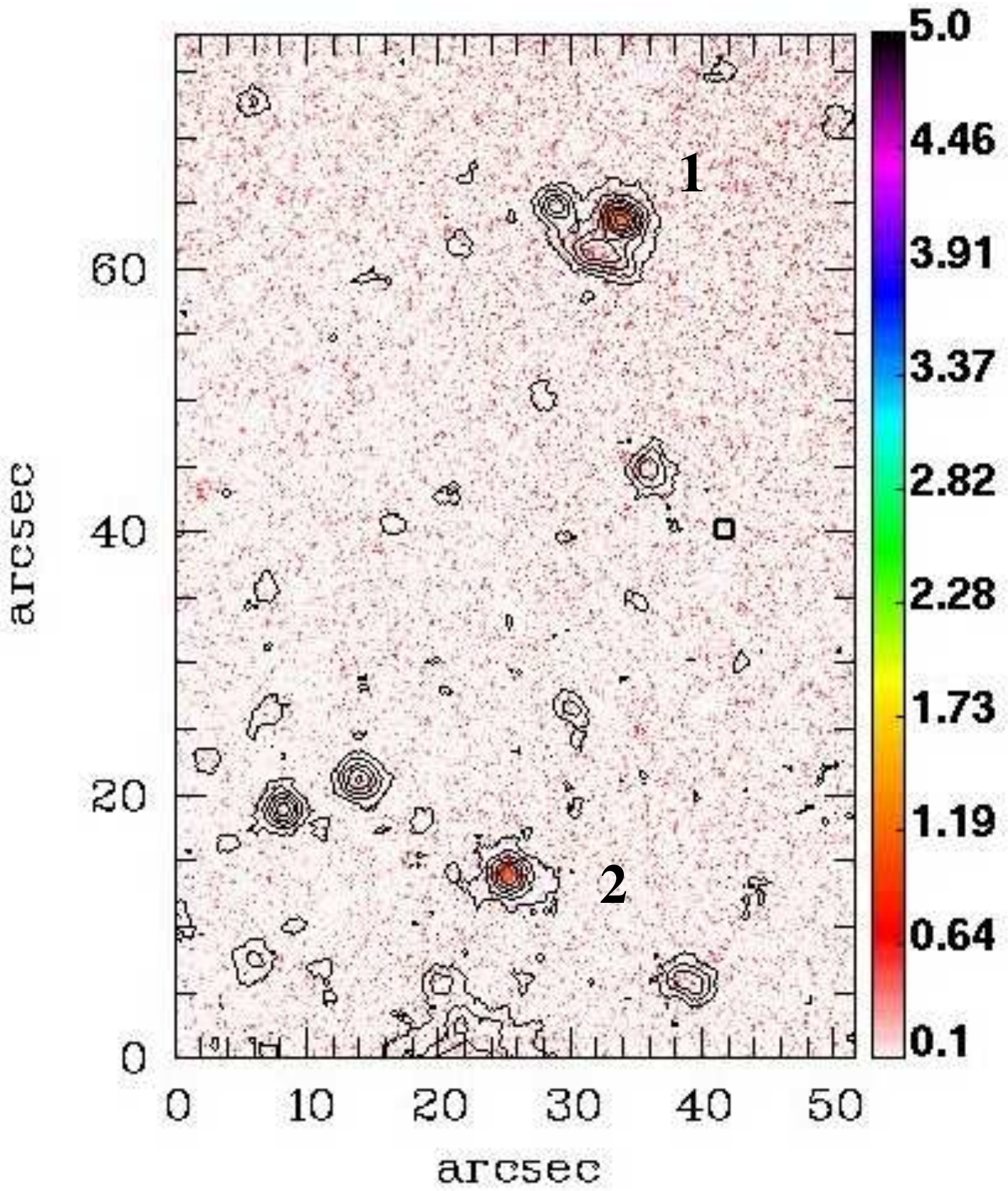}
\includegraphics[width=10cm]{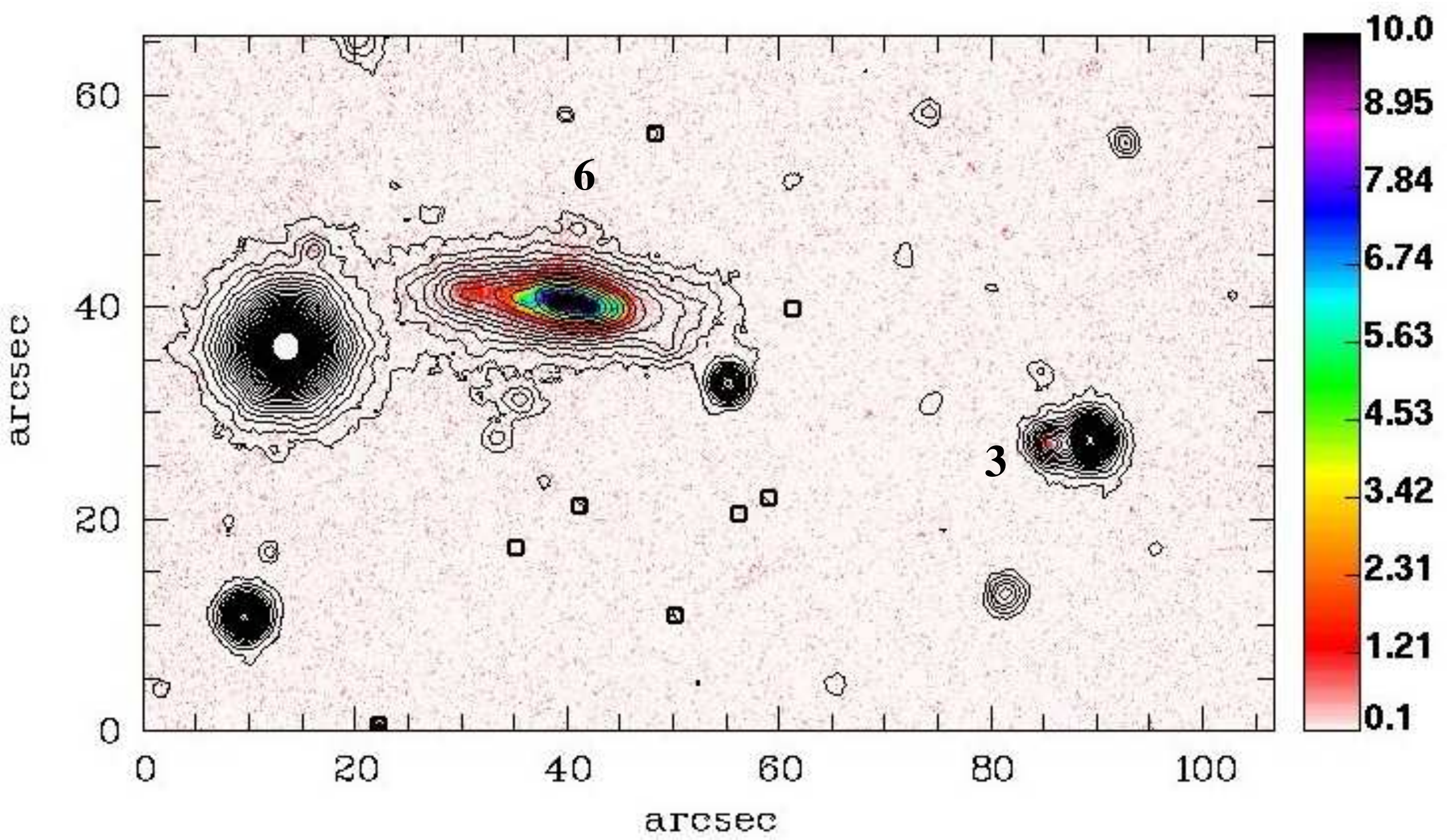}
\includegraphics[width=8cm]{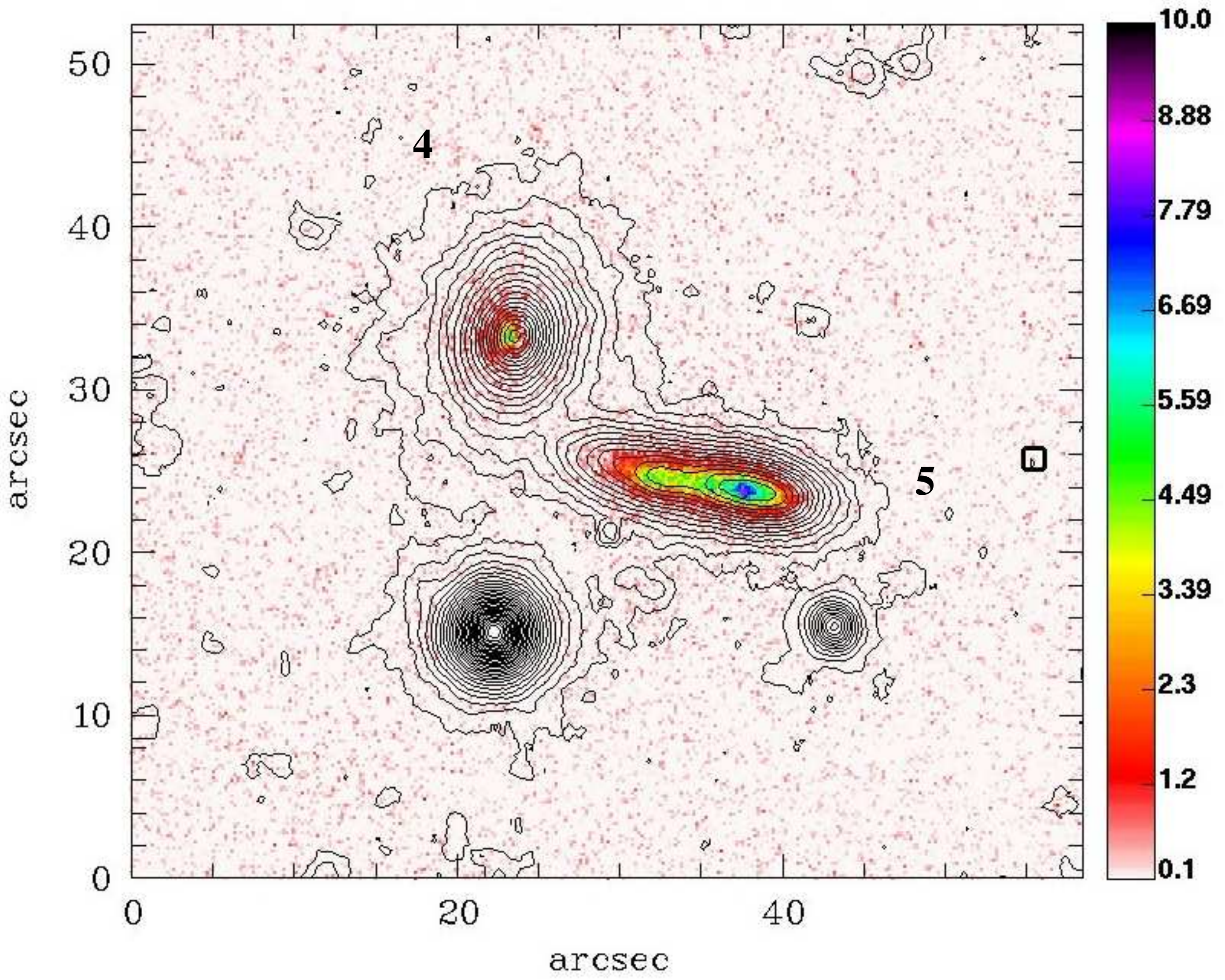}
\includegraphics[width=8cm]{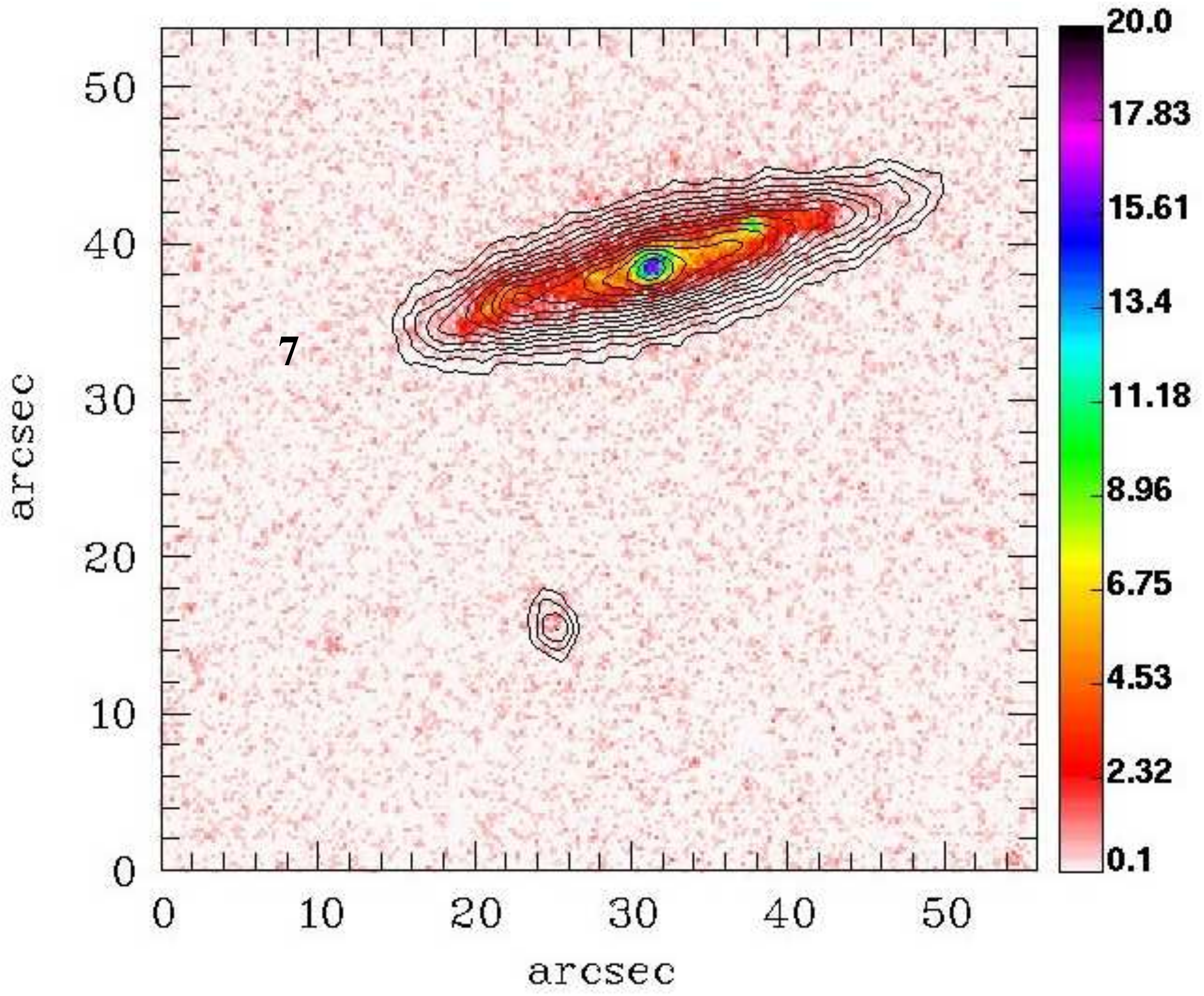}
\caption{H$\alpha$ frames of the A2151 emitting galaxies with the r contours overimposed. The H$\alpha$ surface brightness is indicated by the color scale to the right of each image in 10$^{-16}$erg/cm$^2$/s/arcsec$^2$. The superposed contours are in AB magnitudes arcsec$^{-2}$. The lowest contour is at 24 mag arcsec$^{-2}$ and the increment is 0.2 mag arcsec$^{-2}$}
\label{mosaic}
\end{figure}
\addtocounter{figure}{-1}

\newpage
\clearpage

\begin{figure}
\includegraphics[width=8cm]{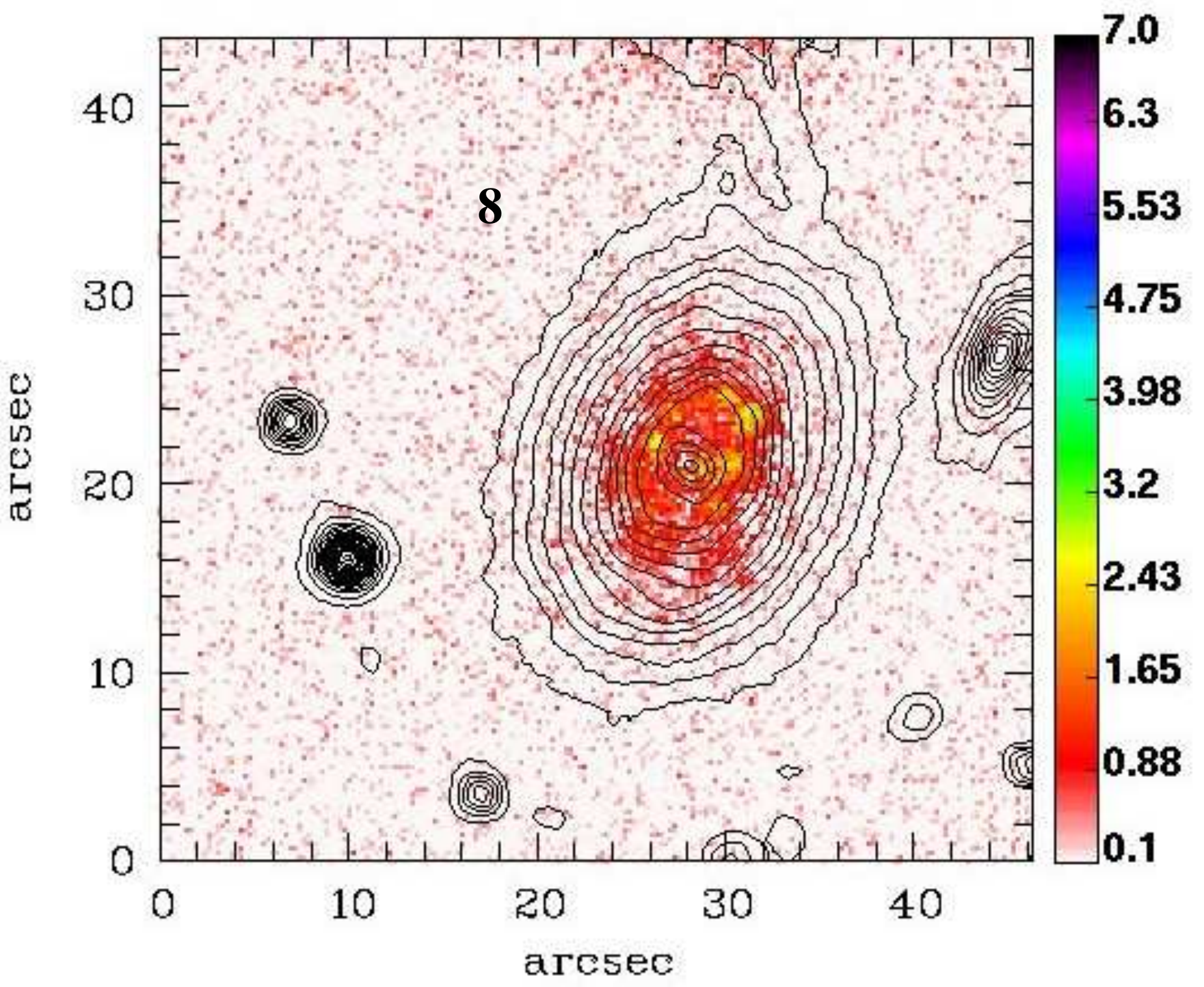}
\includegraphics[width=8cm]{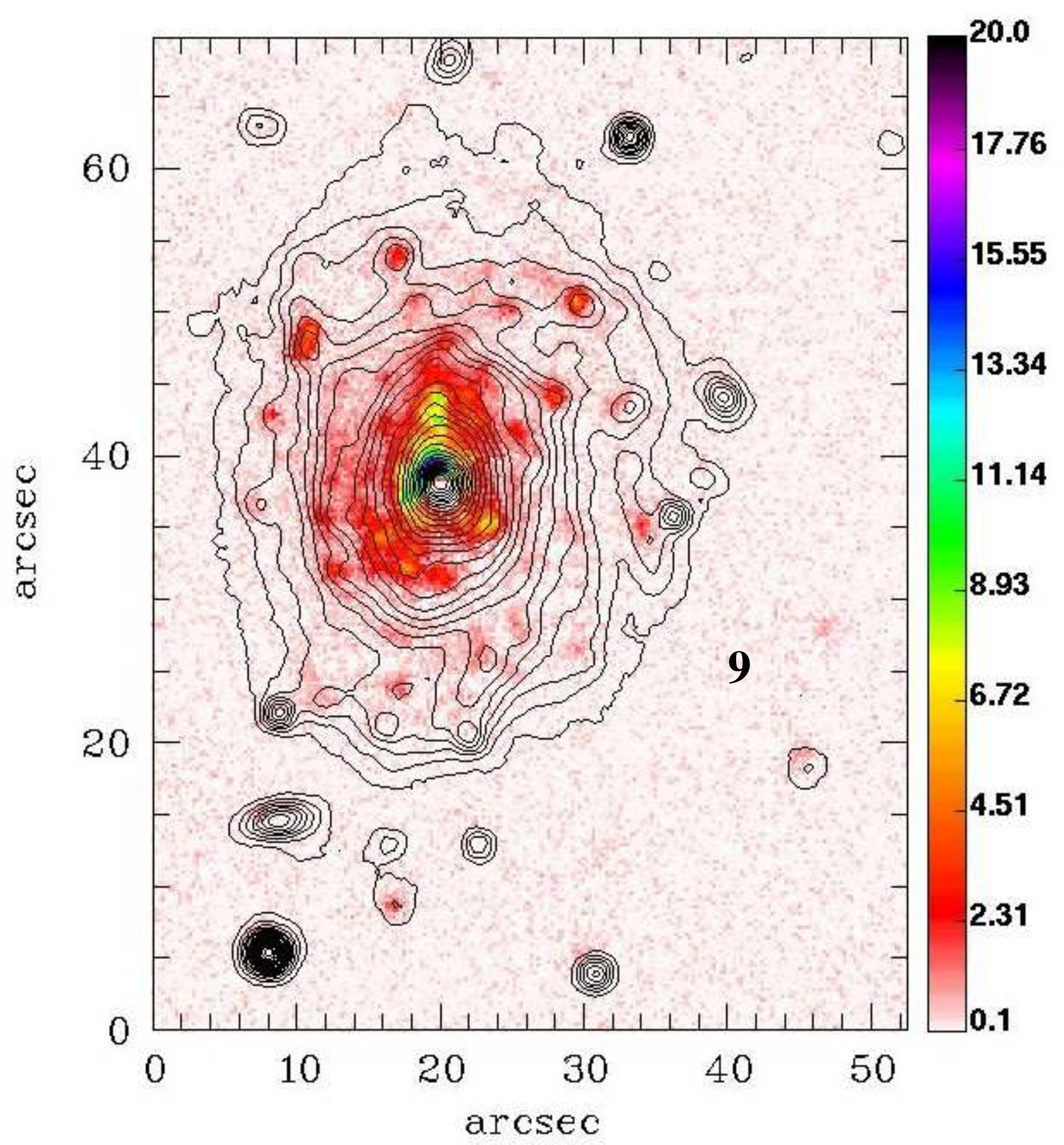}
\includegraphics[width=8cm]{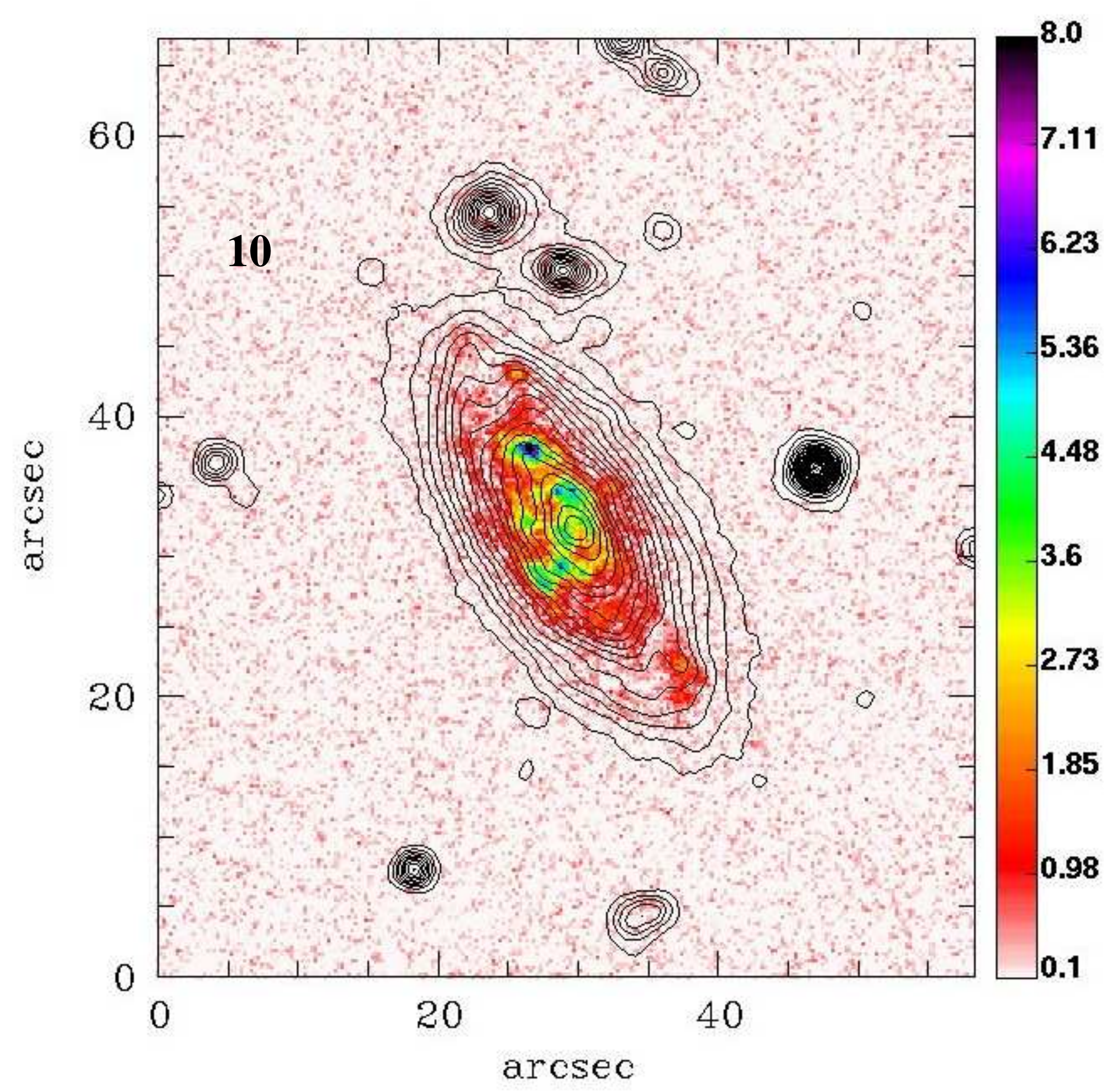}
\includegraphics[width=8cm]{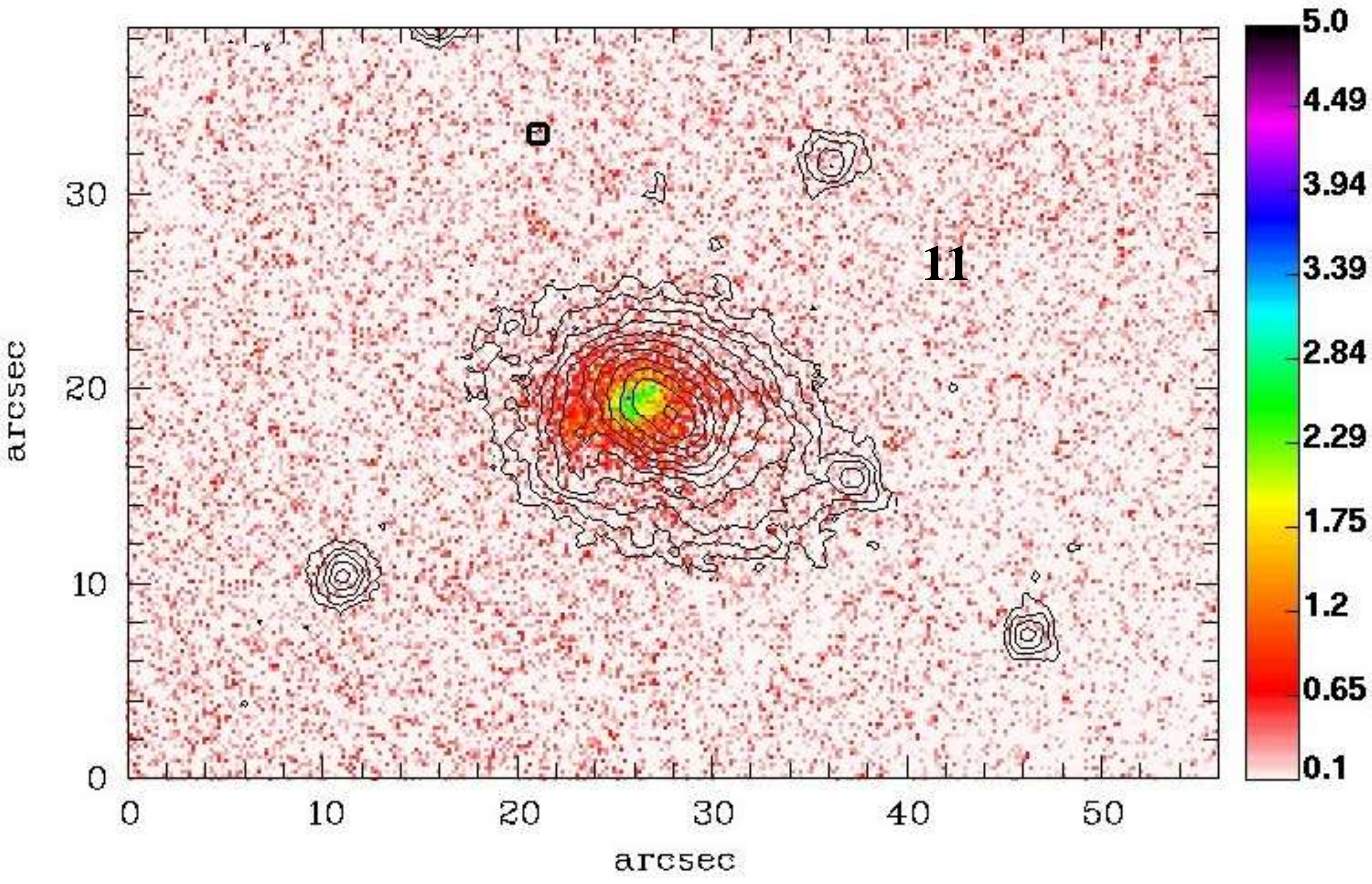}
\caption{Continued}
\end{figure}
\addtocounter{figure}{-1}

\newpage
\clearpage

\begin{figure}
\includegraphics[width=8cm]{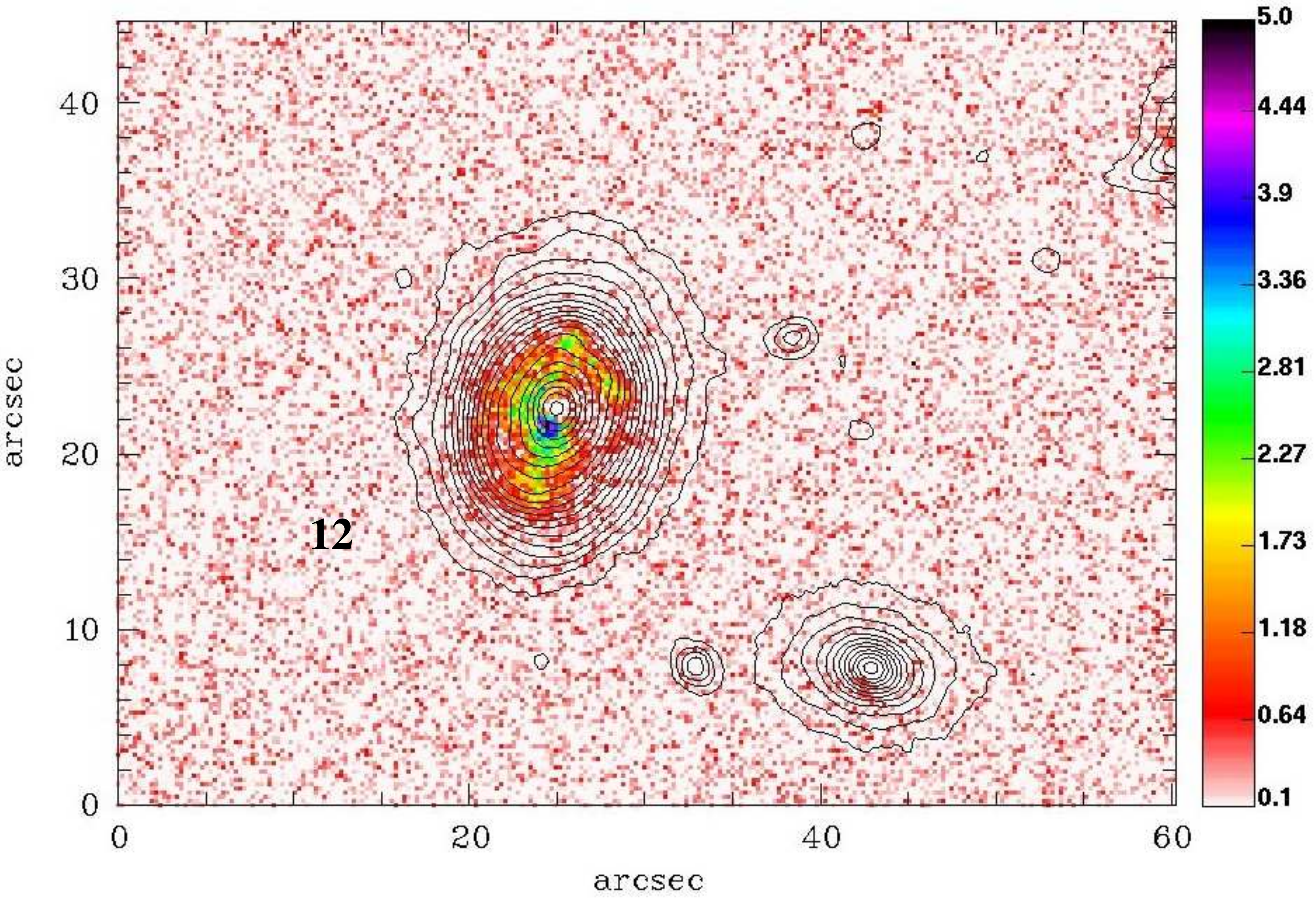}
\includegraphics[width=8cm]{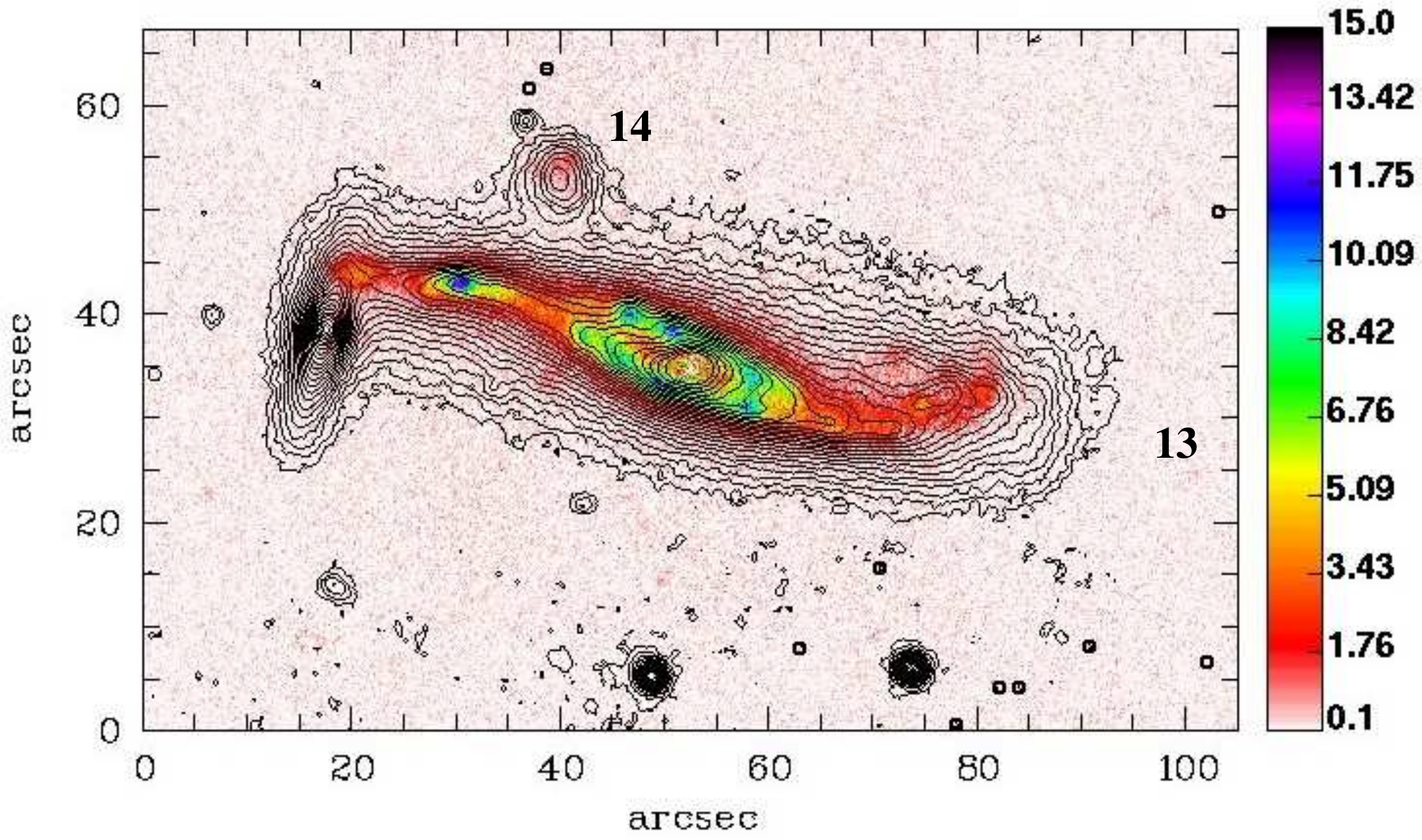}
\includegraphics[width=9cm]{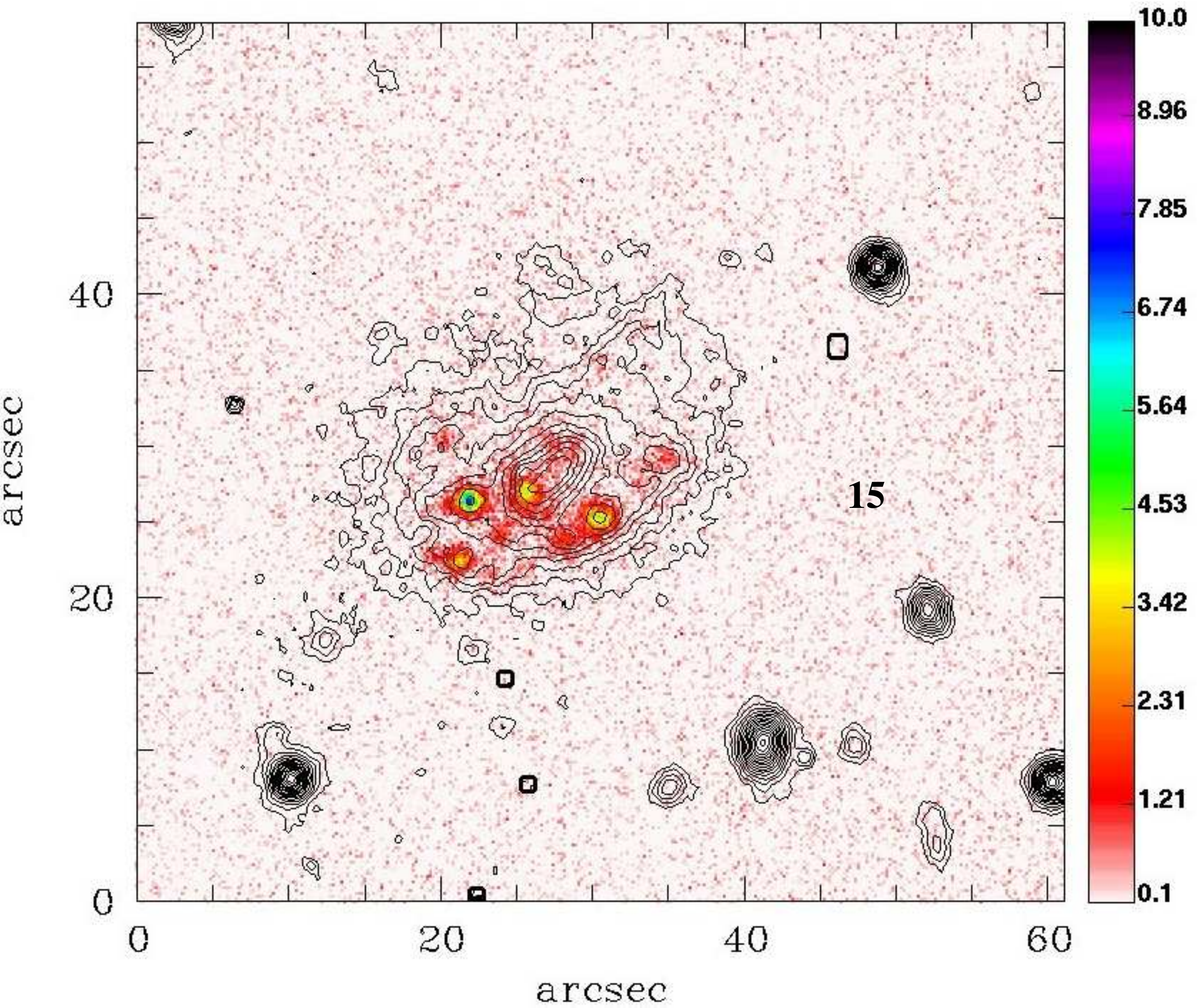}
\includegraphics[width=7cm]{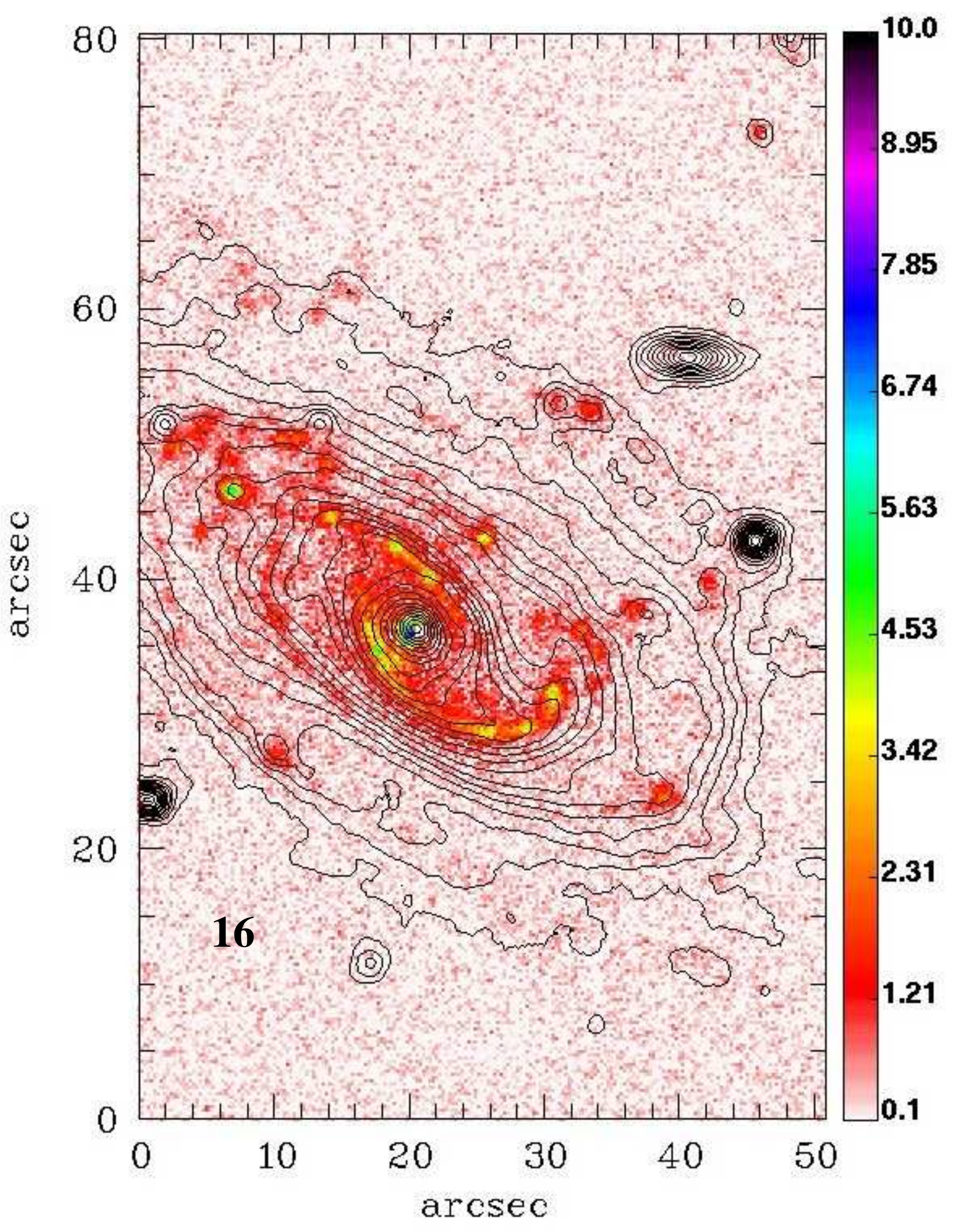}
\caption{Continued}
\end{figure}
\addtocounter{figure}{-1}

\newpage
\clearpage

\begin{figure}
\includegraphics[width=7cm]{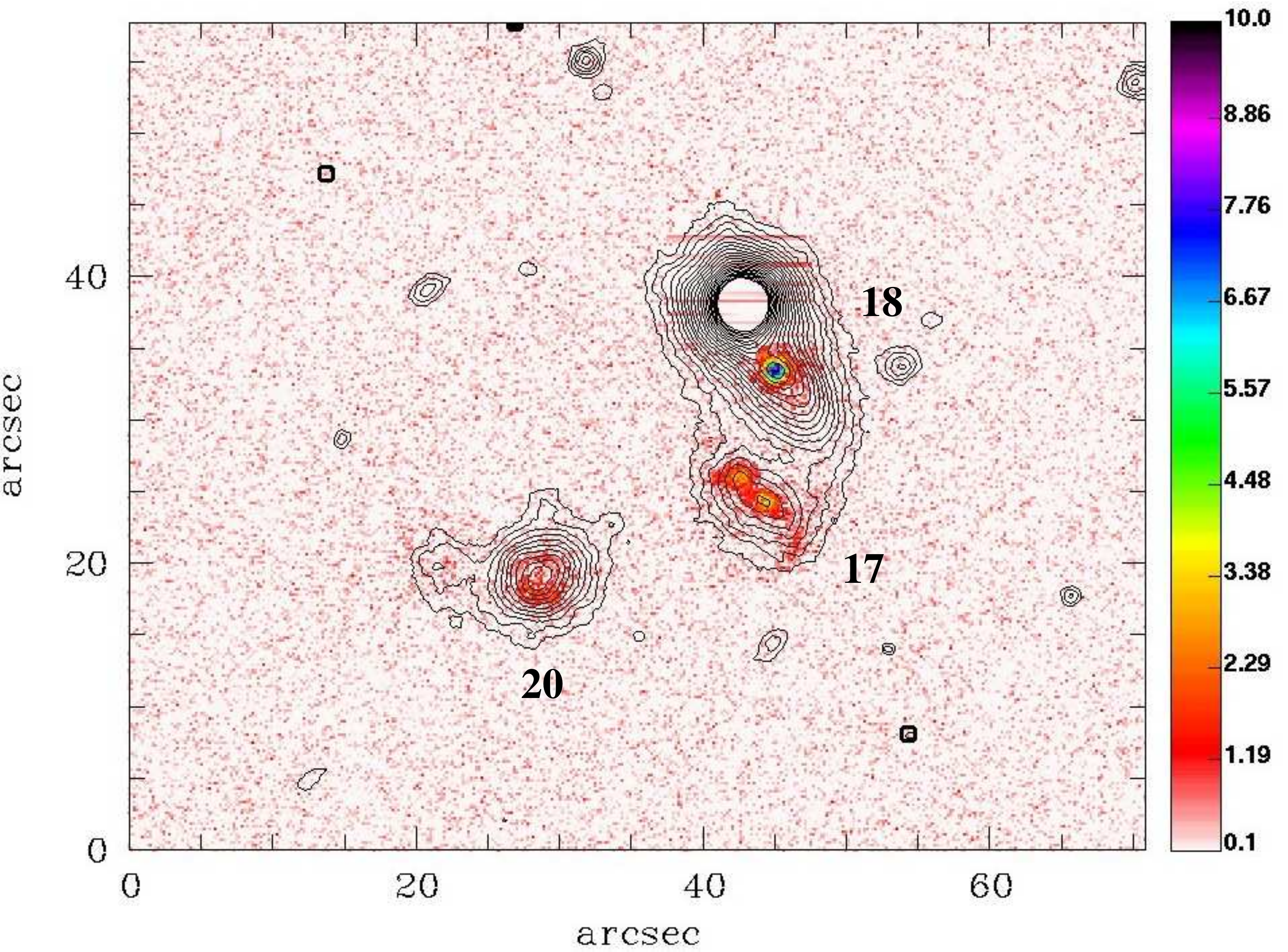}
\includegraphics[width=9cm]{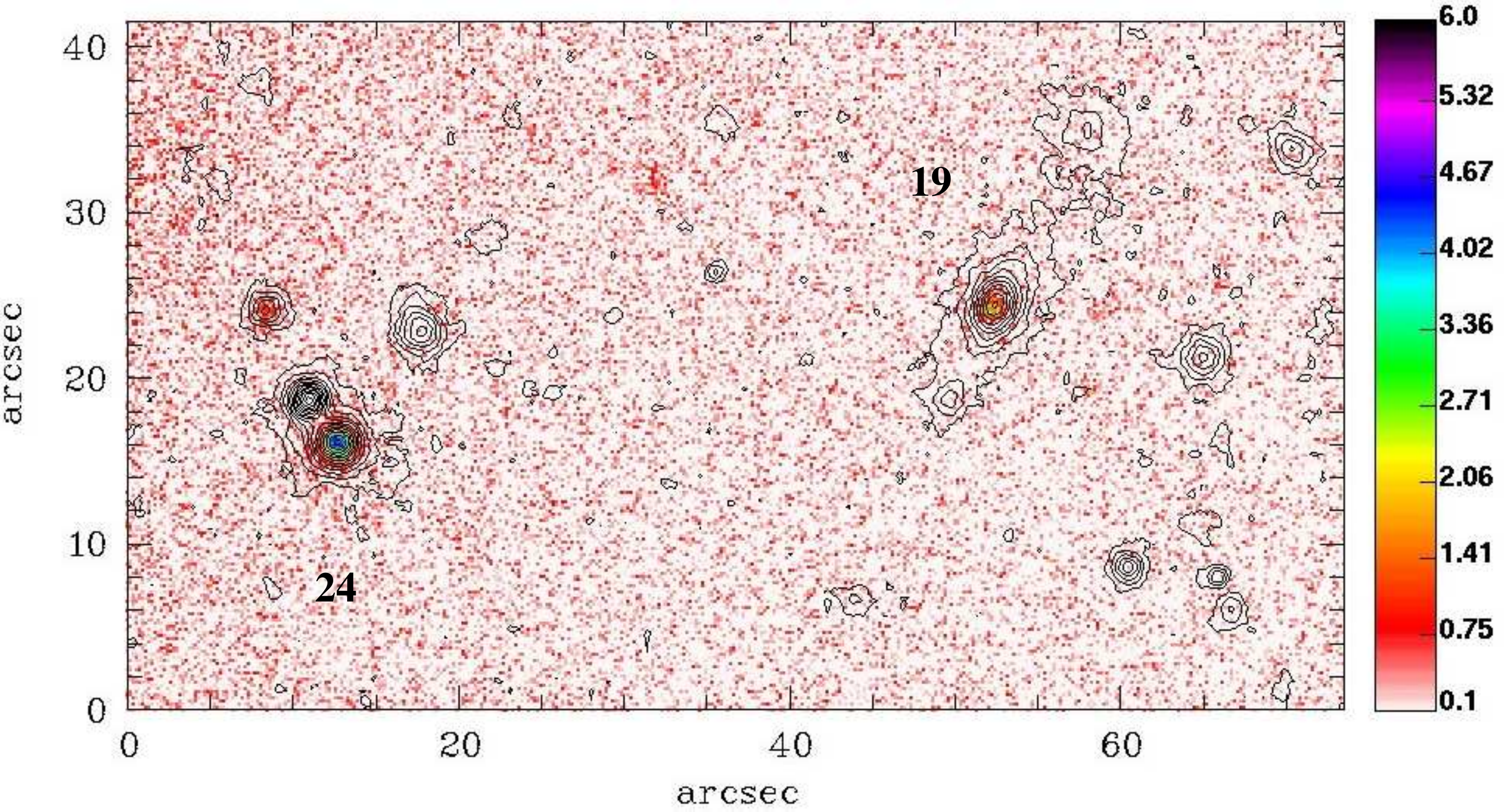}
\includegraphics[width=6cm]{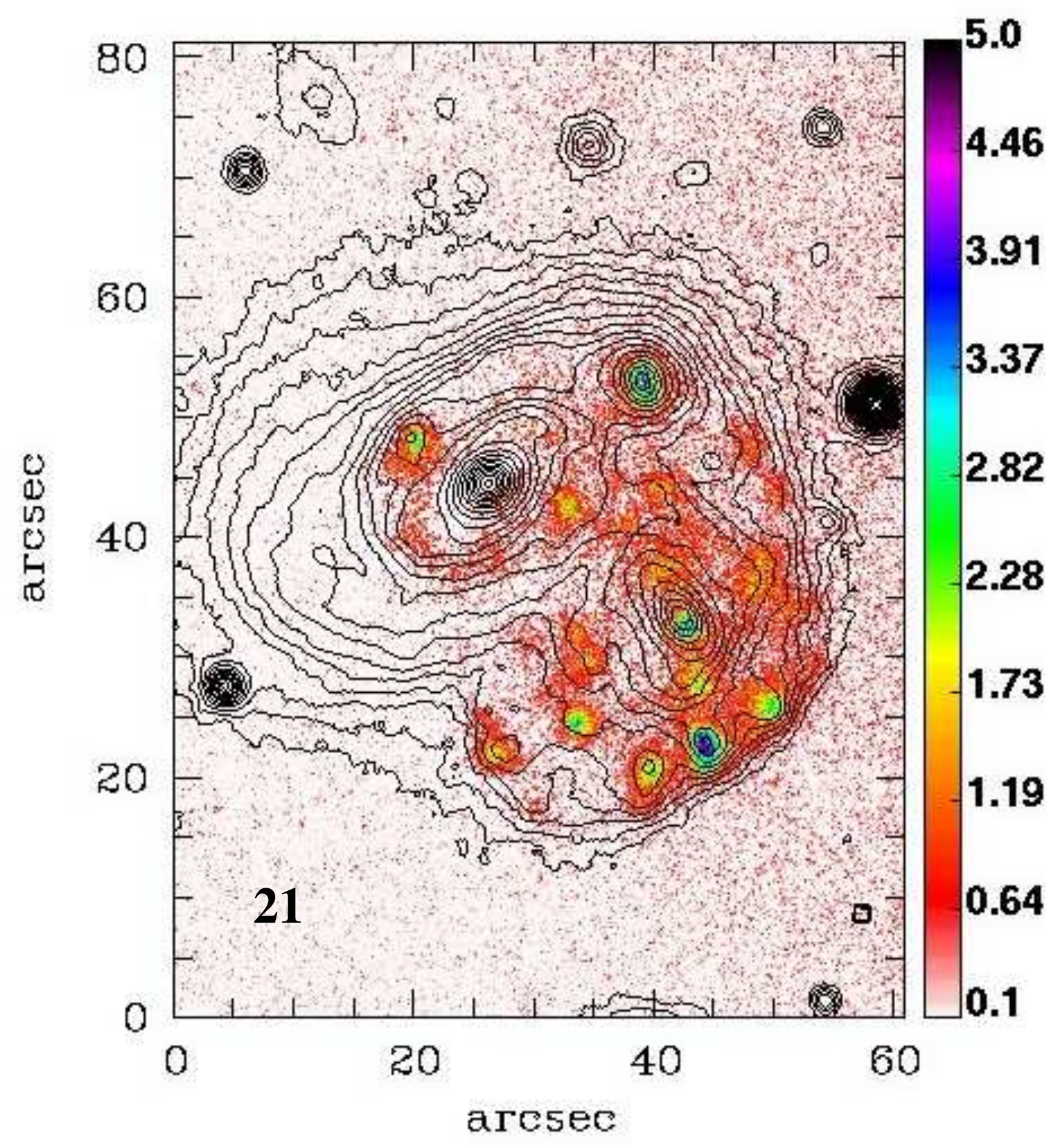}
\includegraphics[width=10cm]{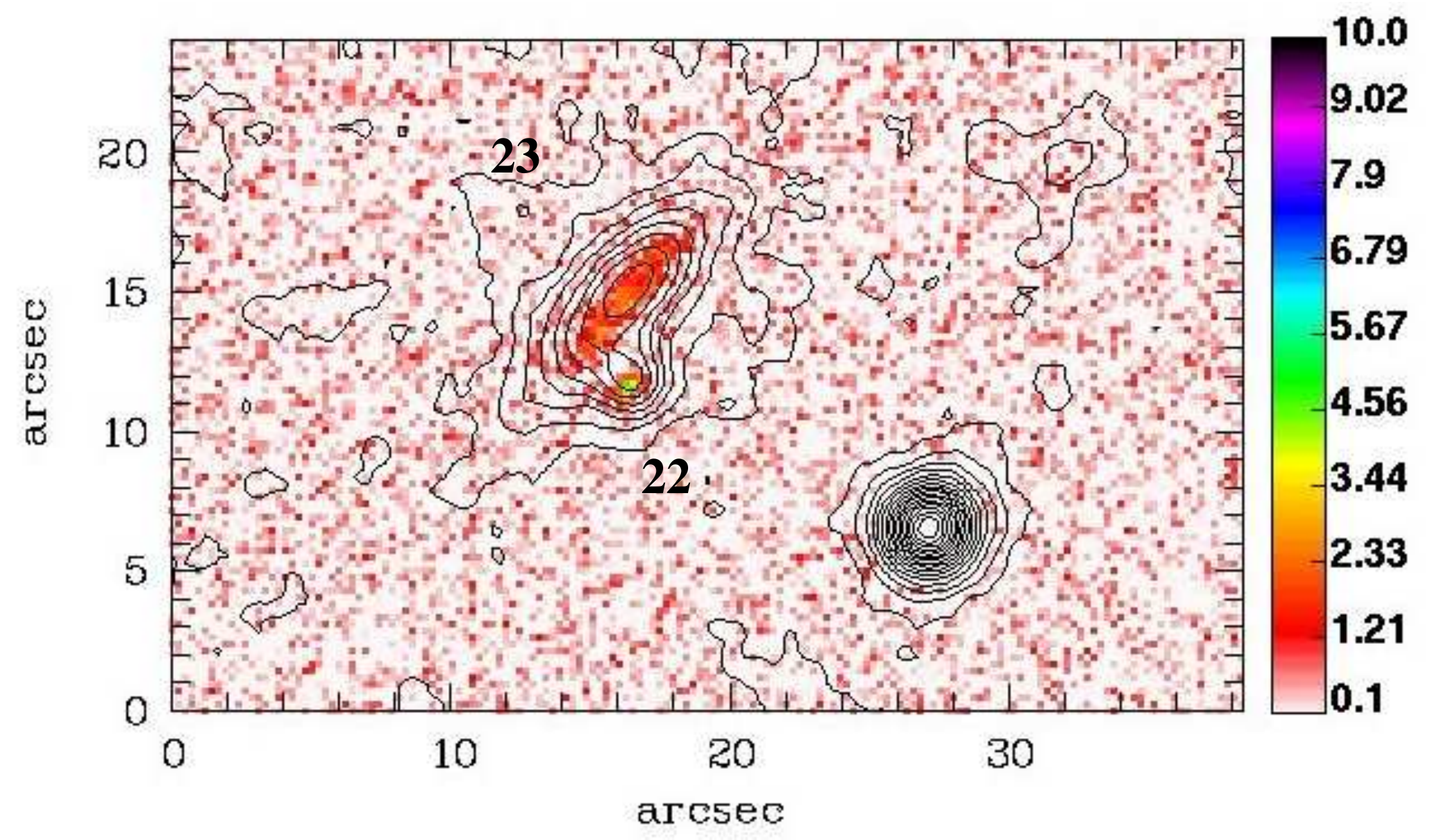}
\caption{Continued}
\end{figure}
\addtocounter{figure}{-1}

\newpage
\clearpage

\begin{figure}
\includegraphics[width=9cm]{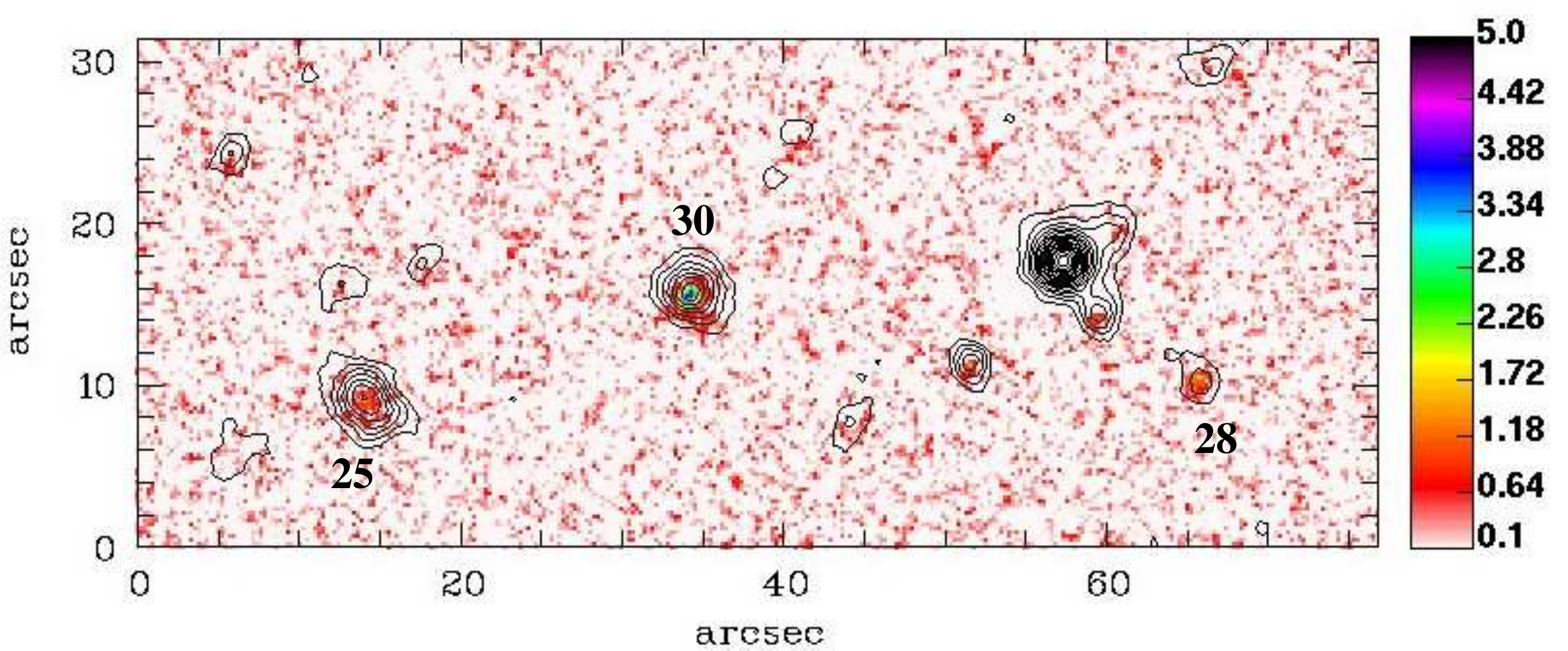}
\includegraphics[width=7cm]{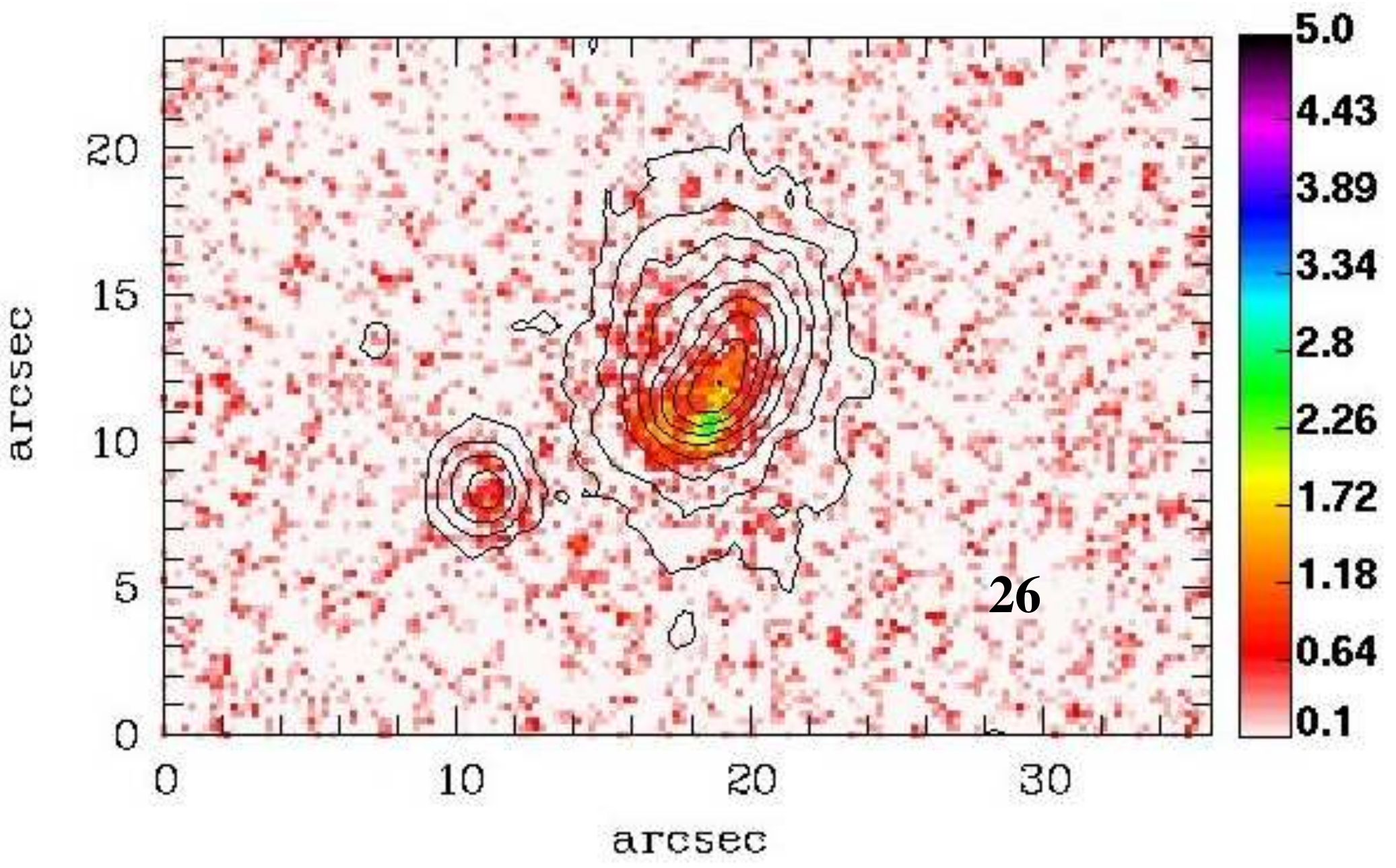}
\includegraphics[width=8cm]{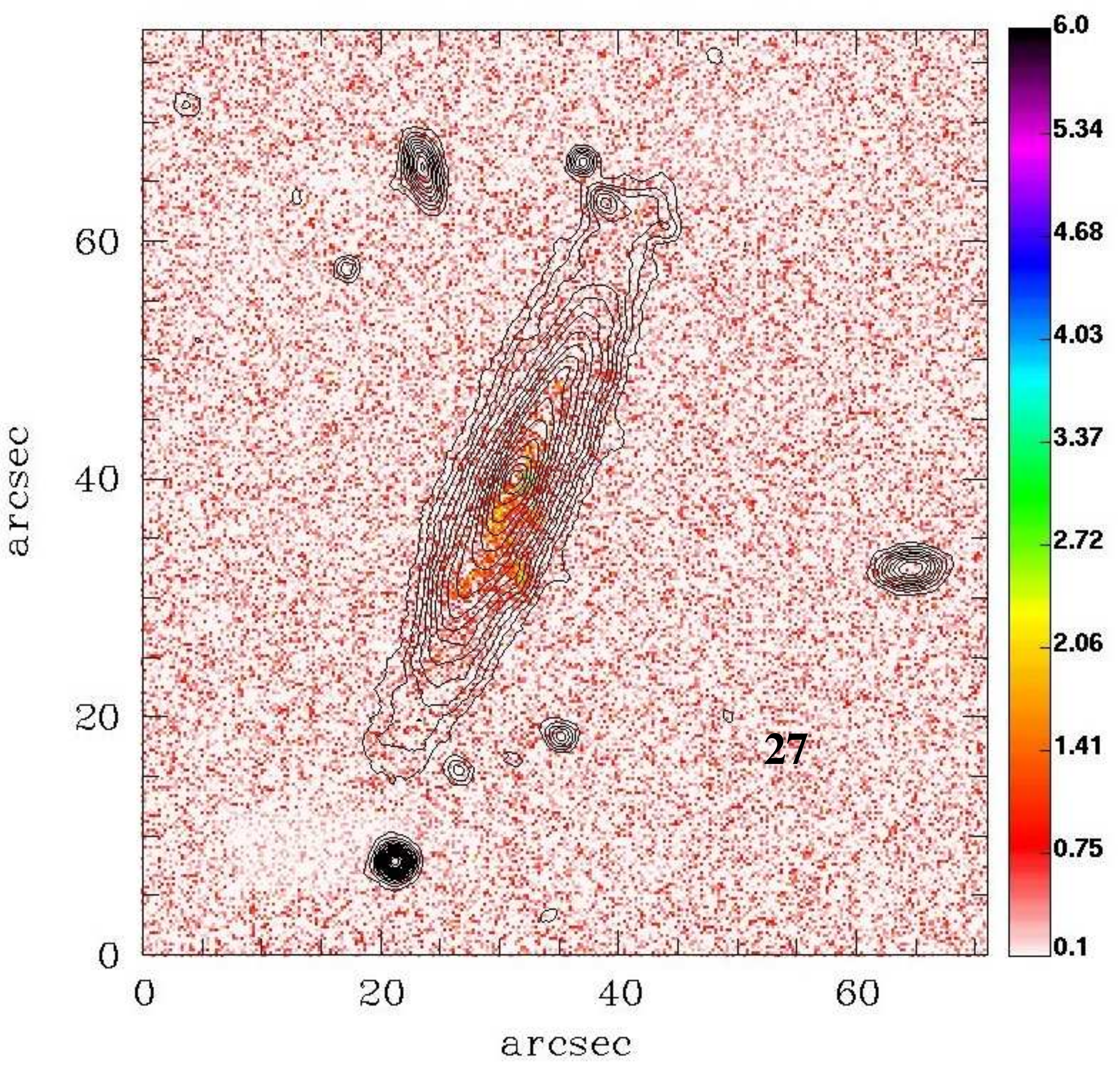}
\includegraphics[width=8cm]{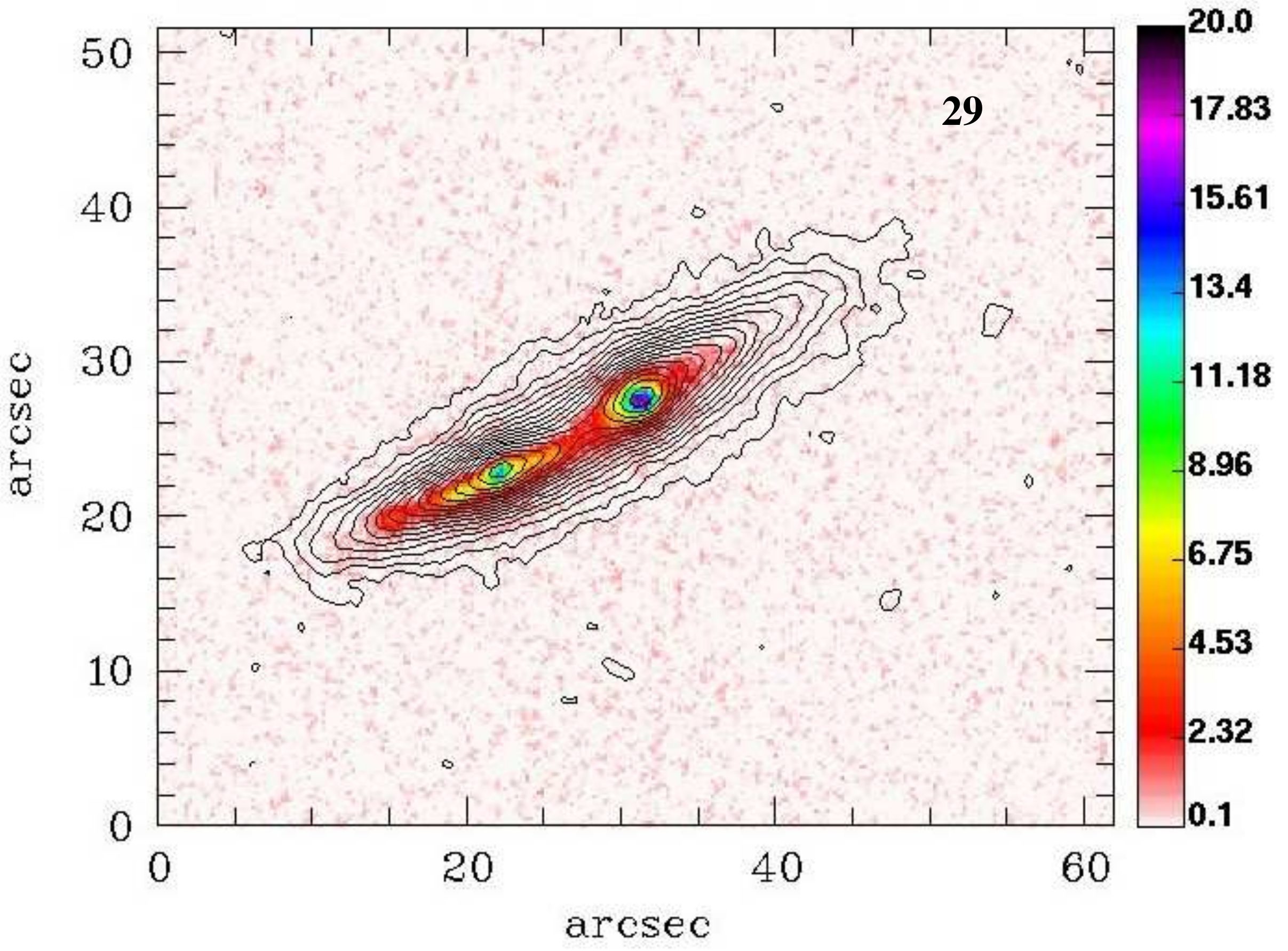}
\caption{Continued}
\end{figure}
\addtocounter{figure}{-1}

\newpage
\clearpage

\begin{figure}
\includegraphics[width=6cm]{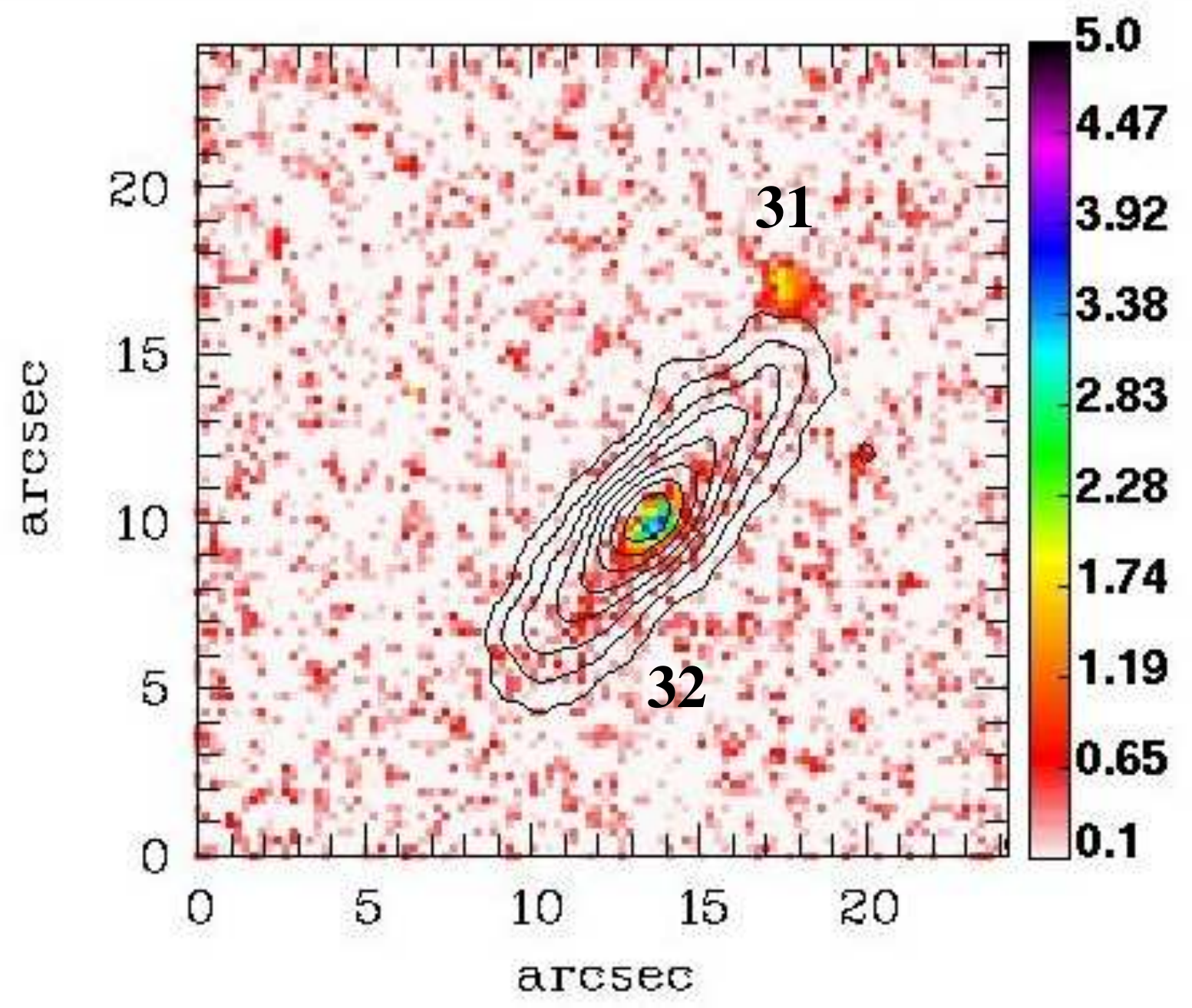}
\includegraphics[width=10cm]{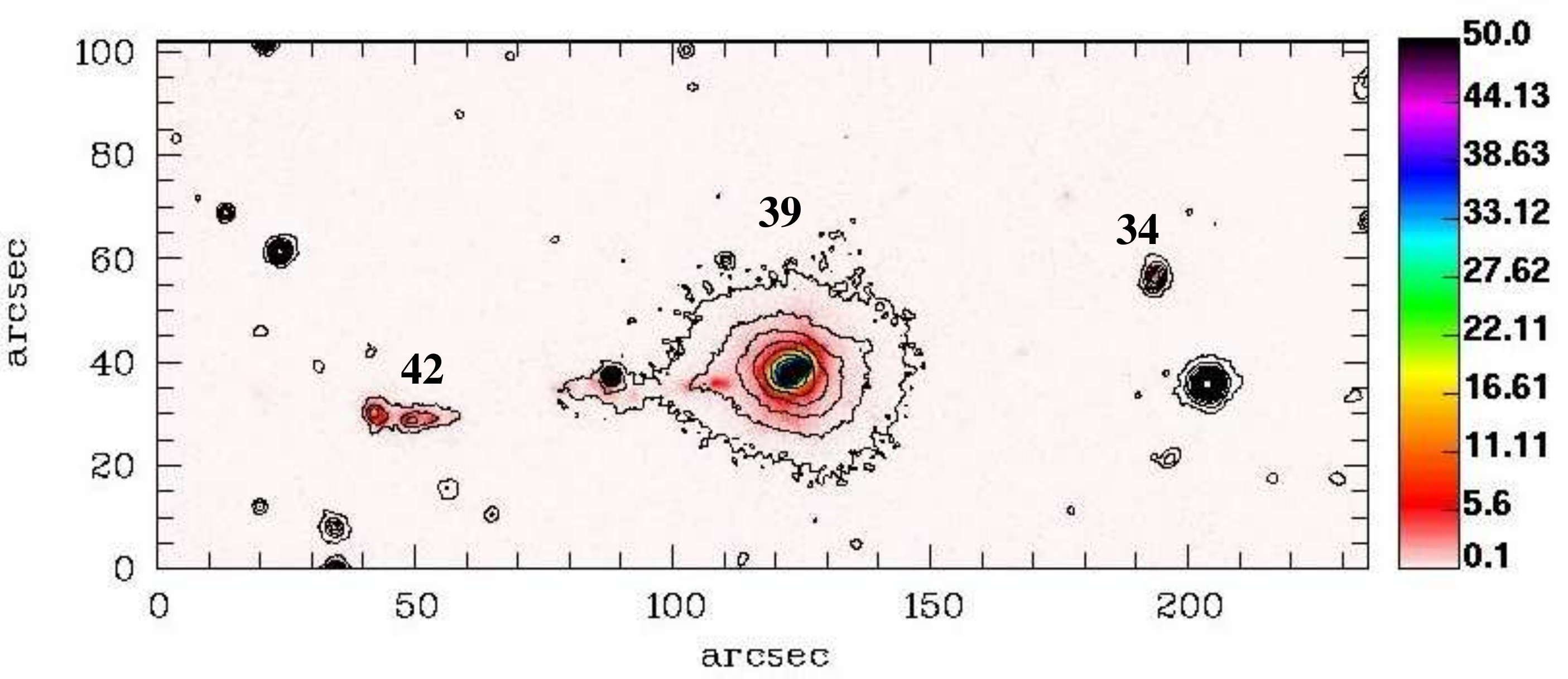}
\includegraphics[width=8cm]{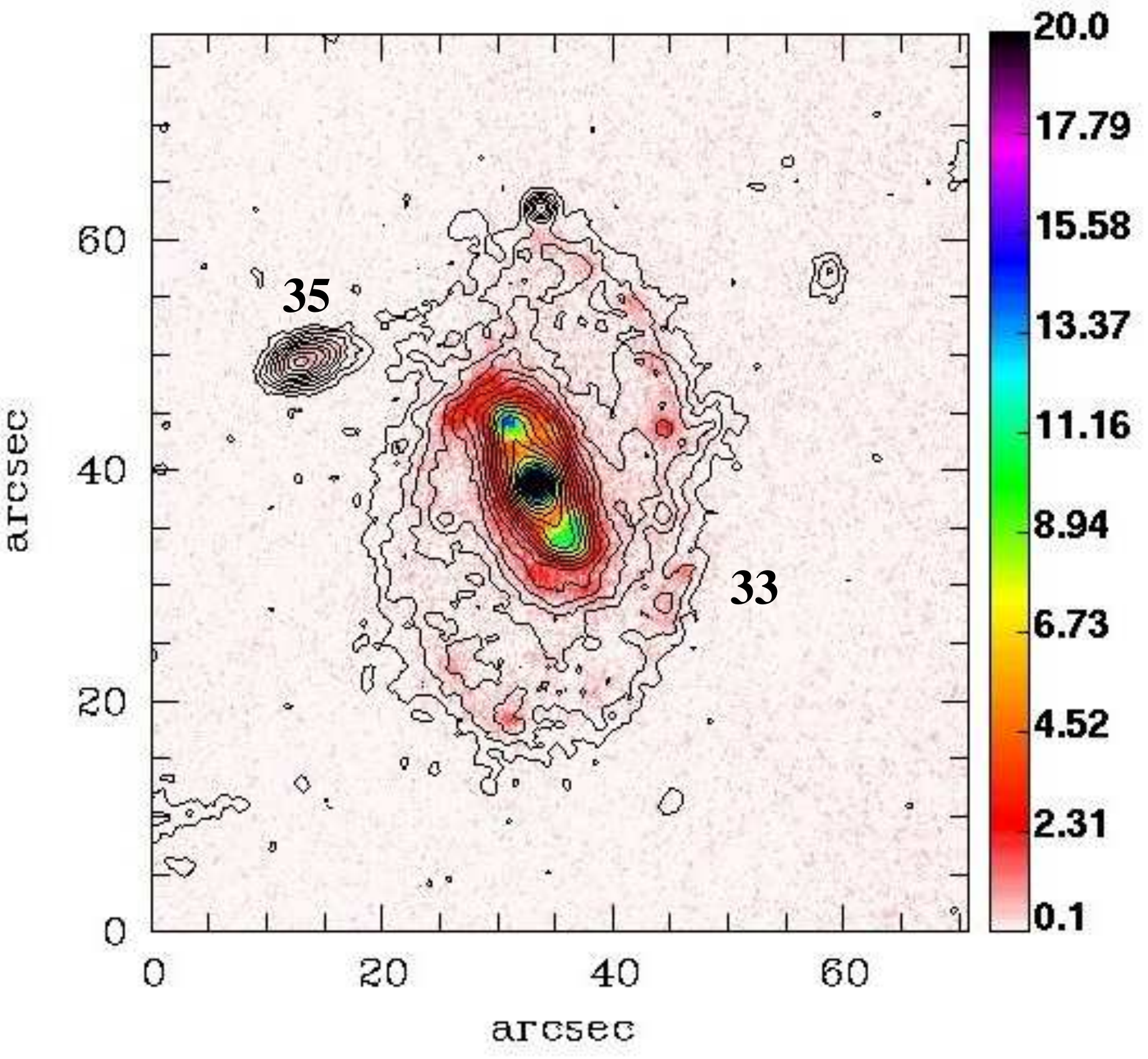}
\includegraphics[width=8cm]{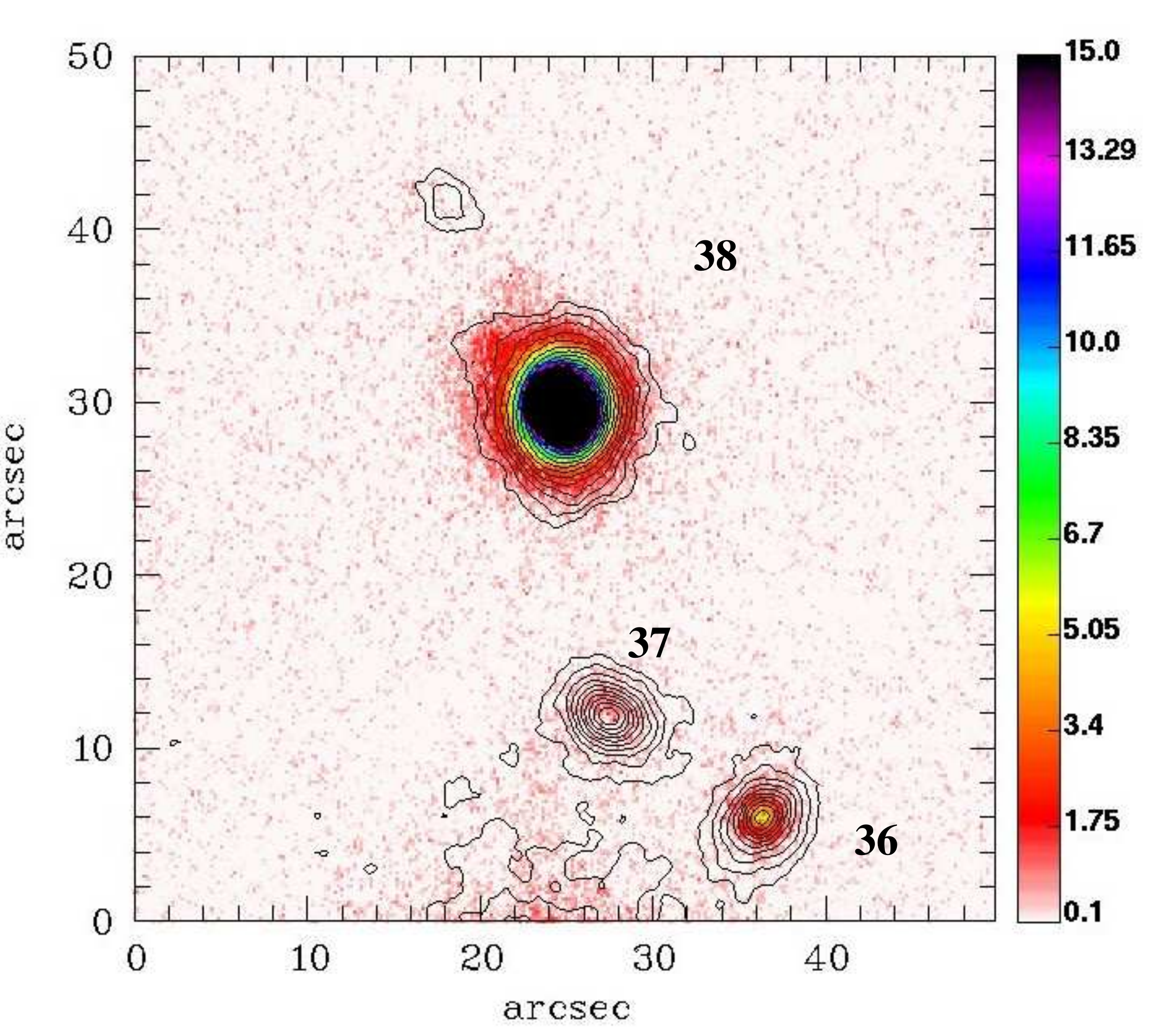}
\caption{Continued}
\end{figure}
\addtocounter{figure}{-1}

\newpage
\clearpage

\begin{figure}
\includegraphics[width=8cm]{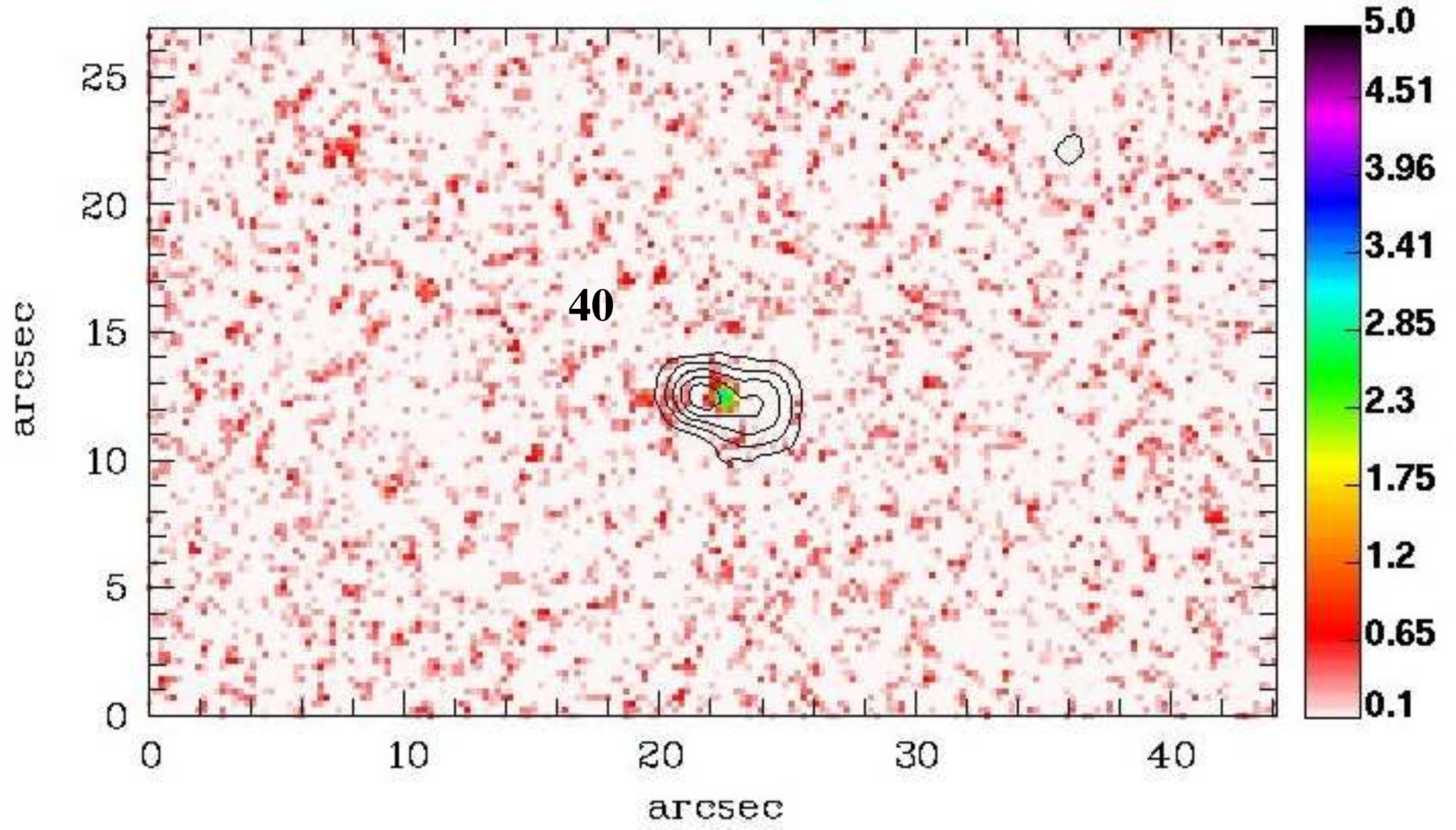}
\includegraphics[width=8cm]{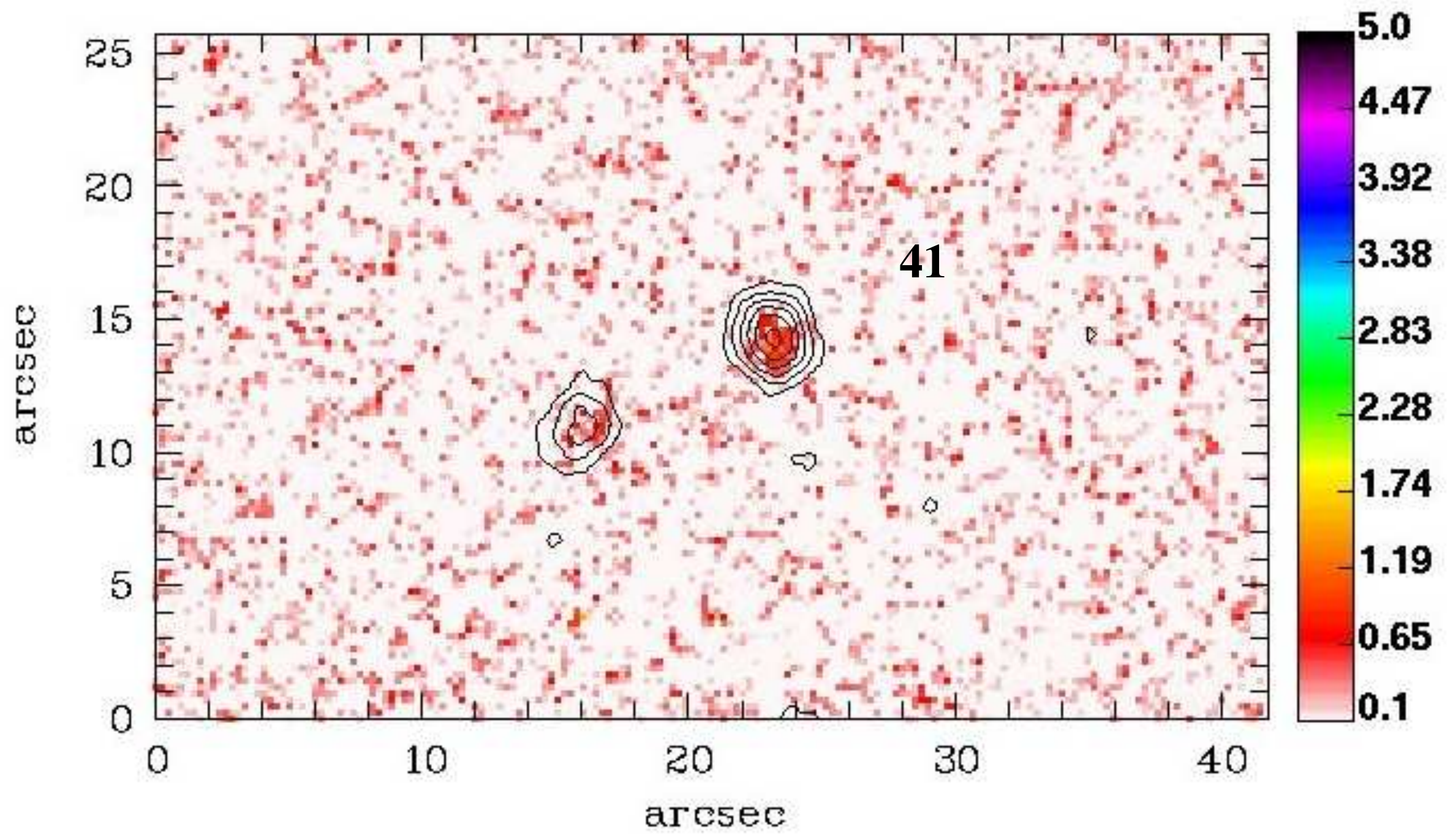}
\includegraphics[width=8cm]{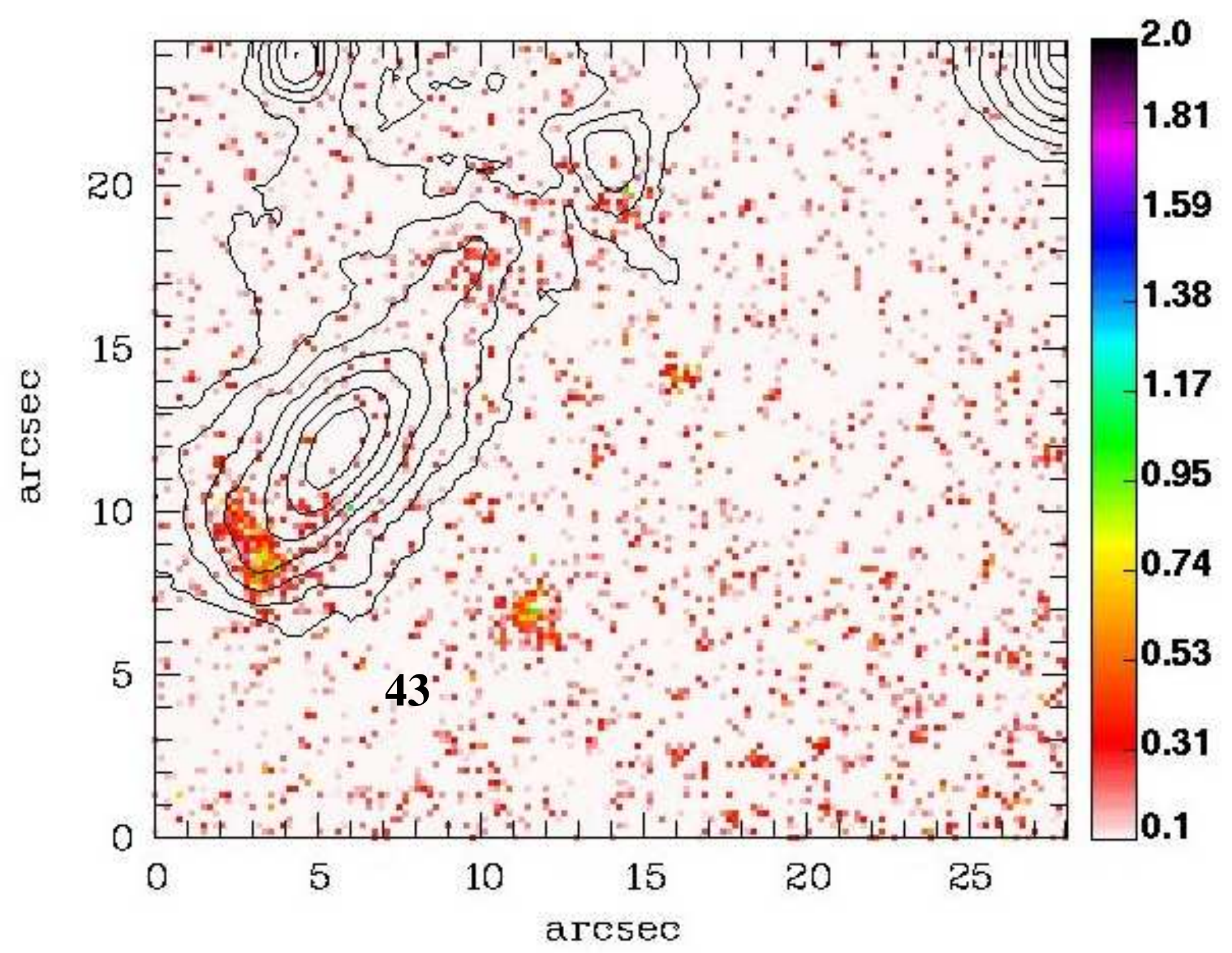}
\includegraphics[width=8cm]{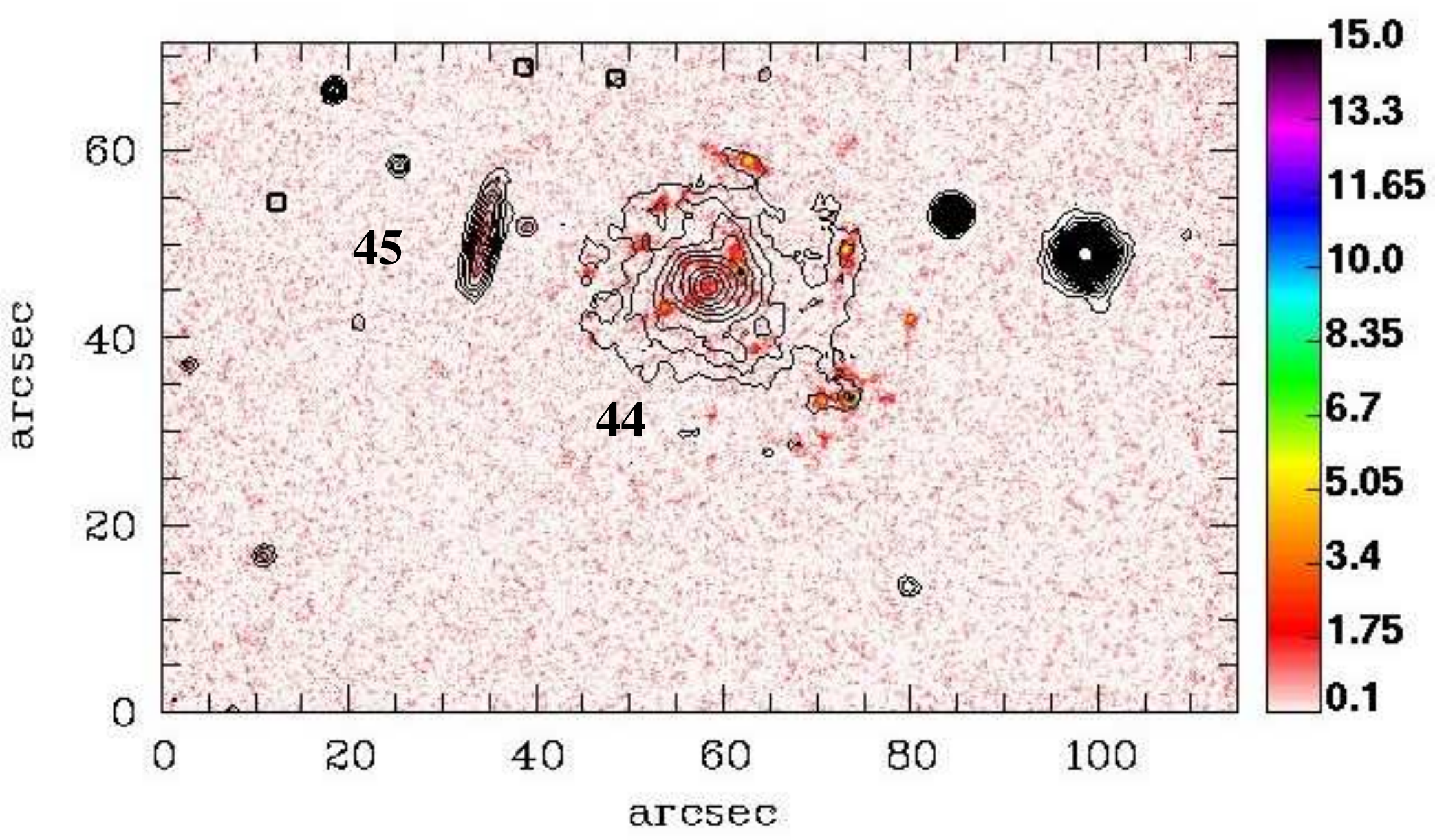}
\caption{Continued}
\end{figure}
\addtocounter{figure}{-1}

\newpage
\clearpage

\begin{figure}
\includegraphics[width=8cm]{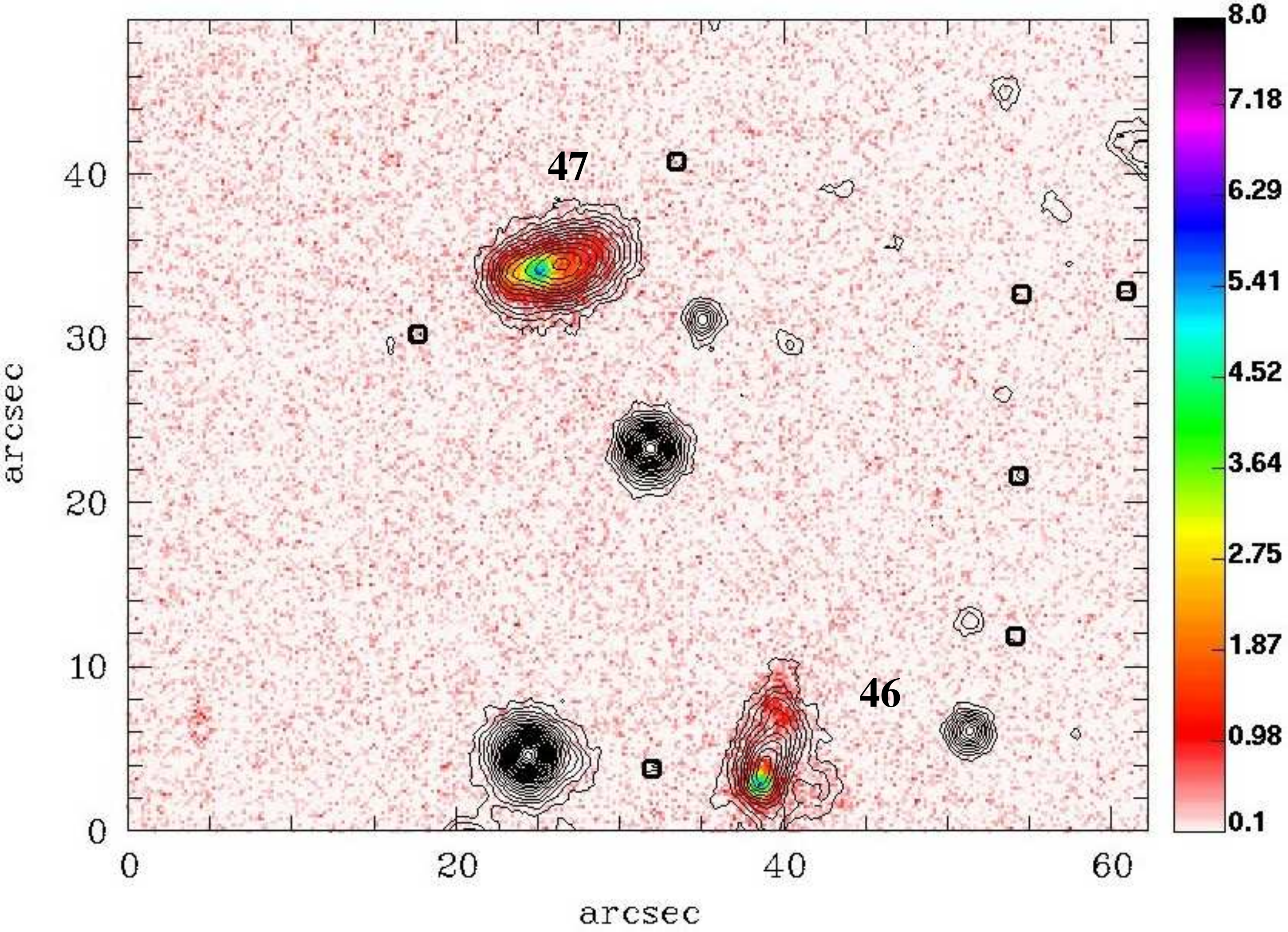}
\includegraphics[width=8cm]{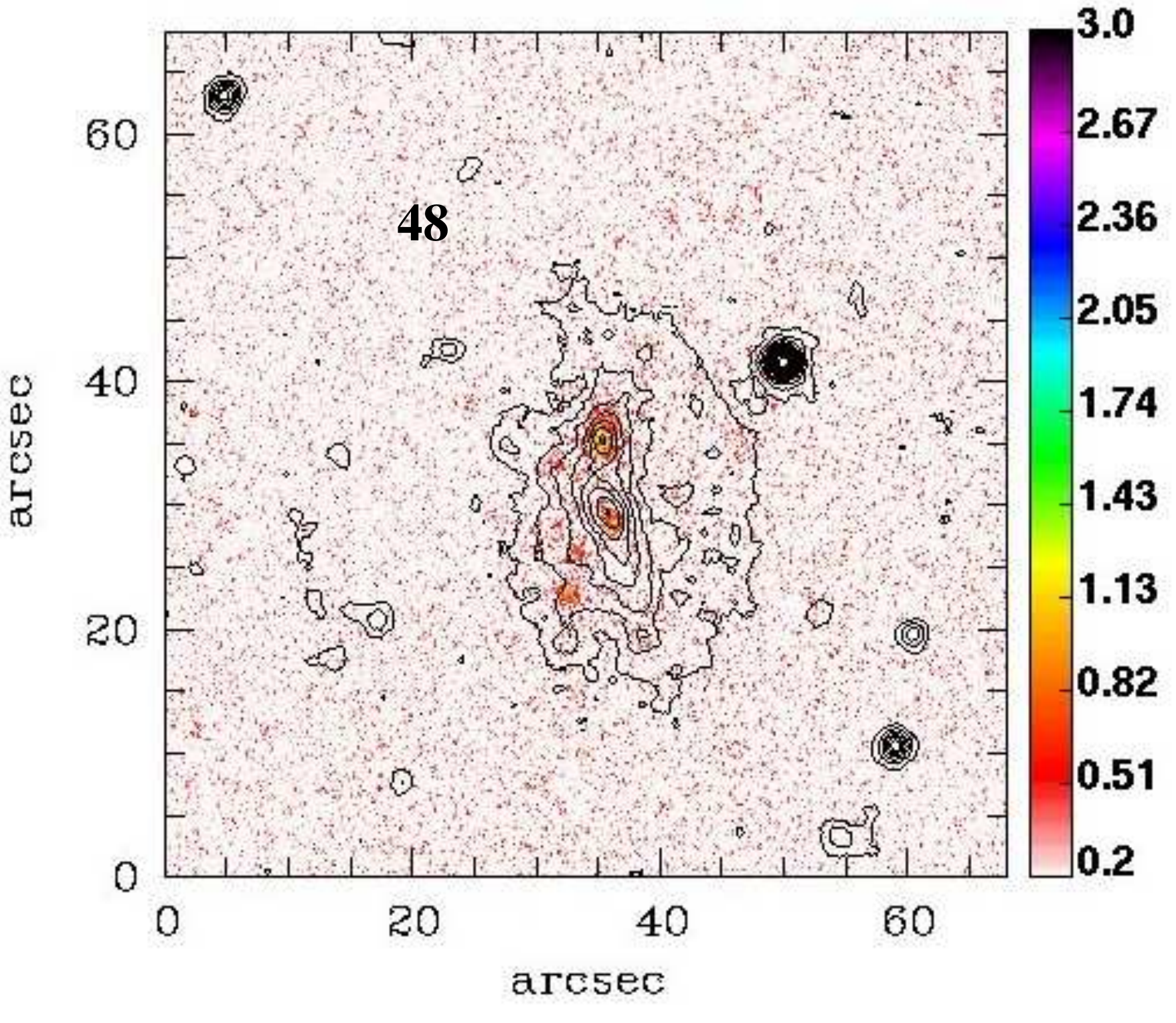}
\includegraphics[width=8cm]{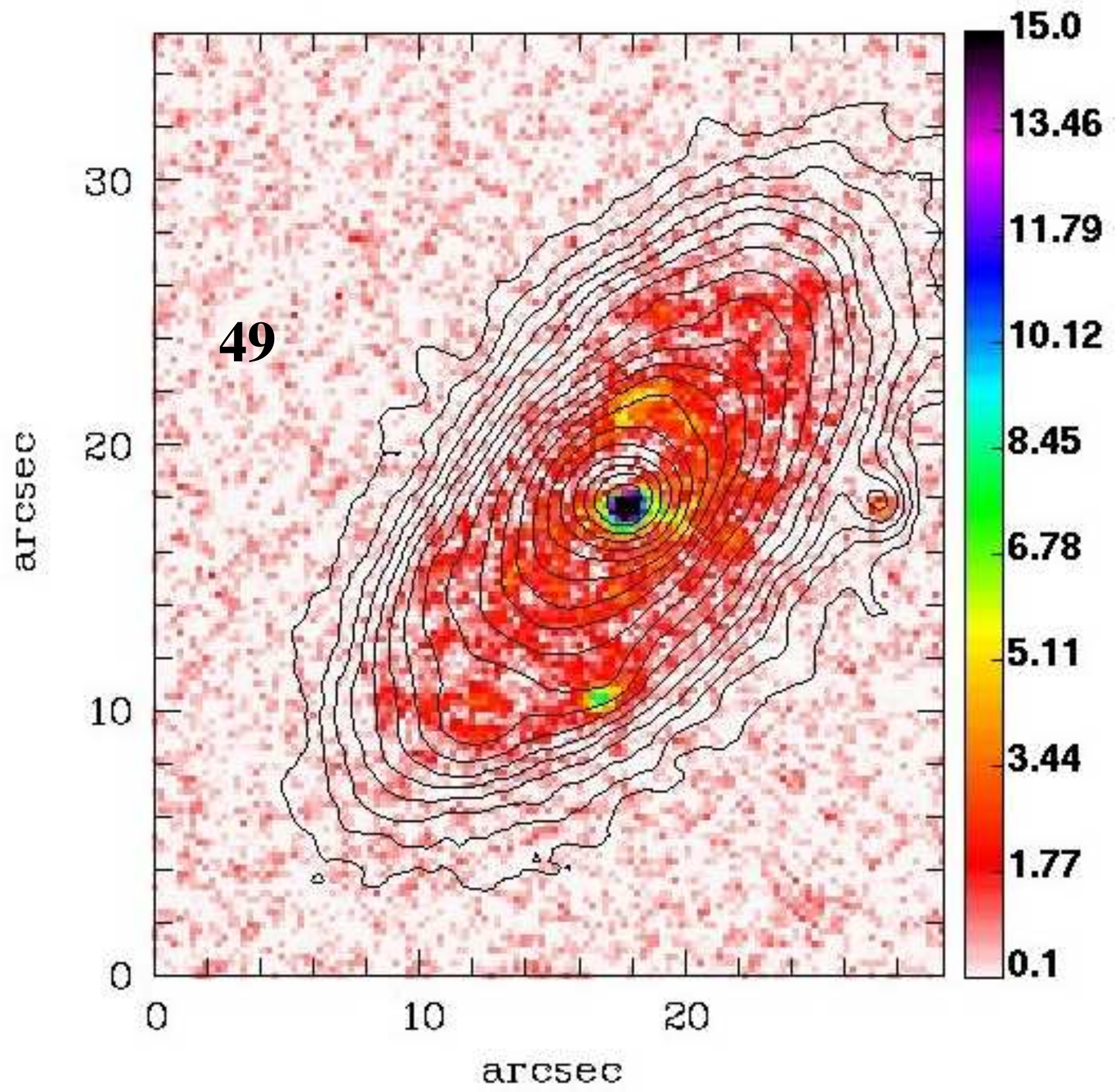}
\includegraphics[width=8cm]{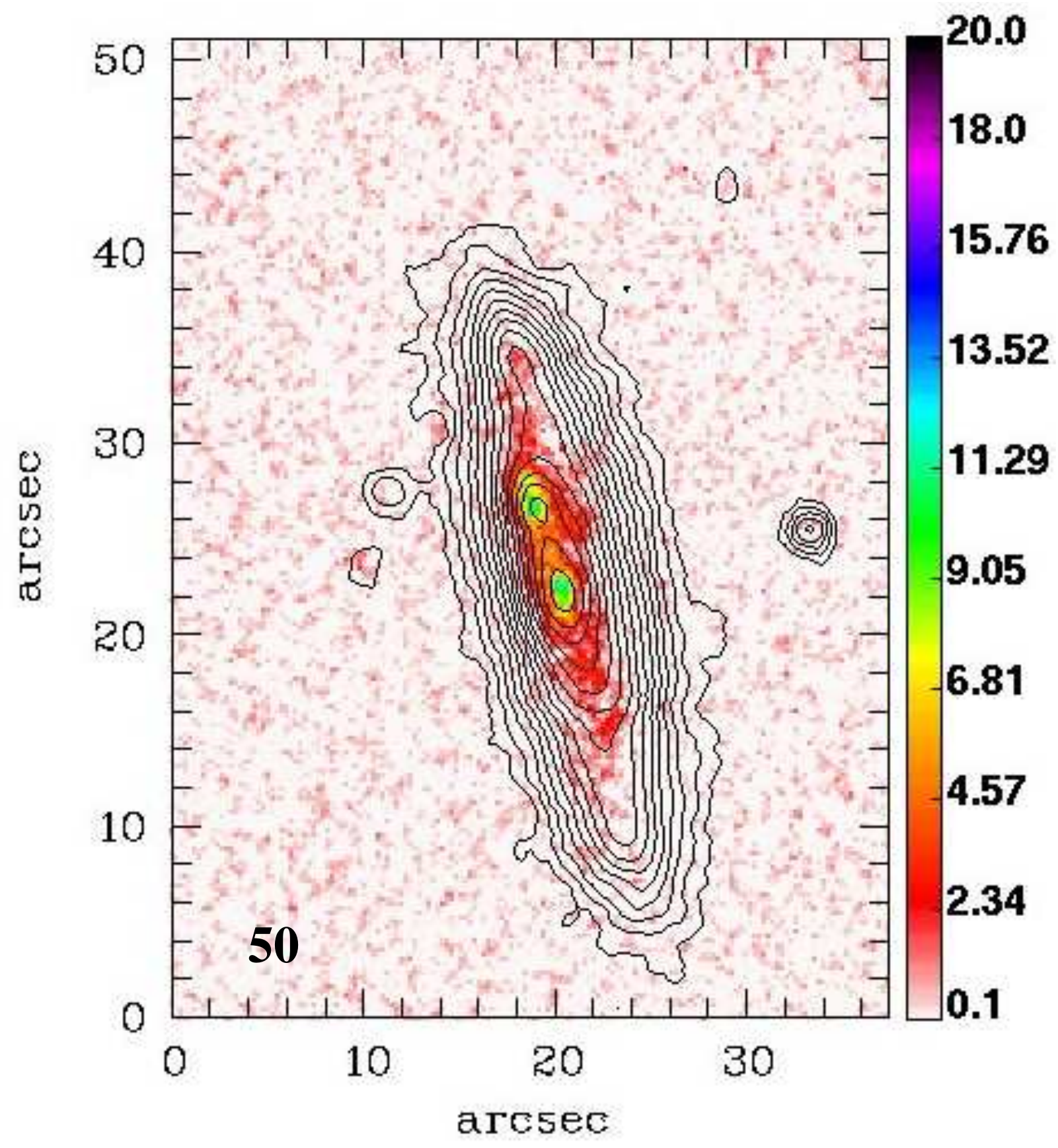}
\caption{Continued}
\end{figure}

\newpage
\clearpage

\begin{figure}
\includegraphics[width=16cm]{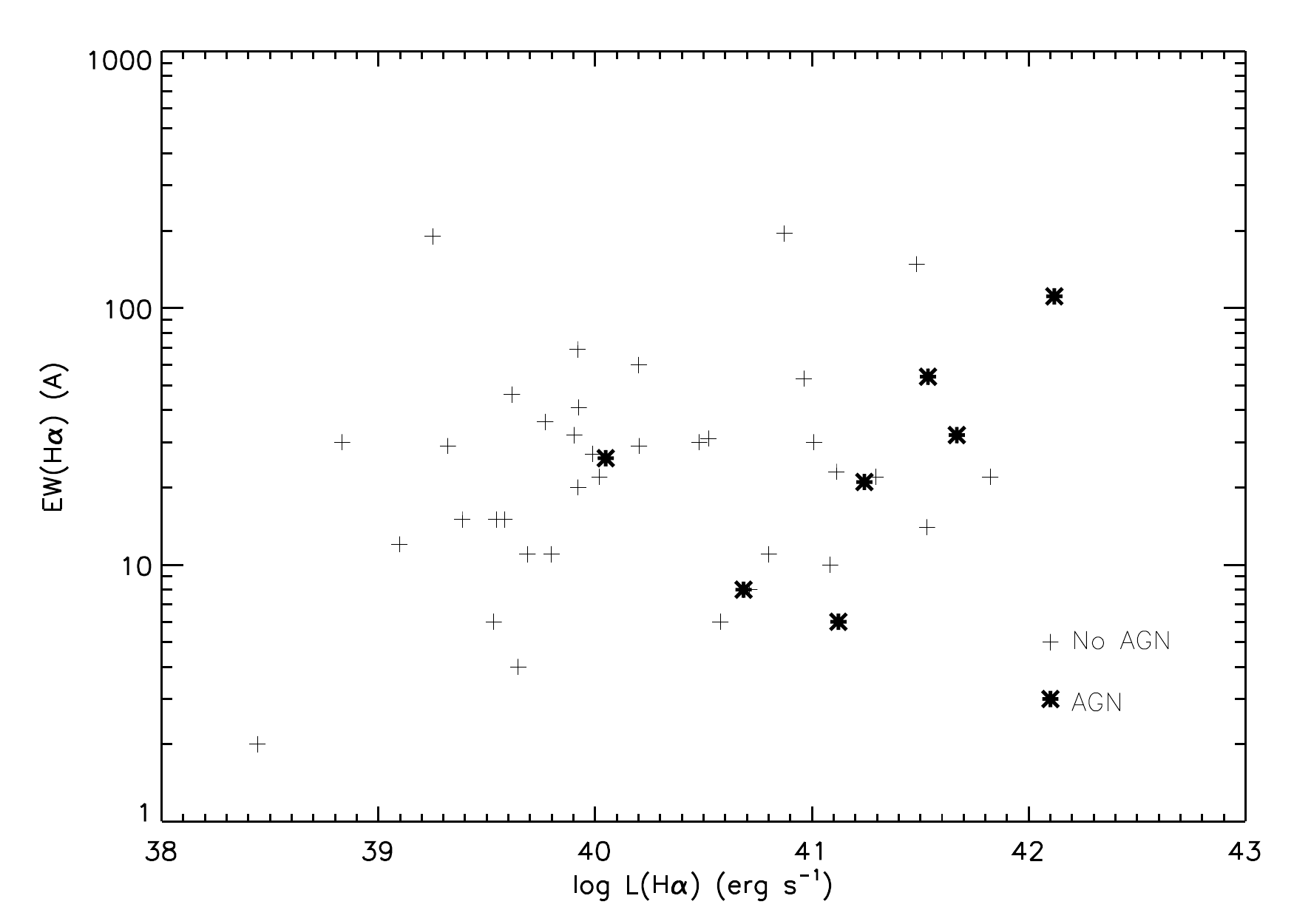}
\caption{$EW$(H$\alpha$) vs. $L$(H$\alpha$) for the A2151 galaxies (crosses). Bold asterisks correspond to the AGN candidates.}
\label{lha_ewha}
\end{figure}

\newpage
\clearpage

\begin{figure}
\includegraphics[width=16cm]{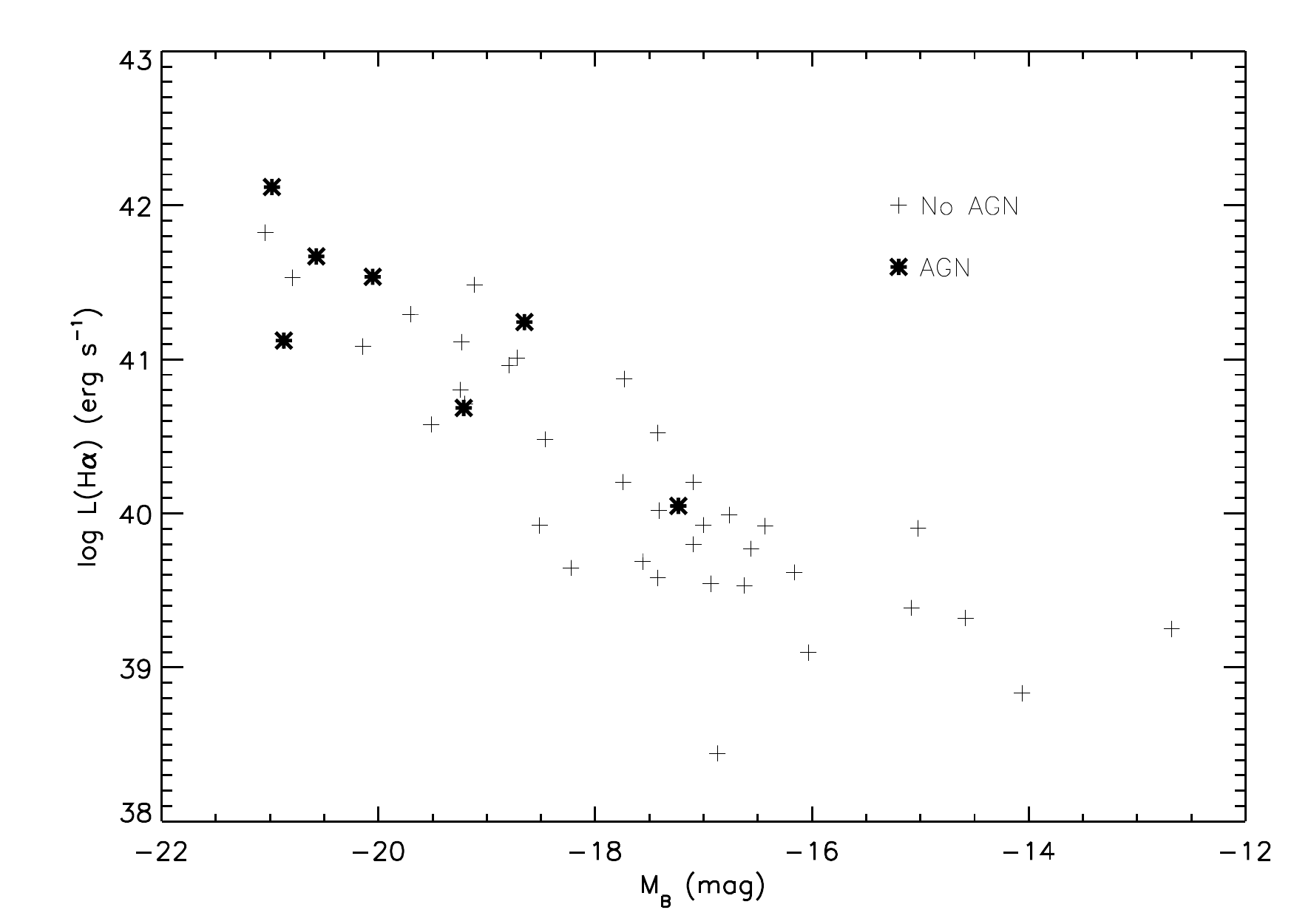}
\caption{$L$(H$\alpha$) vs. $M_{B}$ for the A2151 galaxies (crosses). Bold asterisks correspond to the AGN candidates.}
\label{mb_lha}
\end{figure}

\newpage
\clearpage

\begin{figure}
\includegraphics[width=16cm]{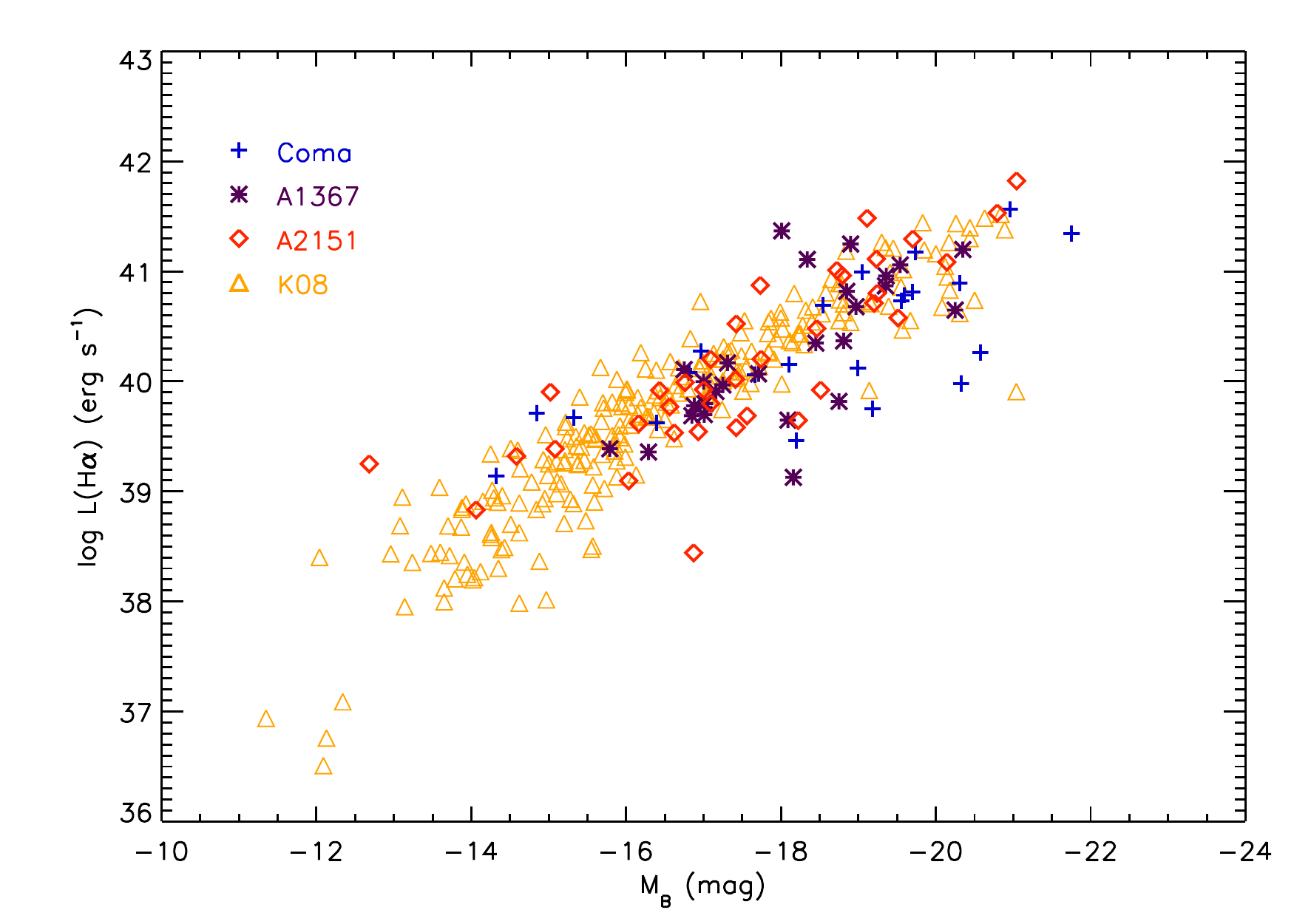}
\caption{$L$(H$\alpha$) vs. $M_{B}$ for galaxies of Coma (blue crosses), A1367 (violet asterisks), A2151 (red diamonds) and K08 (orange triangles).}
\label{lha_mb_all}
\end{figure}

\newpage
\clearpage

\begin{figure}
\includegraphics[width=16cm]{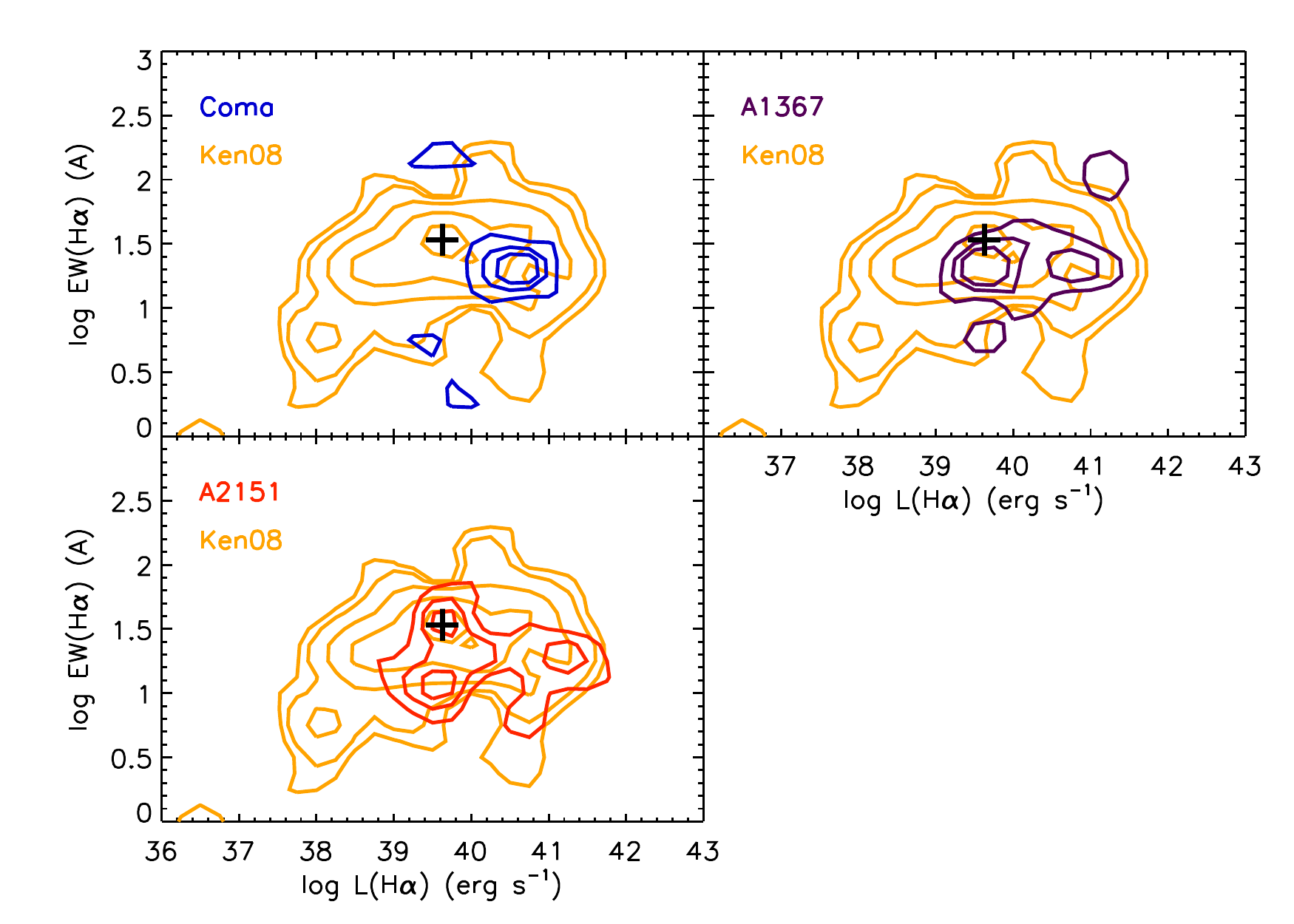}
\caption{Contour maps of the $L$(H$\alpha$) vs. $EW$(H$\alpha$) plot for the clusters. Those of K08 are also plotted in the three panels for comparison. The black cross indicates the position of the maximum of the K08 distribution.}
\label{contour}
\end{figure}

\newpage
\clearpage

\begin{figure}
\includegraphics[width=16cm]{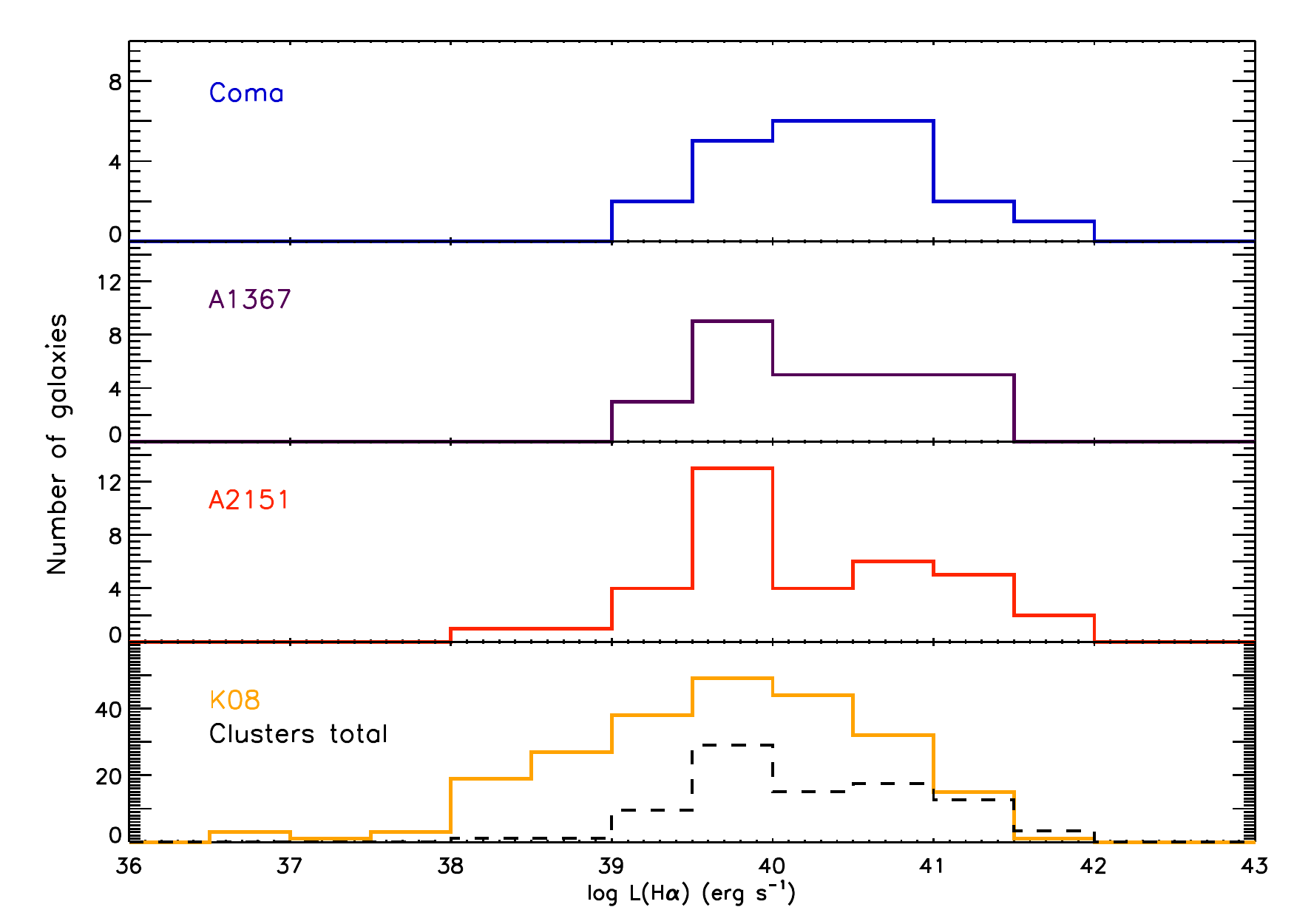}
\caption{Histogram of the logarithm of the H$\alpha$ + [N{\sc ii}] luminosity for the detected galaxies. The black line in the bottom panel corresponds to the combined distribution of the three clusters weighted each one to the corresponding surveyed volume and arbitrarily normalized.}
\label{histo_lha}
\end{figure}

\newpage
\clearpage

\begin{figure}
\includegraphics[width=8cm]{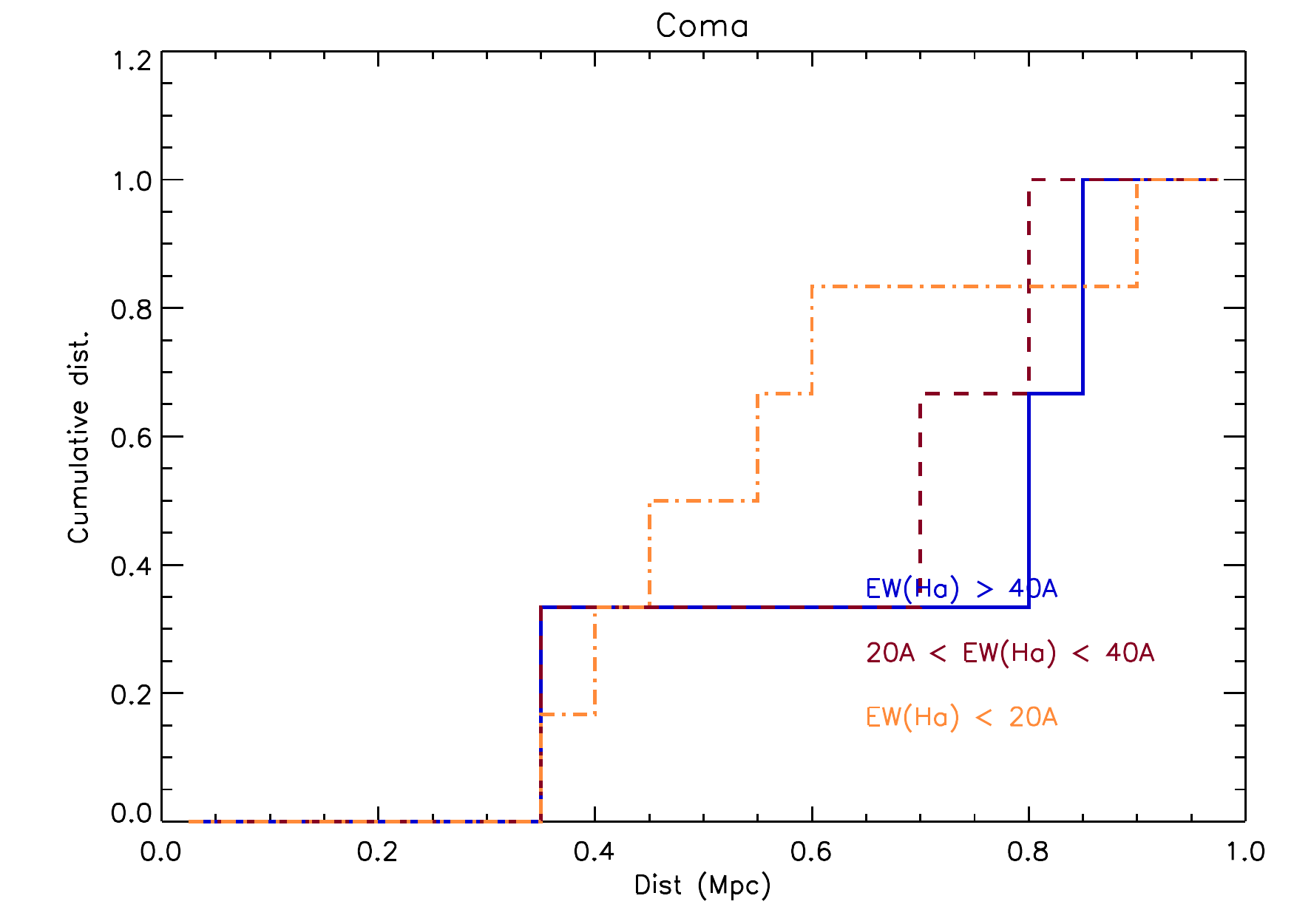}
\includegraphics[width=8cm]{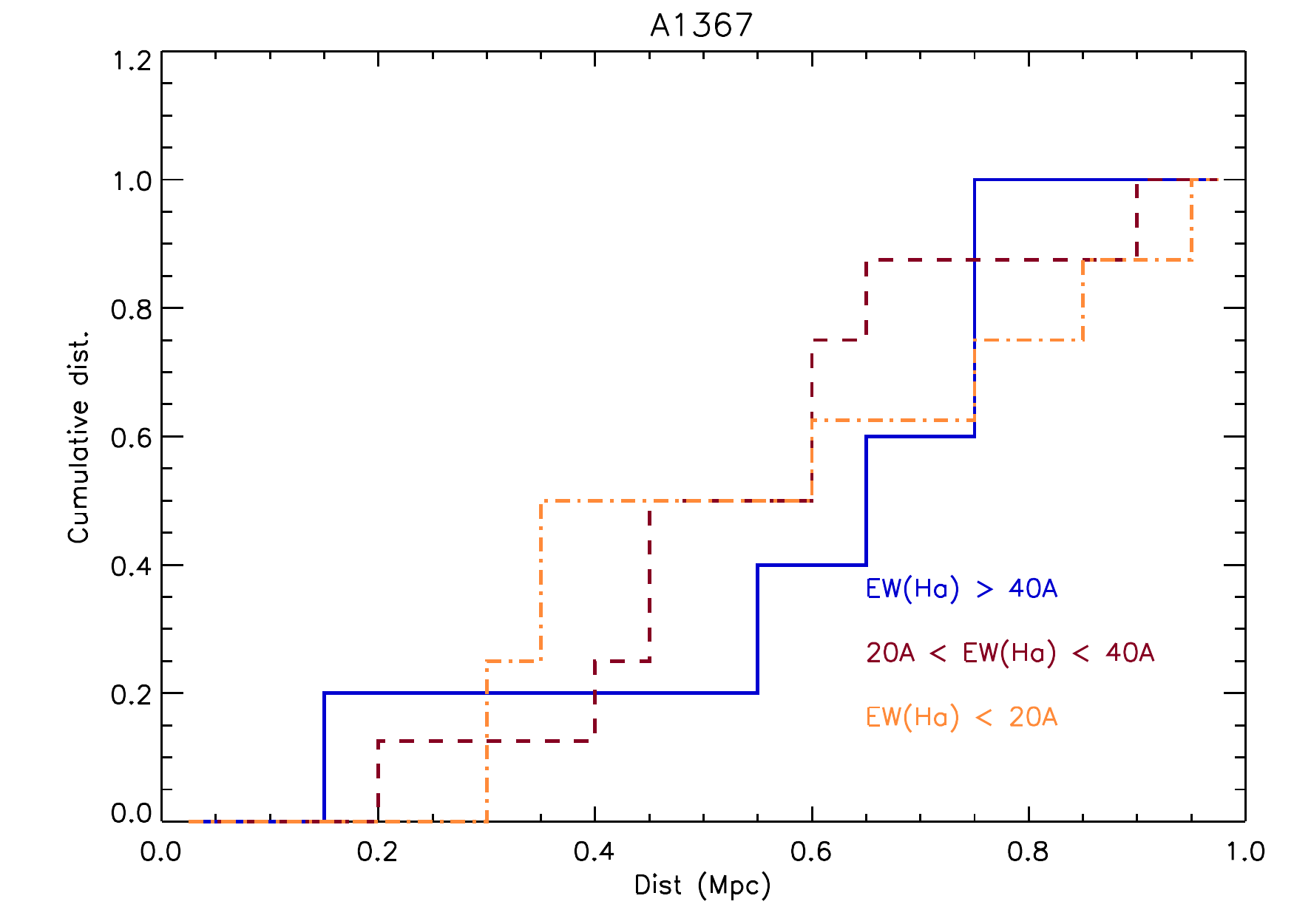}
\includegraphics[width=8cm]{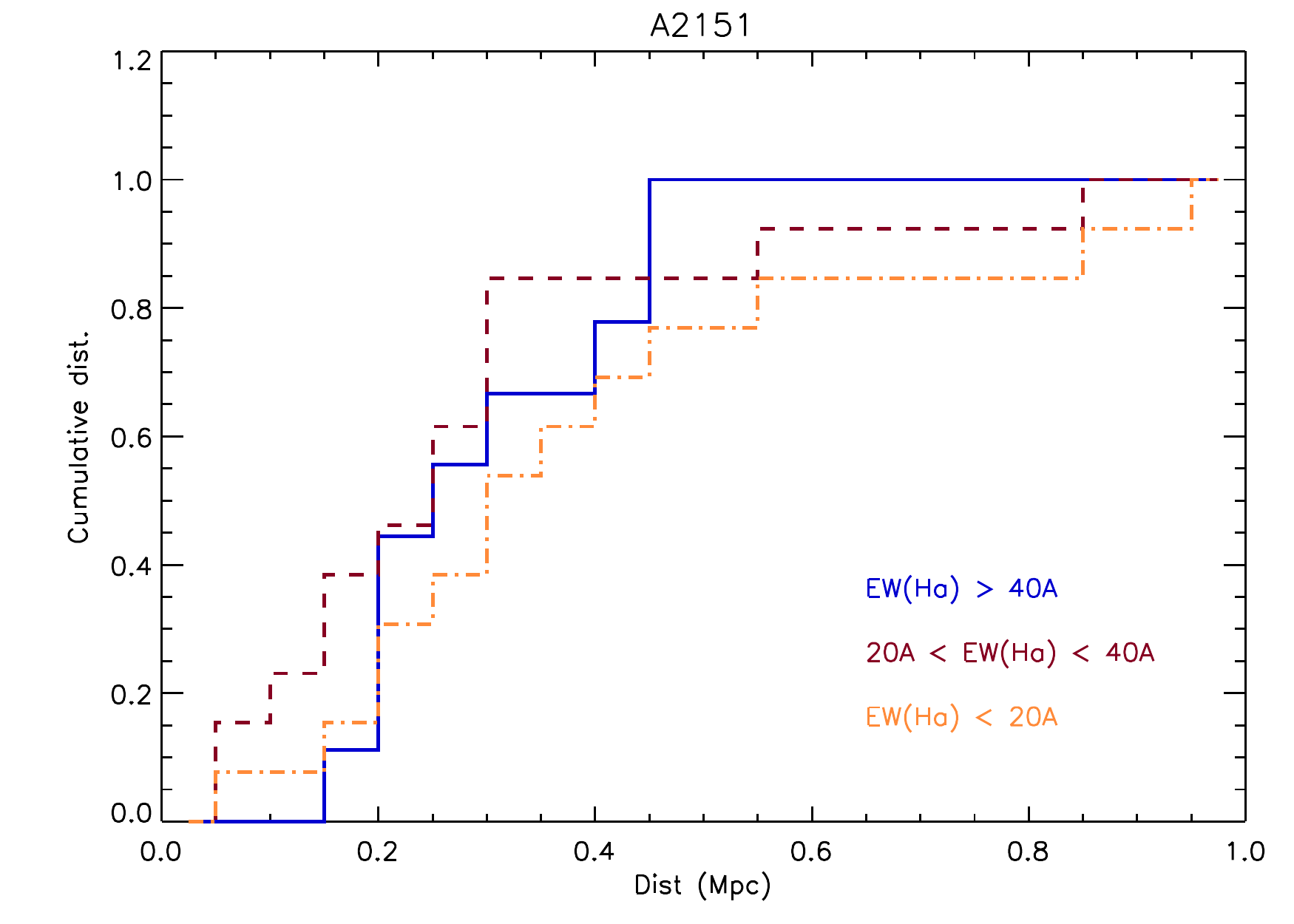}
\caption{Cumulative distribution of the H$\alpha$ emitting galaxies versus the distance to the center of the cluster for Coma, A1367, A2151. Blue, brown and orange lines correspond to galaxies with $EW$(H$\alpha$) $>$ 40\AA, 20\AA\ $<$ $EW$(H$\alpha$) $<$ 40\AA\ and $EW$(H$\alpha$) $<$ 20\AA\ respectively.}
\label{dist_cum}
\end{figure}

\end{document}